\documentstyle[natbib,emulateapj]{article}

\def\etal   {{et~al.\/}}

\slugcomment{To Appear in ApJSupp, Feb 2005}

\begin{document}

\title{SparsePak: A Formatted Fiber Field-Unit for The WIYN Telescope
Bench Spectrograph. II. On-Sky Performance}

\author{Matthew A. Bershady} 

\affil{Department of Astronomy, University of Wisconsin, 475 N Charter
Street, Madison, WI 53706; mab@astro.wisc.edu}

\author{David R. Andersen\altaffilmark{1,2}}

\affil{Department of Astronomy \& Astrophysics, Penn State
University, University Park, PA 16802} 

\author{Marc A. W. Verheijen\altaffilmark{3,4}, Kyle B. Westfall, Steven
M. Crawford} 

\affil{Department of Astronomy, University of Wisconsin,
475 N Charter Street, Madison, WI 53706} 

\author{Rob A. Swaters\altaffilmark{5}}

\affil{Department of Physics \& Astronomy, Johns Hopkins University,
3400 North Charles Street, Baltimore MD 21218; and Space Telescope
Science Institute, 3700 San Martin Drive, Baltimore, MD 21218}

\altaffiltext{1}{CIC Scholar, UW-Madison} 

\altaffiltext{2}{Current Address: Herzberg Institute of Astrophysics,
5071 West Saanich Road, Victoria, BC V9E 2E7, Canada}

\altaffiltext{3}{McKinney Fellow, UW-Madison}

\altaffiltext{4}{Current Address: Kapteyn Astronomical Institute,
Postbus 800, 9700 AV Groningen, The Netherlands}

\altaffiltext{5}{Current Address: Department of Astronomy, University
of Maryland, College Park, MD 20742-2421}

\begin{abstract}

We present a performance analysis of SparsePak and the WIYN Bench
Spectrograph for precision studies of stellar and ionized gas
kinematics of external galaxies. We focus on spectrograph
configurations with echelle and low-order gratings yielding spectral
resolutions of $\sim$10000 between 500-900nm. These configurations are
of general relevance to the spectrograph performance.  Benchmarks
include spectral resolution, sampling, vignetting, scattered light,
and an estimate of the system absolute throughput.  Comparisons are
made to other, existing, fiber feeds on the WIYN Bench
Spectrograph. Vignetting and relative throughput are found to agree
with a geometric model of the optical system. An aperture-correction
protocol for spectrophotometric standard-star calibrations has been
established using independent WIYN imaging data and the unique
capabilities of the SparsePak fiber array. The WIYN
point-spread-function is well-fit by a Moffat profile with a constant
power-law outer slope of index -4.4. We use SparsePak commissioning
data to debunk a long-standing myth concerning sky-subtraction with
fibers: By properly treating the multi-fiber data as a ``long-slit''
it is possible to achieve precision sky subtraction with a
signal-to-noise performance as good or better than conventional
long-slit spectroscopy. No beam-switching is required, and hence the
method is efficient. Finally, we give several examples of science
measurements which SparsePak now makes routine. These include
H$\alpha$ velocity fields of low surface-brightness disks, gas and
stellar velocity-fields of nearly face-on disks, and stellar
absorption-line profiles of galaxy disks at spectral resolutions of
$\sim$24,000.

\end{abstract}

\section{INTRODUCTION}

\subsection{The SparsePak Design}

In Paper I (Bershady \etal\ 2003) we describe the design, construction
and laboratory calibration of SparsePak, a two-dimensional, formatted
fiber array for the WIYN Telescope's Bench Spectrograph.
\footnote{The WIYN Observatory is a joint facility of the University
of Wisconsin-Madison, Indiana University, Yale University, and the
National Optical Astronomy Observatories.}  SparsePak is optimized for
kinematic studies of nearby galaxies, and is designed to be a survey
engine of stellar velocity dispersions in normal galaxy
disks. SparsePak's size and pattern, combined with the versatility of
the Bench Spectrograph, enable a wide range of science programs.

The technical premise for SparsePak was a fiber-optic feed with a
two-dimensional sampling geometry and large light-gathering power for
an existing spectrograph capable of yielding medium spectral
resolutions ($5000<\lambda/\Delta\lambda<20,000$) with large
apertures. Within these constraints, SparsePak's design was optimized
for galaxy kinematic studies to sample a large solid angle with these
performance goals: (a) Velocity dispersions of $\sigma \sim $10 km
s$^{-1}$ are resolvable (for reference, the stellar velocity
dispersion of the old-disk stellar population in the Solar
Neighborhood is roughly 20 km s$^{-1}$, Kuijken \& Gilmore 1989); (b)
such measurements are practical at the low surface-brightness levels
found at several radial disk scale-lengths of nearby, normal spiral
galaxies; and (c) observations are photon-limited at these faint
light-levels -- at or below the continuum sky-brightness.

Meeting these performance goals is difficult. The sky continuum at
550nm yields only $2 \times 10^{-2}$ photons s$^{-1}$ m$^{-2}$
arcsec$^{-2}$ \AA$^{-1}$ at the top of the atmosphere at a dark site
($V = 21.8$ mag arcsec$^{-2}$). Each of SparsePak's 82 fibers subtends
over 17 arcsec$^2$ in solid-angle. This is necessary for achieving
photon-limited performance at medium spectral resolutions on a
4m-class telescope. Still, the demands on the spectrograph are severe:
The spectrograph must be very efficient while achieving high
dispersion and large demagnification. The Bench Spectrograph meets
some, but not all of these requirements.  Since the spectrograph
existed first, its capabilities drove the choice of the SparsePak
fiber size. This, in turn, limits the spatial resolution and
field-of-view of SparsePak, and hence the class of objects for which
SparsePak is well suited.

SparsePak's $72 \times 71$ arcsec sparsely-sampled ``source'' grid
contains 75 of the 82 fibers. In this grid, fibers are spaced
center-to-center by 9.85 arcsec, with the exception of an inner core
of 17 fibers with a 5.6 arcsec spacing. Seven sky-fibers are placed 60
to 90 arcsec away from the center of the grid. This geometry is a
compromise between sampling and field-of-view. The individual fiber
foot-print and grid size sample several radial scale-lengths in normal
spiral galaxies with recession velocities of (roughly) 2,000-10,000 km
s$^{-1}$. Given the hexagonal fiber packing geometry, the grid can be
densely sampled or critically over-sampled with only three pointings.

\subsection{Advantages of SparsePak Bi-Dimensional Spectroscopy for 
Galaxy Kinematics}

SparsePak's two-dimensional fiber geometry provides significant
advantages over long-slit spectroscopic observations of extended
sources for a range of applications. These advantages overcome several
key limitations plaguing studies of galaxy kinematics:

\begin{itemize}

\item {\bf Signal-to-Noise.} Compared to long-slit spectroscopy,
SparsePak samples more area at larger radii (where, for example,
galaxies grow fainter). The linear sampling of long-slit spectroscopy
has been a central problem for the study of disk stellar velocity
fields and dispersions (e.g., Bottema 1997), and has made precision
measurements impractical at radii of 2-3 disk scale-lengths. It is
critical to probe these radii because at these galacto-centric
distances the disk is expected to make a maximum contribution to the
total enclosed galactic mass. In contrast to long-slit observations,
with SparsePak it is possible to co-add fibers within annuli at many
radial intervals. This improves the signal-to-noise in a manner
comparable to measuring surface-photometry in two-dimensional images
(see, for example, Bershady \etal\ 2002 and Verheijen \etal\ 2003).

\item {\bf Position.} Uncertainties in centering and offsets (for
source mapping) can be substantially reduced with SparsePak
observations {\it a posteriori} by reconstructing spatial maps of the
source from the two-dimensional spectral data (see below). This is
either impossible or impractical with long-slit spectroscopy.

\item {\bf Rotation.} Uncertainties in source position angles are
inconsequential for SparsePak observations since all angles are
sampled simultaneously.

\end{itemize}

The uncertainties of centering and position angles (PA) are nagging
problems for optical, long-slit observations of galaxy velocity
fields.  These problems are addressed by Fabry-Perot (FP) observations
of ionized gas, but only for limited samples of galaxies
(e.g. Schommer \etal\ 1993, Beauvais \& Bothun 1999, Barnes \&
Sellwood 2003). For long-slit observations, all random errors in
centering and position angle lead to systematic underestimates of the
true rotation speeds of galaxies. This systematic effect leads to
erroneous, systematic offsets in the Tully-Fisher relation, as derived
from such data. The effects mimic -- and can be misinterpreted as --
an enhancement in the luminosity of a source for a given rotation
speed (mass). Because of the steepness of the Tully-Fisher relation, a
small systematic offset in velocity appear as large offsets in
luminosity (e.g, a 10\% error in rotation speed is equivalent to
$\sim$0.4 mag in luminosity, assuming a typical Tully-Fisher slope of
9-10 in the red--near-infrared; Verheijen 2001). The effects will be
particularly pronounced for galaxies which display significant
non-axisymmetric structure in either their light distribution or
velocity fields. Such galaxies are prevalent among the later-type,
lower-luminosity galaxies in the nearby universe, and appear to
dominate at all luminosities at higher redshifts. Hence, a precise
measurement of the rotation speed is essential for using, for example,
the Tully-Fisher relation as a diagnostic of galaxy luminosity
evolution as a function of mass.

The effects of poor spatial resolution or beam-smearing also degrade
derived kinematic information. HI studies of galaxies can extend
two-dimensional synthesis-maps beyond the optically detectable disks,
but are both expensive in exposure time and limited in spatial
resolution compared to optical observations of ionized gas. FP
measurements are clearly a competitive option if the scientific
premium requires the highest spatial resolution. SparsePak delivers
coarser spatial sampling than most FP instruments, but it still offers
finer sampling than most HI synthesis telescopes, and at much higher
signal-to-noise. Because of its decreased spatial resolution,
SparsePak provides significantly more light gathering power and hence
higher sensitivity and efficiency for low-flux applications. This
advantage is demonstrated in the rapid acquisition of samples for
pilot studies of the ionized gas kinematics of barred spirals
(Courteau \etal\ 2003), low surface-brightness disks (Swaters \etal\
2003), and asymmetric systems (Kannappan \etal\ 2005).

While SparsePak is not a panacea for kinematic studies at all angular
scales, the integral-field approach to spectroscopy is: For extended
sources, high spatial resolution FP or integral-field spectroscopy (IFS) is
superior to long-slit observations since 2-dimensional velocity fields
provide independent spatial information not necessarily gleaned from
photometry. In other words a ``well aligned slit'' may not always be
meaningful. As noted in Paper I the overall efficiency of IFS
vs. stepped, long-slit observations is significantly higher, and free
of spatially-dependent systematic errors due to telescope offset
problems or changing conditions. Unlike centering and PA mismatch, for
SparsePak the problem of spatial scale can be solved a priori by
choosing targets of the appropriate size or working on radial scales
where beam-smearing is unimportant.

As FP observations of ionized-gas velocity fields have come of age
(e.g., Palunas \& Williams 2000, Garrido \etal\ 2002 and references
therein), only one galaxy to date has been similarly mapped in stellar
absorption (Debattista \& Williams 2004). IFS has a greater
combination of simultaneous spectral coverage and spectral resolution
than FP observations. As such, IFS should be superior for stellar
absorption-line kinematic observations of any kind because a
significant range of wavelengths is needed for measuring
cross-correlations. For example, the SAURON IFS instrument has
measured stellar velocity fields for 72 early-type (E-Sa) galaxies,
but only out to one effective radius and at spectral resolutions of
$\sim$1400 (de Zeeuw \etal\ 2002).  The advantage of IFS measurements
is particularly pronounced for obtaining the highest spectral
resolution data, since this requires sampling many weak (hence narrow)
lines. In comparison to FP instruments and SAURON, as of the
completion of this paper, SparsePak has been used to measure over a
dozen stellar velocity and velocity-dispersion fields for late-type
spiral galaxies out to 2-4 radial scale-lengths at spectral
resolutions of $\sim$10,000. This medium spectral-resolution kinematic
mapping out to low surface-brightness is a unique capability of
SparsePak.

\bigskip

This paper reports the on-sky performance of the SparsePak and Bench
Spectrograph. The detailed and systematic nature of the calibration
measurements taken during commissioning allows us to (i) determine if
our design goals are met, and (ii) establish a benchmark of
SparsePak's present capabilities.

We focus on the performance of the Bench Spectrograph for a subset of
gratings relevant to galaxy kinematic studies. Basic performance is
quantified in terms of spectral resolution and throughput, the
trade-offs between the two, and a measure of the absolute throughput
of the Bench Spectrograph. The latter required a determination of the
WIYN stellar point-spread function and the definition of aperture
corrections as a function of internally-estimated seeing conditions.
The final outcome yields a well-calibrated scheme for precision
spectrophotometry with SparsePak.

We use SparsePak data to illustrate superior techniques for
fiber-spectra data analysis.  We give attention to the optimization of
spectral extraction and sky subtraction -- both topics which have
received considerable attention in the literature. We show that
problems with fiber-spectrograph sky subtraction is likely caused by
poorly designed algorithms based on misconceptions about the data. We
argue the results, presented here in the context of SparsePak, will be
of general benefit to a wide range of fiber-fed instruments and
science applications.

Finally, we give examples of commissioning science which highlight the
new capabilities of SparsePak. Four significant results include
stellar and ionized gas velocity fields in nearly face-on galaxies; a
demonstration of the ability to register two-dimensional fiber-data to
broad-band CCD images with sub-arcsecond positioning precision; a
quantitative assessment of the relative merits of different spectral
regions for stellar absorption-line kinematic measurements of systems
with small internal velocities; and estimates of the limiting
performance of SparsePak for stellar-kinematic studies of galaxy
disks.

Commissioning observations and spectrograph configurations are
described in \S 2. The derived system resolution and sampling are
found in \S 3. The total system throughput, spectrograph vignetting,
and fiber-to-fiber variations are presented in \S 4. We examine the
scattered light within the spectrograph in \S5, and define methods of
optimal spectral extraction in the specific context of SparsePak data.
We demonstrate the ability to perform superior sky subtraction without
the use of beam-switching in \S 6.  Examples of early commissioning
science of emission-line and absorption-line galaxy kinematics are
presented in \S 7. We conclude in \S 8 with a summary of the most
significant results. Five Appendices contain spectrograph
characteristics; the system throughput budget; an aperture-correction
protocol for spectrophotometry (including characterization of the WIYN
point-spread-function); a new sky-subtraction algorithm; and an
analysis tool for determining rapidly the kinematic PA of a galaxy.

\section{OBSERVATIONS AND SPECTROGRAPH CONFIGURATIONS}

Verification of SparsePak's performance with the Bench Spectrograph
stem from data obtained during portions of six initial runs. SparsePak
was shipped and installed on the WIYN Telescope during the first week
of May, 2001. First-light and science verification began May 06-07,
2001, immediately after installation. Commissioning proceeded through
the next five runs during May 22-24 2001, June 08-13 2001, two
additional runs in January 2002, and one run in March 2002.

Observations relevant to this paper consist of (i) basic calibration,
i.e., high signal-to-noise (S/N) line-lamp exposures dome-flat
exposures, taken adjacent in time with a suitable set of bias frames
for standard image processing; (ii) on-sky calibration, consisting of
short exposures of bright stars (including spectrophotometric
standards) placed on individual SparsePak fibers; and (iii) long
(sky-limited) exposures of face-on galaxies which filled the SparsePak
grid.  All images were processed and cleaned of detector and cosmic
ray artifacts in a standard way.  Spectral extractions, where needed,
were done with IRAF's ``{\tt dohydra}'' routine.\footnote{IRAF is
distributed by the National Optical Astronomy Observatories, which are
operated by the Association of Universities for Research in Astronomy,
Inc., under cooperative agreement with the National Science
Foundation.}

Using these data we have tested the Bench Spectrograph in 7
configurations suitable for kinematic studies of galaxies. Six of
these configurations use a 316 l/mm (R2) echelle, blazed at 63.4
degrees, in orders 6 and 7 (Ca II near-infrared triplet region near
8600\AA), orders 8 and 9 (H$\alpha$ region near 6650 \AA), and order
11 (in the MgI region near 5100\AA). The seventh configuration uses an
860 l/mm grating, blazed at 30.9 degrees, in second order (near
H$\alpha$). The configurations are detailed in Appendix A.

\section{SPECTRAL RESOLUTION AND SAMPLING}

\subsection{Performance}

We have established SparsePak is able to deliver well-sampled lines at
spectral resolutions of $>$5000 with second order gratings, and
$>$10,000 in on-order echelle configurations.  Spectral resolution is
a combination of dispersion and the detected monochromatic
line-width. The latter depends on optical aberrations and the size and
shape of the spectrograph entrance aperture (here, a series of
circular, 500 $\mu$m apertures). Large demagnification and small
optical aberrations in the Bench spectrograph are some of the
motivations for the SparsePak array: Etendue (the product of area,
solid angle, and system throughput) can be increased by enlarging the
effective slit width without incurring significant degradation in the
spectral resolution. SparsePak's fiber size was set a priori to be
critically sampled in configurations with the largest demagnification,
in the absence of significant aberrations. These assumptions have been
tested.

We use Thorium-Argon and Copper-Argon line-lamp exposures to
characterize the size and shape of monochromatic images and to
determine dispersions for all fibers over the full range of the
detector. We avoided blended or saturated lines, and, where possible,
cross-checked these line-widths and shapes with sky lines for the same
fiber and spectral region. Line-lamp lines are well-approximated by a
Gaussian within several scale-lengths, and have widths comparable to
unresolved night-sky lines from long exposures.

Even with 500$\mu$m diameter fibers the large demagnification yields
small monochromatic full-width at half-maximum (FWHM) in both spatial
and spectral dimensions: $\sim$4 pixels (spatial) and 2.7-3.4 pixels
(spectral), where the latter varies with setup (Table A1, Appendix
A). There is a geometric demagnification of $\sim$3.58 contributing to
the spatial FWHM, and an additional anamorphic factor which varies
with configuration, but yields total spectral demagnifications between
4.75 and 6.34 for the configurations we have considered. Comparable
anamorphic factors exist for our echelle and non-echelle
configurations.\footnote{Anamorphic demagnification is defined as
cos($\alpha$)/cos($\beta$), where $\alpha$ and $\beta$ are
respectively the incident and diffracted angles from the grating
normal. At fixed off-Littrow angle (the camera-collimator angle,
$\theta_{cc} = \alpha - \beta$), demagnification is greatest for
wavelengths where $\alpha$ is maximized and $\beta$ minimized. The
grating equation yields the highest anamorphic factors at the reddest
wavelengths in an order. For smaller grating angles typically used for
low-order gratings, anamorphic factors comparable to the echelle are
achievable because the gratings are used farther off-Littrow
(differences in $\alpha$ and $\beta$ are greater).} Because the
dispersion is higher in the red, and the image quality (monochromatic
FWHM in pixels) is roughly constant or only somewhat worse at both
ends of the spectral range, the best spectral resolution is achieved
within the reddest quartile in wavelength.

The measured spatial and spectral FWHM are less than the predicted,
de-magnified, monochromatic fiber diameter (D) -- as expected. The
precise scaling provides an indication of the level of optical
aberrations. For a uniformly illuminated fiber, and in the absence of
aberrations, the profile's shape will reflect the circular aperture
and yield a profile FWHM of 0.86 D. However, the lines appear
Gaussian. The FWHM for a Gaussian encloses 76.0\% of the total
flux. The corresponding effective slit-width enclosing the same flux
fraction from a circular aperture is 0.646 D. The ratio of column 11
to column 7 of Table A1 in Appendix A yields the monochromatic FWHM =
$(0.80\pm0.05)$D -- intermediate between these two cases.  The
smallest measured FWHM (i.e., the ``best focus'') have values of
$(0.69\pm0.04)$D -- close to the expected value for a Gaussian
profile. The closest agreement is for the spectral widths in the two
cases with the highest spectral demagnification and smallest
line-widths (set-ups for orders 7 \& 9). In the spatial dimension, the
FWHM also is 0.69 D. Hence in both the spatial and spectral dimension
the measured widths of the spectra are close to, but slightly
(10-20\%) smaller than the expected values for a
uniformly-illuminated, circular aperture.

Based on these measured line-widths, we conclude optical
aberrations (including defocus) are below the level of half the FWHM,
or 1.5 pixels (36 $mu$m).  The optical quality of spectrograph is good
relative to the physical size of the entrance apertures and CCD
sampling.

\subsection{High Spectral-Resolution Performance}

It is possible to achieve $>$30\% more than the typical spectral
demagnification and twice the dispersion by working far off-order with
the echelle grating. However, in such configurations it is difficult
to focus the camera over the full range in both spatial and spectral
dimensions; there is strong covariance between spatial and spectral
focus. Further, the high resolution comes at a cost of lowered
throughput because the required large grating angles ($\alpha$) result
in significant grating over-fill and reduced diffraction
efficiency. Compare setups 6 vs 7 and 8 vs 9 for the CaII triplet and
H$\alpha$, respectively, in Table A1. Resolution in the off-order
setups (7 and 9) is a factor of 2-2.5 higher than in the on-order
setups (6 and 8). We estimate from dome-flat data that order 7
throughput is half that of order 6 at comparable central
wavelengths. This is consistent with geometric considerations
concerning the relative grating over-fill in these two
orders.\footnote{The echelle grating is $203\times406$mm (clear
aperture) and the designed collimated beam is 152mm for single
fiber. The total beam footprint at the grating mid-plane is
considerably larger due to fiber FRD and distance from the pupil.}

Nonetheless, when observing at low source light-levels in the red,
where night sky-lines dominate the background, the higher-resolution
modes can deliver superior performance by yielding {\it higher}
spectral resolution while {\it lowering} the background by more than
factors of 4-10 in wavelength regions of interest.

\section{SYSTEM THROUGHPUT}

\subsection{Relative System Throughput}

Flat-field exposures of a quartz-lamp illuminated white spot on the
dome interior are used to determine the relative fiber throughput and
spectrograph vignetting, and to separate these two effects.  The full
2-dimensional CCD image of a flat-field exposure for the echelle
grating (Figure 1) gives a qualitative sense of low frequency
variations present in both spatial and spectral dimensions, and
high-frequency (fiber-to-fiber) variations present in the spatial
dimension. Figure 2 quantifies the relative throughput as a function
of fiber number, normalized for the fibers in the center of the slit
(\# 38-44), for every setup in Table A1. This plot is derived from the
extracted spectra of flat-field images using the IRAF {\tt dohydra}
routine; the values are a mean over the fiber traces. We decompose
this spectrograph slit-function into the low-frequency variations due
to field-dependence of vignetting within the spectrograph, and the
high-frequency variations due to fiber-to-fiber variance in throughput
and/or FRD. Detailed results appear in Appendices A and B.

\subsubsection{Fiber-to-Fiber Variations}

The pattern of fiber-to-fiber throughput variations in Figure 2 are
repeatable at different wavelengths and with different gratings at the
same wavelength. This is seen for DensePak and the Hydra-red cables as
well (Figure 3). The behavior indicates the variations are real, and
wavelength-independent.

To understand what causes these variations, uncharacteristically low-
and high-throughput SparsePak fibers are marked in Figure 2. Of the 7
fibers which also have reliable laboratory throughput measurements,
there is a rough qualitative agreement between their relative
throughput. Referring to Figure 13 of Paper I, fibers 1,3,11, and 22
are consistently low in both plots, while fibers 40, 41, 54, and 70
are at or near the mean. Fibers 19, 37, and 74 are also marked as
relatively low-throughput fibers.  Fiber 37 (and to a lesser extent
fibers 3 and 73) have anomalously low relative throughput, but neither
fibers 3 nor 37 show signs of anomalous FRD (Figure 16, Paper 1).
Differential FRD does not seem to be responsible for low fiber
throughput.

Watson et al. (1994) claim similar variations in fiber throughput with
a multi-fiber spectrograph (FLAIR II) on the 1.2m UK Schmidt are due
to end blemishes. In comparison to SparsePak, DensePak has fewer
fibers with anomalously low throughput (for their slit position), but
a greater dispersion at a given slit position; the Hydra-red cable has
significantly more fibers with anomalously low throughput.  We have
carefully inspected both ends of the SparsePak fiber cable. We find
evidence for blemishes on active fibers, but these do not correlate
with fiber-to-fiber variations in the slit-function. Since FRD does
not appear to be the culprit and there is no evidence for
wavelength-dependence, we can only surmise that these variations are
due to blemishes not readily detectable at the fiber surface.

\subsubsection{Vignetting and Blaze Functions}

To derive the product of the vignetting and blaze function we remove
the high-frequency, fiber-to-fiber variations from the two-dimensional
extracted flat-field spectra via a low-order surface fit. This is
illustrated for the 860 l/mm grating (2nd order) and the echelle
grating (8th order) in Figure 4. Observations of spectrophotometric
standards (\S4.2) reveal flat-field color-terms are $<$3\% over
6500-6850\AA\ using the echelle grating, and therefore are likely to
be minimal over the slightly broader wavelength range observed with
the 860 l/mm grating.

We use these low-frequency maps to calculate the relative mean
throughput across the slit, as well as the relative throughput at the
slit ends for the central, half- and full-wavelength range limits
(columns 13-16, Table A1, Appendix A). Relative vignetting is
significant at the edges (up to a factor of two less light). Low-order
gratings have superior performance in the spatial dimension compared
to the echelle because they are used at smaller camera-grating
distance (see below).  A more significant performance gain for the
low-order gratings is seen in the spectral dimension, largely due to
the blaze function. Echelle-grating setups must be optimized carefully
for the specific wavelength range of interest.

\subsubsection{Geometric Vignetting Model}

We have constructed a geometric model of the spectrograph to match the
observed slit-function. The model is important because it allows us to
place our differential measurements onto an absolute scale. The model
uses the known, clear apertures of the optical components,
obstructions, their physical layout, and the laboratory-measured mean
SparsePak beam profile (Paper I). The low-order curves in Figure 2
represent the models for each of the spectrograph setups; Table A2 in
Appendix B summarizes the vignetting for an on-axis (slit center) and
off-axis (slit end) fiber at the central wavelength of that setup.

The agreement between data and model is excellent. Spatial vignetting
is minimized for the shortest camera-grating distances
(d$_{gc}$). This subtle behavior is recovered in detail by our model.

Without our model one might naively conclude that vignetting losses
(averaged across the slit) are only be 11-18\% (column 13 of Table
A1). The real throughput loss is much larger, taking into account the
on-axis vignetting estimated via the model (column 10, Table A2). We
conclude between 44\% and 77\% of the light is lost through geometric
vignetting in the Bench Spectrograph (not including any transmission,
reflection, or quantum efficiency losses). Clearly there is room for
improvement.

Our model indicates vignetting could be significantly decreased by
placing the spectrograph pupil closer to the grating and camera
objective, or vice-versa.  Currently there is no pupil re-imaging.
Gains for the end-fibers (columns 14-16, Table A1 and column 16, Table
A2) range from factors of 2 to 4.

\subsubsection{FRD Effects}

One subtlety which our model does not match is the slight asymmetry of
the SparsePak vignetting function in Figure 2 about the slit center
(the optical axis). The most likely reason for the asymmetry is due to
the larger FRD for smaller fiber numbers, as shown in Figure 16 of
Paper I, due to the decreasing radius of curvature in the fiber feed
as a function of fiber number. If so, the vignetting profile should
become symmetric about the optical axis for Bench fiber feeds using
smaller fibers since differential FRD due to foot curvature is likely
to be smaller.

To test this hypothesis we have plotted the relative throughput for
SparsePak, DensePak, and the Hydra-red cables for two spectrograph
configurations in Figure 3. These cables have fiber-diameters of
500$\mu$m, 300$\mu$m and 200$\mu$, respectively. Dome-flats for all
cables were measured one after the other with the same spectrograph
configuration. Smooth curves are our models for these specific setups
using the laboratory-measured SparsePak {\it mean} beam profile in all
cases.

Low-frequency throughput variations (the slit-function) are different
for each cable, but the differences are subtle. For example, the
Hydra-red cable, with the smallest (most flexible) fibers shows the
most symmetric slit-function of all the cables. This is consistent
with our hypothesis. DensePak has the most asymmetric slit-function of
all, with extremely low throughput at one end (the top) of the
slit. This may indicate the presence of additional stress for the
top-end fibers (for example, these fibers are at the edge of the array
-- see discussion in \S 6.3 of Paper I), or another source of
cable-specific vignetting, e.g., a blockage or improper placement of
the fiber slit in the fiber-feed toes.

The similarity of the gross vignetting function in Figure 4 for all
cables is consistent with our findings in Paper I that the beam
profiles for the various fiber cables are similar. This, coupled with
the success of our geometric spectrograph model, indicates the
measured beam profiles are correct and our geometric throughput model
is accurate.

\subsection{Absolute System Throughput}

SparsePak's large fibers enable reliable spectrophotometric
calibrations using standard stars. Aperture corrections are relatively
small and well-estimable using the near-integral core of the SparsePak
object grid.  Extended-source spectrophotometric standards exist and
have been used with DensePak and Hydra (e.g., NGC 7000 for H$\alpha$
calibrations, or Jupiter for continuum calibrations; C. Anderson,
private communication). The advantages of spectrophotometric standard
stars are their sky coverage and multi-wavelength calibration.  A
successful measurement near 6700\AA\ of a stellar spectrophotometric
standard allows us to estimate the total system throughput of
telescope, fiber cable and spectrograph combined.

\subsubsection{Aperture Corrections}

SparsePak aperture corrections are described in Appendix C. These
corrections are calibrated as a function of PSF FWHM between 0.45 and
2.5 arcsec based on $R$-band CCD images taken at the same telescope
port. This is suitable for our SparsePak spectrophotometric
calibration measurements. The fraction of light enclosed within a
central fiber is between 97\% and 93\% for the lower and upper
quartile in currently-delivered image quality at WIYN (0.7 and 1.0
arcsec FWHM, respectively); the median value is 96\% at 0.8 arcsec
FWHM. Under very poor conditions of 2 arcsec FWHM, the encircled
energy is $\sim$75\%. Based on the ratios of flux in central and ring
fibers, we estimate we can determine the 3-25\% aperture corrections
within 1-2\% overall.

\subsubsection{Spectrophotometric Observations}

A 600 second observation of the spectrophotometric standard Feige 34
(Massey \etal\ 1988) was made with SparsePak on the night of March 25,
2002 in good conditions at 1.12 airmasses. The spectrograph was
configured in order 8 of the echelle grating (Table A1), covering
6500-6850\AA. The star was centered on fiber \#52, the central fiber
within the source grid. Light is detected in this fiber and in the
surrounding 6 fibers, with a ratio of $0.0344 \pm 0.0003$ for the sum
of these 6 fibers relative to fiber \#52. The light distribution in
the surrounding fibers is azimuthally uniform to $<1$\%, indicating
good centering. Based on the results of Appendix C, the derived
aperture correction is $11.5 \pm 1.5$\%.

\subsubsection{Derived Calibration}

System throughput is calculated using the effective
telescope aperture of 7.986 m$^2$ (17.1\% central
obstruction of the 3.5 m diameter primary by secondary and
baffle), and an extinction coefficient of 0.097, appropriate for
6700\AA\ at KPNO.
% NB: extinction in mags is given by A = dexp( 0.4 * 0.097 * airmass )
There is a small correction of 2.5$\pm 0.5$\% to account for light
lost in the spectrograph CCD focal plane due to the finite extraction
aperture equivalent to $\sim$9 unbinned pixels (see \S 5.2 and \S 5.3;
a comparison of Figures 5 and 9 indicates scattering at 5131\AA\ and
6687\AA\ is comparable). Finally, we converted the throughput estimate
from fiber \#52 to the on-axis and spatially-off axis (slit-edge)
values using the vignetting function defined by dome flats at the
central wavelength near 6687\AA. In this echelle setup the peak
efficiency is slightly redward due to the blaze function. The
efficiency on-axis is 94\% of peak.

Our results yield a mean efficiency of 4.1\% and a peak efficiency of
7.0\% from the top of the atmosphere, discounting light lost outside
of the fiber in the telescope focal-plane and outside the spectral
extraction aperture. Including losses from the atmosphere (1 airmass)
and apertures lowers the values to 3.2 and 5.5\% for mean and peak,
respectively. The mean is over all SparsePak fibers and wavelengths.
The uncertainty is of order a few percent of the estimate, with the
largest contribution from the extinction correction. The peak
efficiency of this echelle configuration compares favorably to the
estimate of 5\% peak efficiency using the 860 l/mm grating and Hydra
cables, quoted in the Hydra Manual. Assuming the relative efficiencies
of the gratings are 50\% and 65\%, respectively, the slight increase
in efficiency is significant. Some of this gain may be due to the
decreased vignetting from the more open design of the SparsePak toes,
the larger SparsePak fibers, and perhaps a more careful treatment of
the aperture corrections. 

Appendix B presents an estimated throughput budget for SparsePak,
spectrograph, telescope plus atmosphere. Our calibration is used to
establish the transmission of the spectrograph camera optics.  A
general discussion is found in the concluding section of the paper.

\section{FIBER SPACING, SCATTERED LIGHT, AND OPTIMAL EXTRACTION}

\subsection{Fiber Spacing}

Based on identifying and tracing apertures with high S/N dome flats,
we find the fiber spacing to be 10.61$\pm$0.16 pixels at 6600-6700\AA\
(each pixel is 24$\mu$m). The spacing uniformity, apparent in Figures
1, 5 and 6 is better than 1.5\%. Because the all refractive camera has
a chromatic focus dependence, so too does the spectrograph
demagnification. We measure a fiber separation of 10.69 pixels at
8675\AA, and 10.42 pixels at 5131\AA. Given the 902$\mu$m
outer-diameter of the stainless-steel micro-tubes used to house fibers
in the slit, and $\leq$ 0.5-1 $\mu$m of glue thickness between
micro-tubes, we derive the spatial demagnification changes from 3.52 to
3.61 between 8700 and 5100 \AA. These values bracket the nominal
ratio of the collimator and camera focal-lengths.

\subsection{Scattered Light}

SparsePak's fiber-to-fiber placement was designed to yield minimum of
waste along the slit, but provide sufficient separation to avoid
significant cross-talk between fibers in a spectrograph where the
coherent internal reflections are small. These attributes are
illustrated in Figure 5. Note the faint arc lines continuing to the
right of the slit. The curvature and blueward shift of this arc
indicates this is a reflected portion of a more inner region of the
slit; the reflection deflects the image in the spatial
dimension. Since the reflection amplitude is $\sim$0.1\%, it will not
be considered further.

A cross-section of a portion of Figure 1, plotted in Figure 6, shows
fibers are well separated relative to the FWHM of their spatial
profiles. A low level of signal is present in the trough between
fibers ($\sim$2\% of peak). The flat-field image was bias-subtracted,
so counts are due to photon flux, and represent scattered light. The
scattering profile evident in the right-half of Figure 6 appears to be
a power-law. We will show this is indeed the case. However, since this
profile is a superposition of the profiles from all fibers, it is not
possible to determine if the scattering properties of the fibers are
uniform. We therefore turn to independent measurements where the fiber
light-profiles are individually accessible.

\subsubsection{Quantitative Measurement}

For quantitative measures of scattering and cross-talk between fibers,
we utilize exposures of single bright stars down individual
fibers. Exposures were sufficiently short (5 sec at 8650\AA\ and 30
sec at 5125\AA) that sky counts are undetected, and hence yield
high-contrast, continuum spectra in single fiber channels. While
filling factors at the fiber inputs are not the same as for
dome-flats, the output far-field pattern -- relevant for the estimate
of cross-talk from adjacent and more distant fibers -- will not depend
on these details of the fiber illumination in the telescope focal
plane. Some cross-talk in the telescope focal plane is present at very
low light levels of 0.1-1\% when using fibers within the SparsePak
source grid. Despite the large fiber size and separation, this
cross-talk is due to the tails of the PSF. Since the 7 sky fibers
offer the most isolated fibers in the telescope focal-plane, these
were used for our primary measurements.

Figure 7 shows the spatial profiles for the two spectral regions
investigated with the echelle grating: 5125\AA\ and 8650\AA.
Scattering at 5125\AA\ is 10 times lower than at 8650\AA\, for reasons
which are currently under investigation. The raw spectra exhibit very
low levels of ghosting far away from the exposed fiber at the 0.03\%
level, but only occur for fibers illuminated near the lower edge of
the slit (i.e., high fiber numbers), closest to the semi-reflective
surface of the optical bench.  There is no discrete scattering
detected at a significant level. The spatial cuts of the light
profiles in Figure 7 can be characterized well by a Gaussian core
going down to 0.04\% or 1\% of their central intensity, respectively
for 5125 \AA\ and 8650 \AA\, followed by a break into a power-law
tail.\footnote{This is reminiscent of the profiles observed for
stellar point-spread-functions observed in direct images, e.g., King
(1971). In this case, certainly, the scattering is not coming from the
atmosphere, but is coming directly from the optical components within
the spectrograph.}

\subsubsection{Coherence}

The spectral signal in the scattered light quickly degrades such that
the scattered signal is a smooth continuum devoid of features found in
the original spectrum. This appears to occur within 6 pixels away from
the central peak. We could check this because our illumination source
is a K1 III giant star (HR 4335) with deep, sharp spectral features,
e.g., the near-infrared CaII triplet. The degradation in features
within the scattered light is due to the fact that the scattering is a
two-dimensional process. This is desirable because the cross-talk
contributes a featureless continuum that does not introduce
high-frequency spectral structure.

\subsection{Optimal Extraction} 

\subsubsection{Random vs Systematic Errors}

Given the presence of scattered light, it is essential to investigate
the trade-offs between different spectral extraction schemes in terms
of random versus systematic errors.  The considerable literature on
this topic (e.g., Marsh 1989, Hynes 2002) mostly focuses on cases not
relevant to SparsePak or Bench Spectrograph data. The IRAF {\tt
dohydra} package offers two options which suffice to span the range of
viable approaches. The first represents a unweighted spectral
extraction within some relative surface-brightness threshold, as
defined by the dome-flat trace. The advantage of this algorithm is
that it is simple. The second algorithm is a weighted extraction based
on the scheme of Horne (1986), designed in the context of long-slit
spectra where the weights are defined from the source spectrum itself,
suitably smoothed or otherwise averaged.  For multi-fiber data, the
dome flat is used to define the weights. In the source photon-limited
regime, the weighted and unweighted schemes are identical. In our case
the ``source'' is the flux through the fiber.  What Horne refers to as
the ``background-limited regime'' is, in our case, the read-noise, or
detector limited regime. In the detector-limited regime the weighted
extraction has superior S/N, particularly if the extraction aperture
is large (i.e. the surface-brightness threshold is low). The relevant
issue however, is to weigh the relative gains in total S/N (decreased
random error) against the inclusion of additional spurious signal from
scattered light (increased systematic error) as the aperture width
is increased.

\subsubsection{Optimal Threshold}

To investigate the trade-offs of random vs systematic errors, we
calculate the normalized surface-brightness profile, integral
``source'' counts, integral ``scattered'' light, and the integral S/N
versus the extraction half-aperture. The latter two integrals are
calculated both for weighted and unweighted extraction using the
measured surface-brightness profiles of the stars. 
Measurements were made in the source-limited regime, from which we
estimate the S/N profiles for a variety of cases from
source-limited to detector-limited regimes.

Figures 8 and 9 show the results for the 5125\AA\ and 8650\AA\
spectral regions. Because focus variations exist, the details of these
plots depend on the location of the spectrum on the CCD, but is
relatively independent of fiber. The Figures show results for two
fibers at their central wavelength. One has above-average throughput
near the edge of the slit; the other with below-average throughput
near the center of the slit. Their behavior is essentially identical.

While the scattering amplitudes are different at 5125\AA\ and 8650\AA,
at both wavelengths the scattered light contributions to a given fiber
continue to be significant for up to 10 fibers distant on either side
of the fiber under consideration.  Further, the half-light radius
(i.e., extraction thresholds of 0.5) defines a nearly optimal
extraction.  Within this aperture 85-90\% of the light is contained,
the S/N peaks at this radius in the read-noise-limited regimes, and
the scattered light contribution is 0.8\% and 10\%, respectively for
5150\AA\ and 8659\AA. This corresponds to extraction apertures of 4
pixels, considerably smaller than the $\sim$10-pixel fiber separation.
At lower extraction thresholds (larger extraction apertures) the S/N
increases by less than 5\%, but only for the source-dominated regime,
while the scattered light contribution continues to rise. At higher
extraction thresholds (smaller extraction apertures), the
scattered-light can be diminished by 20-30\% of its value at the
half-light radius, but the S/N decreases rapidly.

\subsubsection{Weighted vs. Unweighted Extractions}

Figures 8 and 9 also shows that the S/N and scattered light profiles
are improved for the weighted extraction, but only for the
detector-limited case at large extraction apertures, which is
uninteresting. 

Figures 10 and 11 show a broader picture of the trade-offs between
systematic and random error as a function of extraction threshold.  In
these figures, the integrated scattered light and S/N is plotted as a
function of the extraction threshold and position on the detector. The
extraction threshold is normalized to the peak intensity of the fiber
spatial profile. The lower the threshold, the larger the aperture, as
defined in Figures 8 and 9. In Figures 10 and 11, scattered light is
calculated to contain contributions from the nearest 10 pairs of
neighboring fibers, normalized to the total light from the fiber as
determined from large apertures in from the profiles in Figure 7. S/N
is normalized by the peak value for the {\it weighted} extraction for
each of the two regimes: photon-limited and detector-limited. In
general, the first and last 20\% of the recorded spectrum have larger
amounts of scattered light due to degradation in the spatial focus,
while the S/N profile is essentially constant. Greater than 95\% of
peak S/N can be obtained while keeping the total scattered light to
$\sim 1$\% of the signal. Over the entire CCD (at all wavelengths), an
insignificant improvement can be made by considering a weighted
(optimized) extraction.

For these reasons we advocate using an unweighted extraction with a
threshold near the half-maximum level.

\section{SKY SUBTRACTION}

Sky subtraction with fiber-fed spectrographs historically has been a
difficult task; the literature is littered with examples.
There also is an alarming divergence in
the discussion of why fiber-fed spectrographs perform poorly in this
regard. However, some work has shown convincingly that a primary
contribution to systematic errors in the subtraction of spectrally
unresolved sky lines are the field-dependent optical aberrations
present in spectrographs (e.g., Barden \etal\ 1993 for the Mayall 4m
RC spectrograph, and, indirectly, Watson \etal\ 1998 for the 2dF
system). 

Recent literature begins to address this issue of field-dependent
optical aberrations in the context of refined sky-subtraction
algorithms, e.g., Viton \& Milliard (2002) and Kelson (2003). We
concur that the practical difficulty of achieving good sky subtraction
for data from fiber-fed spectrographs has been the use of sky spectra
far from the source spectra in the spectrograph focal plane. This is
something that typically would not be done with multi-slit data. The
basic problem with sky subtraction with fiber-fed spectrographs, then,
is not the fibers nor the spectrographs they feed, but the way in
which the data is processed.  Where our argument differs from others
is in the nature of solution to the sky-subtraction problem, and its
efficiency.

\subsection{Inefficiencies in Observational Methods}

The frequently-used method for optimizing background-limited
observations with fibers is to modify the data-gathering procedure,
namely to adopt a beam-switching strategy. In this observing mode each
fiber alternates in time between sampling sky and source such that a
sky spectrum can be built up out of spectra taken through the same
fiber and spectrograph path as the source observations. This has been
shown (e.g., Barden \etal\ 1993) to improve sky-subtraction
performance.  Nod-and-shuffle (Glazebrook \& Bland-Hawthorn, 2001) is
a significant technical improvement along these same lines. The
advantage of the latter is the time-averaged simultaneous nature of
the source and sky sampling, and scanning of the signal over many
pixels. A compelling argument for why the dominant source of
systematic error with sky subtraction in fiber-fed spectrographs is
due to field-dependent spectrograph optical aberrations is that both
of these techniques dramatically improve the quality of sky
subtraction.

While beam-switching and nod-and-shuffle go a long way to solving the
sky-subtraction problem, they incur a significant (50\%) efficiency
penalty into the data gathering process. Alternatively, a judicious
design and allocation of sky fibers within the spectrograph slit,
coupled with the proper handling of sky subtraction, should permit
{\it shot-noise--limited performance with no penalty in observing
efficiency for a wide variety of programs.} Instead of a solution
involving an inefficient data acquisition strategy, we have found a
less wasteful solution using a better instrument design and approach
to data processing.

\subsection{A New Sky Subtraction Algorithm}

We consider here the problem of subtracting night sky lines and
continuum where the source signal of interest consists of narrow
emission lines. Our solution works equally well for source
signals consisting of well-separated narrow absorption lines.  The
broader issue of subtracting night sky lines from data where the
source signal consists of broad or blended absorption lines is
deferred to future papers.

There are a number of software packages available for reducing
multi-fiber data, but we focus on IRAF's {\tt dohydra} routine and
SparsePak data to illustrate the problem and solution.  The {\tt
dohydra} routine is a complex script calling several independent tasks
which serve to identify on each CCD frame the multi-fiber apertures,
trace them in the spectral dimension, extract an optimally weighted
spectrum for each fiber for matching source, flat, and
wavelength-calibration frames, and finally, gain-correct, wavelength
calibrate and rectify the source spectra. At this tertiary stage, the
sky spectra can be identified, combined, and subtracted from the
individual source spectra. Herein lies the key issue.

The basic problem in sky subtraction arises when the fiber data is not
treated like long-slit or multi-slit data, despite the fact that
fibers are typically fed into a spectrograph in a pseudo long-slit.
Because the optical transfer function (OTF) varies over spectrograph
field angles, the width and shape of spectrally unresolved lines
changes along the (pseudo-)slit. This is illustrated in Figures 12 and
13, which demonstrates the pitfall of using all of the sky fibers to
define a mean sky spectrum. While this example uses Thorium-Argon
line-lamp spectra, this high S/N data of uniformly distributed,
unresolved lines make the point about the performance of subtraction
of strong sky-lines: It is the high-frequency spectral information
that is the dominant problem -- i.e., the sky-lines, not the
continuum. It is these unresolved sources whose profiles change as a
function of field-angle within the spectrograph. For this reason one
would like to have sky fibers placed evenly along the slit, and indeed
this was one of the criteria in our mapping of fibers from the
telescope to spectrograph focal planes (see Paper 1). In this
situation, one can assign a sky fiber or the proper interpolant to
each source fiber. Unfortunately, such a simple algorithm introduces
significant shot-noise since, for any given source, fewer fibers are
being used to measure sky.

The solution is to use the fact that the variation in the OTF is a low
order function of slit position, and is usually symmetric about the
optical axis. All of the sky fibers can be used in concert to
constrain a low-order function for the spatial modulation of the sky
level at each spectral channel (wavelength), in a mode similar to what
is done typically for sky subtraction with long-slit spectra. If the
spectral features of interest do not occur at the same wavelength in
each fiber, then many fibers can be used to fit the ``sky'' at any
particular wavelength. This greatly increases the signal-to-noise of
the sky estimation.

The sky subtraction method we propose takes advantage of situations
where emission lines of interest (i) only occur in a limited number of
spectra or (ii) occur at different wavelengths in different spatial
channels (henceforth fibers) due either to kinematic sub-structure in
the object or a range of radial velocities in multiple objects. In
case (ii), the fiber-to-fiber change in wavelength of the emission
feature should be larger than the line-width. A prerequisite is that
the continuum levels (sky plus object) have been removed and the
spectra are all wavelength calibrated and rectified to the same
dispersion relation. Consequently, the spectra should only contain
high-frequency spectral features (emission or absorption lines from
sky and object).  

A synopsis of the specific sky-subtraction algorithm we propose is
this: (1) rectified fiber spectra are put into a spatially sorted,
two-dimensional format; (2) spectral continuum is subtracted (sky and
source) via low-order polynomial fits to each fiber channel; (3) sky
lines are subtracted using a suitable, clipped mean; (4) continuum is
replaced; and (5) line-subtracted sky fibers are used to subtract sky
continuum from source fibers. The most complicated step is the removal
of the sky lines (3). The detailed algorithm is found in Appendix D.

\subsection{Algorithm Performance}

The first three steps are illustrated in Figure 14 for three sources
carefully chosen to have emission features on strong sky lines, but
with a range of internal-velocity spreads. (The last two
steps are only necessary to provide continuum information for, e.g.,
determining line equivalent-widths.)  For comparison, we have shown
three implementations of step 3, including (i) a straight mean of the
seven sky fibers (fitting a zeroth-order baseline; this is method 3a
in Appendix D); (ii) a second-order fit to the seven sky-fibers, with no
clipping; and (iii) a second-order fit to all 82 fibers, with an
interactive-clipping of fibers (this is method 3c in Appendix D).  A
visual inspection shows the latter implementation is significantly
superior in terms of random error.

This preferred algorithm (Appendix D, item 3[c]: ``iterative clipping
using a wavelength-dependent noise estimate'') works in the regime
when the majority of fibers have source features that are separated
from each other by more than the sources' line-width. For redshift
surveys and high-spectral resolution kinematic studies, this regime is
usually achieved.  (See UGC 7169 and 4256 in Figure 14 as two examples
of the latter.)  In this case there are significant decreases in
random noise by using all of the fibers and {\it there is no need to
assign any fibers to sky.}

When should the algorithm fail? Consider the following two types of
observations: (a) An extended source with no velocity structure or N
sources at the same radial velocity within the spectral
resolution. This fails because all of the source features are aligned
in wavelength -- like the sky. UGC 4499 is a good example of this
case, and indeed a close inspection of the spectrum for UGC 4499 in
Figure 14 reveals a systematic over-subtraction at the bright H$\alpha$
line which straddles a bright sky-line. (b) An extended source with
spatially-resolved velocity structure comparable to the (unresolved)
velocity width for each fiber. This will systematically over or
underestimate the sky level (emission and absorption lines
respectively) and hence modify the profile shapes.

When only a restricted set of the ``sky'' fibers can be used to
estimate the night-sky line emission, as is the case for UGC 4499 in
Figure 14, a straight average of all sky fibers appears to yield a
result superior to fitting a 2nd-order function to these restricted
channels. This is true even though the sky-fiber locations are chosen
carefully to span the spectrograph slit. The 5th
panel from the top in Figure 14 reveals spatial curvature in the
sky-subtracted continuum which is not seen in the 4th panel. The
poorer result of using a low-order polynomial as an interpolant stems
from having insufficient points to constrain the fit, hence resulting
in a noisy fit.

In summary, we find the sky-subtraction performance of multi-fiber
spectra using {\tt dohydra}'s algorithm with SparsePak's sky-fiber
slit-mapping geometry is, to first-order, as good as long-slit
data. Under certain conditions, this simple algorithm can be replaced
with a more aggressive procedure which yields superior
performance. This scheme also can be applied to long-slit data
provided certain conditions are met on the spatial and spectral
distribution of source emission.

\subsection{Discussion}

Four ancillary conclusions stem from the above analysis:

1. The above result concerning the poor, second-order fits to
seven sky-fibers modifies the arguments made by Wyse \& Gilmore (1992)
and by us (Paper I) on the optimal number of sky fibers. Specifically,
the optimum number is probably larger than the analytic formulae cited
in these references. Before designing future fiber systems, the
optimum number of sky fibers should be explored in the context of the
sky-subtraction algorithm presented here.

2. Figures 12 and 13 do not reveal significant fiber-to-fiber
differences in line profiles. There are no apparent high--frequency
variations superimposed on the smoothly-changing mean profile shape
across the slit. This indicates that beam-switching or nod-and-shuffle
will not significantly improve the sky-subtraction performance beyond
the algorithm outlined here, and would increase random errors for an
equal amount of observing time.

3. Our proposed algorithm cannot be applied to multi-slit data in
its current form because slitlets are typically offset in both spatial
and spectral dimensions. Similar wavelengths in different slitlets are
subject to discontinuous ranges of optical aberrations. Hence while
fibers introduce entropy losses via FRD, the ability to use fibers as
light-pipes to remap focal planes may have significant advantages for
efficient, high-performance sky subtraction.

4. The result that the straight average of the sky fibers does a
good job for SparsePak sky-subtraction (e.g., Figure 14) is
surprising. Perhaps the spectrally-unresolved lines are
still well-sampled in the spectral direction due to the large fiber
size; the shape of the spectrally-unresolved lines are dominated by
the demagnified fiber image, and not the field-dependent OTF.
However, one must reconcile the apparent lack of residual structure
seen for real on-sky spectra (Figure 14, 4th panel) with what is seen
for Thorium-Argon calibration-lamp spectra (Figures 12 and 13).

One possibility is that the results of Figures 12 and 13 are not
representative of ``on-sky'' data. It is not known, for example, if
the illumination of the calibration lamps onto the fibers is the same
as the sky and source illumination, nor if this lamp illumination is
constant across the fiber array. However, since the residual pattern
seen in the lamp spectra are continuous across the slit, while the
mapping from the telescope to spectrograph focal plans is not, it is
hard to imagine that this explanation is valid.

Another possibility is that the $<$25\% residuals seen in Figures 12
and 13 occur primarily at low levels in, i.e. the wings of,
emission-lines. Hence the corresponding features in the on-sky data of
Figure 14 are difficult to detect because they are below the level of
the random noise. Indeed the panel 4th from the top does show
low-levels of spatial variation in the sky-line residuals. Therefore
we believe it is warranted to dismiss our concerns about differences
of illumination. Our conjecture remains plausible that SparsePak's
large fibers yield relatively invariant resolution elements, thereby
simplifying sky-subtraction.

\section{EXAMPLES OF COMMISSIONING SCIENCE}

A hall-mark of SparsePak IFS is the ability to achieve spectral
resolutions $\sim$10,000, work at low surface-brightness, and create
kinematic or spectrophotometric maps that can be registered reliably
to imaging data. SparsePak resolution and sensitivity limits enable
the study of gas and stellar kinematics in otherwise unfavorable
geometric projections, such as face-on disks, and create the ability
to probe velocity dispersions in these dynamically cold
systems. Spatial registration of kinematic and photometric properties
is critical for deriving dynamical information. Illustrative examples
are given below.

\subsection{Velocity-Fields at Echelle Resolutions}

Andersen's (2001) study of the photometric and kinematic properties of
face-on disks showed that optical, bi-dimensional spectroscopy with
DensePak (Barden \etal\ 1998) yields efficient determinations of
kinematic inclinations for nearly face-on systems -- in an inclination
regime that has been claimed unmeasurable with radio-synthesis
observations. The precision of these measurements enables, for
example, a Tully-Fisher relation to be measured for galaxies with
inclinations between 15 and 35 degrees (Andersen \& Bershady 2002,
2003). While these studies focused primarily on ``normal'' disks with
Freeman-like central surface-brightness, several sources approach
the low-surface-brightness (LSB) regime. One such source is PGC 56010.
Re-observation with SparsePak reveals the relative merits of this array
compared to DensePak.

The pointing maps for PGC 56010 in Figure 15 show SparsePak signal
detection is much more extensive and complete than for DensePak. Both
arrays were used for the same amount of time and in comparable (good)
conditions. Further quantification is derived from the line profiles
for a single fiber from each array, taken to lie near the same
position (Figure 16). The measured H$\alpha$ flux ratio is 3.34
(SparsePak/DensePak), higher than the expected ratio of 2.78, based on
fiber sizes and assuming a uniform surface-brightness source. The
enhanced SparsePak performance is comparable to the gains noted in
\S4.2, possibly due to the decreased vignetting in the SparsePak fiber
toes.

We also compared the radial light-profile of the spectral continuum
for this source, measured using both arrays.  These profiles, rendered
in units of detected electron per hour per fiber, should be offset by
the factor of their relative area and throughput. The comparison does
not require any two fibers to be spatially coincident, and yields a
scaling consistent with our emission-line measurement.

The net effect of the enhanced SparsePak etendue is seen in Figures 15
and 17: The velocity field, derived rotation curve and estimate of disk
inclination are significantly improved using SparsePak. This success
has lead to further study of LSB systems with SparsePak (Swaters
\etal\ 2003).

Despite SparsePak's sparse sampling, the fiber-packing geometry still
enables the spectra to be sorted for quick assessment of the radial
extent and spectral coherence of the emission-line data, as described
in Appendix E. Such sorting is invaluable for assessing the data
quality at the telescope, or for quickly estimating kinematic position
angles of barred spirals (e.g., Courteau \etal\ 2003).

\subsubsection{Absorption-line vs. emission-line velocity fields}

The velocity fields of disks traced by stars and ionized gas can have
systematic differences due to the fact that the gas is collisional
while stars are not. SparsePak can probe these differences even in
nearly face on systems. NGC 3982 serves as an example, chosen for its
size, high surface-brightness, and extensive existing data (Verheijen
1996). SparsePak footprints for observations in the H$\alpha$ and CaII
triplet spectral regions are shown in Figure 18.

Because the stellar absorption-line observations were only sparsely
sampled, we used the H$\alpha$ data to explore sampling effects on the
derived velocity field. We find the basic shape and position angle
remains unchanged over a factor of three range in sampling (Figure
19). We also find the stellar and gaseous velocity fields are similar,
with nearly identical kinematic PAs, as expected if the gas and stars
are co-planar. However, there is less curvature of the stellar
absorption-line iso-velocity contours. This is evidence for asymmetric
drift, which we quantify elsewhere.

\subsection{Absorption-line Velocity Dispersions at Echelle Resolutions}

As part of the primary SparsePak commissioning science, we undertook a
pilot survey to measure the stellar kinematics of nearly face-on
spiral disks to directly estimate their mass from the amplitude of the
vertical stellar velocity dispersions (Bershady \etal\ 2002; Verheijen
\etal\ 2003, 2004). This survey consists of SparsePak H$\alpha$
observations of several dozen galaxies of suitable size and apparent
inclination; MgI and CaII-triplet observations of a kinematically
regular and suitably inclined subsample; and MgI and CaII-triplet
observations in identical configurations for a library of several
dozen template stars for cross-correlation. In our initial
commissioning run we explored the viability of the high-resolution
order-7 echelle setup for the CaII triplet, which achieves
$\lambda/\Delta\lambda \sim 24,000$. In the remainder of \S 7, we
evaluate data taken with this configuration by comparing them to
similar observations taken on a subsequent run using the
lower-resolution, but higher throughput order-6 echelle setup and
order 11 setup for the MgI region.

\subsubsection{Resolution Performance From Stellar Line Profiles}

We have explored two observational modes for acquiring a spectral
library for application to extended sources.  Stars were observed in
``stare'' mode and by drifting them across a row of fibers. For ``drift''
mode, the illumination typically was uniform to about 15\% for 16
fibers. Because of azimuthal scrambling, drifting stars across the
fiber faces (i) samples all of the fiber modes, and (ii) yields a
time-averaged fiber image which more closely approximates the
near-uniform fiber illumination of extended sources, e.g., galaxies.
A priori, we thought such drifting was important to minimize
slit-illumination systematics in cross-correlation analysis of stellar
templates with galaxy spectra. However, our preliminary inspection has
not revealed significant systematic differences in line-width or shape
between stellar spectra obtained in ``stare'' or ``drift'' modes.

Observations of a K-giant, shown in Figure 20, illustrate one of the
advantages of the MgI region. In the order 6 and 7 spectra the
CaII-triplet lines dominate over a number of other weaker, but
narrower lines (primarily from Fe I). The intrinsic widths, $\sigma$,
of the CaII triplet are 15-35 km s$^{-1}$ in the high resolution
spectrum, and 27-43 km s$^{-1}$ in the lower-resolution spectrum,
where the width correlates with the line-strength.  The MgI lines in
order 11 have comparable $\sigma$'s between 25 and 35 km s$^{-1}$, but
the spectrum in this region is filled with strong, narrow lines.
These narrow (mostly iron) lines have $\sigma$'s of 7-8 km s$^{-1}$
(about 4 pixels) in the high resolution spectrum, and are nearly
unresolved. 

An auto-correlation analysis of these spectra, shown in Figure 21,
reveals a significantly narrower peak with $\sigma = 13.5 $ km
s$^{-1}$ in the MgI region, compared to 31.5 and 34.5 km s$^{-1}$ in
the high- and low-resolution CaII triplet region spectra,
respectively.  The higher-resolution CaII cross-correlation spectrum
has a narrower core than the lower resolution spectrum, but no
narrower than then MgI spectrum.  Hence, despite the strength of the
CaII triplet lines, their large intrinsic width, coupled with a
paucity of narrower lines, makes this region less attractive than the
MgI region for kinematic analysis of systems with intrinsically small
velocity dispersions.

\subsubsection{Galaxy Spectra and Sky Line Emission in the Red}

One advantage of high spectral resolution is the ability to separate
the plague of emission lines dominating the sky background redward of
650 nm. Spectra in Figure 22, corresponding to 1 hour of integration
in the high-resolution, order-7 configuration for the CaII triplet,
illustrates the situation. The galaxy spectrum is from a region close
to the center of NGC 3982, sampled by fiber \#52.  The sky spectrum is
an average over 6 sky fibers from the same exposure (the seventh sky
fiber was contaminated by a faint star). A comparison of the sky
spectrum with Dressler's (1984) seminal paper on the use of the CaII
triplet for galaxy kinematic studies reveals an entirely new
background terrain at spectral resolutions of 24,000. Despite the
remarkable resolution of the sky's molecular bands and atomic
doublets, it is clear that judicious choice of source redshift is
required to keep the CaII triplet lines at 8498, 8542, and 8662 \AA\
free of the sky's strong-lined regions.

\subsection{Source Registration and Continuum Calibration}

Based on an analysis of the spatial distribution of NGC 3982's
spectral continuum relative to the $I$-band light profile from
Verheijen (1996), we estimate the spectrum of fiber \#52 in Figure 20
is centered 2.25 arcsec east and 0.75 arcsec south of the optical
center. Similar offsets were found for the order-6 data. The high
quality of the $\chi^2$-minimization of a SparsePak spectral continuum
map to the $I$-band surface-brightness, shown in Figures 23 and 24,
yields an offset precision of under 0.25 arcsec, and an $I$-band
surface-brightness of 17.75 mag arcsec$^{-2}$ for fiber \#52's
observed continuum in order 7.

Another registration method we have developed convolves the fiber beam
with a CCD image to generate a broad-band continuum map that can be
directly compared with the fiber spectral data, fiber-by-fiber. For
intrinsically axisymmetric systems suffering contamination from, e.g.,
foreground stars, the one-dimensional approach used here may have
advantages; azimuthal averaging provides filtering for the foreground
source contamination. The two-dimensional method is better equipped
to register data for more irregular sources, and has been used
effectively with low-surface-brightness galaxies, e.g., DDO 39
(Swaters \etal\ 2003).

% See Notes.N3982_obs for detailed notes on calculations in this section.

\subsection{Overall Performance}

The above calibration permits several useful calculations. First,
based on the order-7 data, the sky continuum in the 848-868 nm region
is roughly 18.2 mag arcsec$^{2}$ near full-moon (based on the CCD
calibration of the spectral continuum); the total sky background
(continuum plus lines) is about 0.4 mag brighter still. (A separate
pointing, offset by roughly 26 arcmin to a blank region of sky, but
taken immediately following the on-target exposure, yields the same
continuum background levels.) In dark time, based on the order-6 data
in the 845-884 nm region, the sky continuum drops to 19.6 to 20.2 mag
arcsec$^{2}$; the total sky background is roughly 18.7 mag
arcsec$^{2}$.\footnote{Estimated sky levels appear 1.2 to 1.4 mag
brighter than typical $I$-band sky backgrounds in full and new moon,
respectively. This is partly because we observe redward of the nominal
$I$-band, where the sky background is significantly brighter. We
estimate band-pass effects amount to 0.5 and 0.2 mag in the wavelength
regions sampled in orders 6 and 7, respectively (Turnrose 1974).  Some
additional background is due to large zenith distances of 25$^\circ$
and 42$^\circ$, respectively, for order 6 and 7 observations. Massey
\& Foltz (2000) find between 0.3 and 0.5 mag brightening at Kitt Peak
in the $V$-band at zenith distances of $\sim$60$^\circ$
degrees. Assuming more brightening occurs at longer wavelengths (OH
gets stronger), this leaves less than a factor of 2 in increased
sky-brightness in our observations unexplained, with the worst case
being our bright-time observations.  Moon illumination was 100\% for
order-7 observations, but the moon-source distance was over
70$^\circ$. However, it is possible that scattered light off the dome
floor, etc., is a cause of the high observed continuum background
levels.} These values can be compared to the normal-disk central
surface-brightness of 20.2 mag arcsec$^{-2}$ in the $I$-band for a
galaxy of this color. Hence even in dark-time galaxy disks lie below
the sky continuum in the red.

Second, despite the high background continuum level in the order-7
data, within the 4-pixel, unbinned optimum extraction aperture, the
sky spectrum is 1:1 photon-to-detector noise-limited (i.e., photon
shot-noise is equivalent to the detector read-noise). This can be
improved with on-chip binning of pixels, and increased exposure
time. On this basis, we estimate the photon and detector noise
contributions during dark-time for several setups assuming a $2 \times
1$ pixel binning in the spatial direction and a limiting exposure time
of 1 hour. (The spatial binning yields no loss of spectral information
or beam separation; longer exposures suffer from too many cosmic
rays.) Our calculation takes into account the relative dispersion and
spectrograph throughput (Tables A1 and A2), CCD quantum efficiency, and
sky brightness (19.9 mag arcsec$^{-2}$ in the $I$ band and 21.8 mag
arcsec$^{-2}$ in the $V$ band). Of the two CaII-triplet region setups,
the higher-resolution order 7 setup remains 1:1 photon-to-detector
noise-limited, while the lower resolution (order 6) setup has a ratio
of 1.8:1, i.e., the photon shot noise is almost twice as large as the
detector read-noise.  In the MgI-setup, at an intermediate resolution
and improved ($\times1.7$) CCD quantum efficiency, but darker sky, the
ratio is also 1:1. System throughput increases or improved detector
read-noise of at least a factor of 2 are needed to significantly alter
this situation.

Third, these data provide an additional check on the throughput of the
system, which we estimate is 1.2\% from the top of the telescope for
fiber \#52. This result is in close agreement with our throughput
estimates at 6700\AA\ described in \S 4.2 by taking into account (1)
the relative CCD quantum efficiency at 6700\AA\ and 8600\AA\
(0.80:0.47); (2) the relative on-axis vignetting in respective order 8
and 7 setups (0.69:0.41), based on the laboratory-measured SparsePak
fiber exit-beam profile summarized in Table 2; (3) the relative
grating efficiencies used off-blaze, (we estimate a 10\% relative
difference between order 8.41 versus order 6.53 with a standard
echelle grating 5.5 degrees off-Littrow); and (4) the loss from
scattered light (15\%, cf Figures 8 and 9).

With this information in hand, we estimate the exposure time required
to achieve a usable S/N for measuring absorption line-widths in a
``normal'' disk ($\mu_0(I) = 20.2$ mag arcsec$^{-2}$) at a radius of
2.2h$_{\rm R}$. This radius is where the rotation curve is usually
fairly flat (little asymmetric drift) and the maximum disk circular
speed is achieved (Sackett 1997).  We adopt S/N = 15 per resolution
element as a practical limit for measuring reliable line-widths.  We
assume 12 fibers are averaged within a radial bin\footnote{A galaxy
with a scale-length between 10 and 20 arcsec will have between 6 to 18
fibers sampling annulus at R/h$_{\rm R}$ = 2.2.}, and each fiber has
been corrected for projection.  Nearly face-on galaxies are
advantageous here, but in general high S/N velocity-fields from
ionized-gas emission-lines can be used to deproject the stellar
velocity field if the asymmetric drift is zero or otherwise
understood.

The NGC 3982 spectrum in Figure 22 has an apparent continuum S/N of
19.6 per resolution element (2.5 pixels), but the true S/N is actually
somewhat smaller ($\sim$16) due to the presence of correlated noise in
this wavelength-rectified (i.e., re-sampled) spectrum. The latter
value agrees with a first-principles calculation of the expected
signal-to-noise based on the photon shot-noise (object plus sky) and
detector read-noise. Again assuming the same observing conditions as
in the previous calculation of photon-to-detector noise ratios we
estimate 7, 48, and 10.5 hours of total integration are required in
orders 6, 7, and 11 respectively.

The numbers presented in this sub-section in isolation indicate the
order-6 CaII-triplet setup yields superior performance for measurement
of stellar kinematics in extended, low-surface-brightness systems,
while the higher-resolution order-7 setup is suitable only for the
highest surface-brightness systems.  However, the order-11 MgI setup
has lower sky-continuum levels and little contamination from strong
sky-lines, $10\times$-lower scattered light, intrinsically narrower
lines, higher instrumental resolution than the order 6 setup, and
[OIII]$\lambda$5007 for estimating the projected velocities. On
balance, and despite the limited band-width of the Bench
Spectrograph's sampling of this order, the MgI region likely provides
a superior region for stellar kinematic work with SparsePak.

\section{SUMMARY}

We have presented the capabilities of SparsePak and the WIYN Bench
Spectrograph in several configurations relevant for the study of
galaxy kinematics, established procedures for conducting precision
spectrophotometry of extended sources, and quantified and understood
the throughput efficiency of the Bench Spectrograph. We have used
SparsePak to implement significantly improved methods for sky
subtraction. In this regard, performance for this fiber-fed system is
comparable to long-slit, imaging-spectroscopic instruments. Finally,
we have demonstrated SparsePak's on-sky capabilities based on
observations of two, nearly face-on galaxies. Here we summarize each
of these components.

The Bench Spectrograph offers a wide range of possible spectral
resolutions across the visible band-pass, with trade-offs between
resolution and system efficiency. We have focused on the high-end of
the resolution range achievable with SparsePak:
$\Delta\lambda/\lambda\geq5000$. Resolutions between 9,700 and 12,000
are typical with the echelle grating for wavelength between
500-900nm. Resolutions as high as 20,000-24,0000 are possible using
the echelle in off-order (high incidence-angle) configurations.  These
modes are 50\% less efficient due to overfilling the grating and lower
diffraction efficiency. Gain factors in resolution roughly equal loss
factors in throughput. Resolutions of 4,000-6,000 are possible with
2nd-order gratings. Due to the smaller camera-grating distances used
for the low-order gratings, made possible by larger camera-collimator
angles, off-axis vignetting is $\sim$20\% smaller than for typical
echelle configurations. This, combined with higher diffraction
efficiencies make the low-order grating configuration the most
efficient, but at the price of lower spectral resolution. The above
resolutions are specific to SparsePak, but the trade-offs between
spectral resolution and signal-to-noise are generic for the
spectrograph.

In all cases we have explored, the spectral sampling is only 2.5-3.5
pixels (FWHM) even with SparsePak's large fiber diameters due to the
large geometric and anamorphic demagnifications in the spectrograph.
Spatial sampling is $\sim$4 pixels (FWHM). The fiber spacing (roughly
10 pixels) is such that on-chip binning by a factor of 2 can be used
in this dimension without signal degradation. This is useful for
low-light-level applications where detector noise is significant.

The combined absolute throughput of the telescope, SparsePak and
spectrograph has been established, based on measurements of a
spectrophotometric standard star, to be 7\% peak at 670nm, and 4\%
mean (over all used field angles -- fibers and wavelengths).  Our
ability to measure reliably the absolute throughput is due to
SparsePak's large fibers and array geometry. Large fibers result in
small slit-losses for observations of stellar spectrophotometric
standards. The two-dimensional format of the array has allowed us to
develop a second-order, empirical formulation of aperture corrections
with a precision better than 2\%.

The modest system throughput stems in part from the exhibited, strong
spatial vignetting function. We have well-matched this vignetting
function with a geometric model that traces a realistic
(laboratory-measured) fiber-output beam-profile through the optical
system. On this basis we have been able to conclude (a) the vignetting
is due to the lack of proper pupil placement within the spectrograph
and large distances between collimator, grating and camera; and (b)
averaged over all fibers, typically half of the light within the
spectrograph is lost to vignetting at the central wavelength.

It is remarkable to ponder where the photons are lost overall. A
detailed breakdown is documented, and plausibly understood.  The fiber
introduce only a 20\% loss, and are a minor contributor to the overall
budget. By our estimate, half of the light is lost before it gets to
the spectrograph (``top end'' losses due to atmosphere, 3 aluminum
surfaces, and fiber surface losses and internal attenuation); half of
this light is lost in geometric vignetting within the spectrograph
(above); another factor of $\sim2$ decrease comes from filters and
gratings; and a final factor of $\sim2$ decrease comes from
surface-losses on spectrograph optics, dewar window, and CCD quantum
efficiency.  We have grouped these losses together in this way
purposefully. There is little that can be done to any single component
in the first and last groupings to significantly increase the
throughput; attention to improving many components is needed to make
appreciable gains. In contrast, the middle two groupings offer the
possibility of making significant throughput gains by shortening the
collimator focal-length, properly placing the pupil (see Paper I), and
introducing more efficient gratings (e.g., volume-phase holographic
gratings, for which order-blocking filters are generally not
needed). While collimator changes may result in decreased spectral
resolution, the amplitude of the decrease would likely be of order or
less than 30\%, which would still be sufficient for galaxy kinematic
studies. Since photon starvation is the critical limit, the trade of
efficiency for spectral resolution is desirable.  A program to improve
spectrograph throughput along these lines is underway and will be
reported elsewhere.

We have also explored the implications of the fiber spacing for
scattered light and optimal extraction of the spectral signal. The
fibers are regularly spaced and sufficiently well separated for a
clean beam extraction, with some degradation in the far red,
CaII-triplet region at 860nm. At bluer wavelengths, cross-talk is
below 1\% when using extraction apertures capturing $\sim$90\% of the
desired signal. At 860nm, the cross-talk rises to $\sim$10\% for the
same captured signal. For SparsePak data, we find that weighted
extractions offer little gain over an unweighted extraction, and
therefore we advocate the unweighted extraction. An extraction
threshold near 50\% the peak is optimum.

One of the most striking results from the present work is the quality
of the sky subtraction with the SparsePak array. This is due partly
to the even placement of the sky fibers along the SparsePak slit, and
possibly because the large fibers are well resolved and their
monochromatic images are not dominated by the spectrograph OTF. Also
important is our development of a new algorithm for data handling and
fitting of the sky-fibers. The algorithm includes subtraction of
spectral continuum and sky line-emission in separate stages. In the
highest-performance implementation of this algorithm, a low-order
function is fit to all fibers in each spectral channel to model the
effects of field-dependent optical aberrations. By including an
interactive clipping algorithm to remove source flux, we have shown
that in certain cases almost all of the fibers can be used for sky
subtraction, resulting in significant improvement in the S/N of the
data. These cases include any situation where discrete source emission
in the fibers has a large range of apparent Doppler shifts, e.g., in
redshift surveys or galaxy kinematic studies where the spread of
internal velocities is large. Here the performance appears to be as
good as one might expect from beam-switching or nod-and-shuffle, but
our method is twice as efficient.

Two other significant implications stem from our results on sky
subtraction. First, there is a need to revisit the optimum
number of sky fibers for survey work. We conclude more
sky-fibers are needed than traditionally calculated from simple
random-error analysis models because such analyses do not consider the
effects of field-dependent optical aberrations. This is a critical
issue for subtracting unresolved sky-lines. Second,
we have realized the advantage of fiber-fed spectrograph over
slitlet systems for ``multi-object'' spectroscopy.  The ability of the
former to control the mapping of the telescope to spectrograph
focal-planes allows for better -- or at least more efficient --
control over optical aberrations for sky-subtraction purposes. In a
very general sense, this offsets the information loss
introduced by FRD in the design and use of a spectrograph.

We have also given several examples of the types of science for which
SparsePak was designed.  We have shown SparsePak easily generates data
yielding high-resolution emission-line and stellar absorption-line
velocity fields of nearly face-on galaxies. It appears SparsePak has a
higher throughput than DensePak by about 20\% -- above and beyond the
3$\times$ gains from increased fiber solid-angle. A simple tool can be
used to examine such SparsePak data and quickly assess, by eye, the
kinematic position-angle of a galaxy.  SparsePak's sampling geometry
also permits the {\it a posteriori} registration of spectral data to
normal CCD images via analysis of the spatial distribution of the
spectral continuum. The precision of this registration is better than
one-tenth the fiber diameter.

Our primary motivation for building SparsePak, a large-etendue,
two-dimensional fiber array, capable of achieving resolutions
($\sigma$) of 10 km s$^{-1}$ or better, is to measure the stellar
motions and velocity dispersions in galaxy disks.  An analysis of
kinematic data of a K giant star and the high-surface brightness blue
galaxy (NGC 3982) indicates that this is indeed feasible.  To measure
stellar velocity dispersions at radii of 2-3 disk scale-lengths in a
``normal'' disk requires $\sim10$ hours. Given the paucity of such
data, even these lengthy integrations will yield important results
worthy of the effort. Bench Spectrograph throughput improvements will
enable large-scale surveys. Due to the combination of larger intrinsic
line-widths, higher backgrounds, and greater scattered light in the
CaII-triplet region, we conclude the MgI region is preferable for
stellar kinematic studies.  For emission-line galaxies the MgI regions
also offers the [OIII]$\lambda$5007 line within the same spectral
window. Therefore in one setting it is possible to trace stellar and
gaseous velocity fields, and hence the asymmetric drift, as well as
line-of-sight velocity dispersions. Such measurements offer great
opportunities for studying the dynamics and mass distributions of
spiral galaxies.

SparsePak is now a facility instrument available to the public on the
WIYN telescope. A web page on this instrument is maintained at: \\
http://www.astro.wisc.edu/$\sim$mab/research/SparsePak/.

\acknowledgments

We wish to thank C. Corson, D. Bucholtz, S. Buckley, and G. Jacoby for
making SparsePak a reality at WIYN; D. Harmer and C. Harmer for
frequent, expert assistance with the Bench Spectrograph and
consultation on the optical model; and the anonymous referee for
constructive comments. We also thank J. Hoessel and A. Glenn for
allowing us to use WIYN Mini-Mosaic data in advance of
publication. Support for this project is from NSF AST/ATI-9618849,
AST-9970780, AST-0307417 and the UW Grad School.

\clearpage

\begin{appendix}

\section{A. SPECTROGRAPH CONFIGURATIONS AND CHARACTERISTICS WITH SPARSEPAK}

Table A1 summarizes the spectral and spatial sampling, resolution, and
vignetting properties for all setups discussed in this paper. A
dewar-azimuth angle of -0.128$^\circ$ was used for all configurations
associated with reported measurements.  Columns 2-5 list the physical
configuration parameters (standard order-blocking filters are not
included).  Columns 6-12 describe spectral resolution characteristics
(\S3).  Expected spectral demagnification and re-imaged monochromatic
fiber diameter are given in columns 6 and 7, based on the geometric
properties of the spectrograph. Measured central wavelength,
wavelength range, dispersion, sampling (FWHM), and derived spectral
resolution are in columns 8-12, based on line-lamp
exposures. Dispersions are calculated via line-fitting in {\tt
dohydra}, yielding high precision measurements typically better than
one part in 10$^{-4}$. Tabulated values for FWHM and spectral
resolution are for the central fiber at the central wavelength.
``Errors'' on the dispersion, FWHM and spectral resolution map the
range on these values across the detector.  Columns 13-16 describe
vignetting characteristics (\S4), measured from dome-flat
exposures. The measured relative throughput in Column 14 can be
directly compared to the modeled vignetting ratio in Column 16 of
Table A2 in Appendix B.

\section{B. SYSTEM THROUGHPUT BUDGET}

Table A2 presents a break-down of the geometric vignetting produced by
different optical components, as estimated via our model described in
\S4.1.3. Columns 1-5 are repeated from Table A1 for
reference. Columns 6-16 contain the model vignetting values used to
estimate the throughput budget of the spectrograph. The model contains
no vignetting from the toes and filter (appropriate for SparsePak but
{\it not} the other fiber feeds), and no vignetting from the camera
enclosure, which should be minimal or non-existent for the camera
back-distance used in these setups.

Table A3 itemizes a complete throughput budget for the WIYN Bench
Spectrograph and SparsePak cable, starting from the top of the
atmosphere, and compared to observations described in \S4.2. The
spectrograph configuration is for the echelle, order 8, as listed in
Table A1, using the X19 interference filter. The ``On Axis'' budget
corresponds to the spectrograph optical axis and central fiber while
``Off Axis'' refers to the spectrograph light-path for the slit-edge
fiber.

Notes to Table A3 document the source of each estimate.  Components we
have measured directly, and with high confidence, are given a rating
of ``Excellent''; older measurements we did not make but reported in
the Hydra Manual are given a rating of ``Good.'' One exception is the
grating response, where the estimate quality is designated ``Fair''
because the Hydra Manual does not report a measured value for the
actual Bench Spectrograph echelle.  Estimates of aluminum reflectance
are given a ``Fair'' rating because actual measurements are not
available. Camera throughput (T$_{cam}$) is given a ``Poor''
rating because there is no available information on laboratory
measurements or manufacturer specifications on coatings.  We have used
the measured throughput to derive T$_{cam}$ in row 20. Given the
number of optical elements, the derived value of $\sim$75\% is
reasonable. T$_{cam}$ values for on- and off-axis agree to within
5\% -- a confirmation that our geometric model is accurate to within
this margin.

\section{C. WIYN POINT-SPREAD-FUNCTIONS AND SPARSEPAK APERTURE CORRECTIONS}

The stellar profiles observed on the WIYN telescope's Nasmyth imaging
port were analyzed to establish aperture corrections suitable for
spectrophotometric calibration of SparsePak data using stellar
standards. The imaging port has no corrector, and is used by
SparsePak, DensePak, the Mini-Mosaic Camera, and the WIYN Tip-Tilt
Module. SparsePak is suited for spectrophotometric calibration using
standard stars because of its large fibers. In the inner region of the
SparsePak array are 17 contiguous fibers. Three of these are
surrounded by 6 contiguous fibers (\# 31, 47, and 52; refer to Figure
1, Paper I). These three fibers are particularly well-suited for
establishing a spectrophotometric zeropoint because the ring of
surrounding fibers can be used to check for good centering (uniform
ring illumination), and to determine aperture corrections.

The basic aperture correction scheme is to determine what fraction of
the total light is contained within the central fiber. This is a
function of the seeing, which can be estimated from the ratio of the
flux contained within the central fiber to the surrounding fiber
ring. Knowledge is required of the fiber geometry and the detailed
shape of the point-spread function (PSF). The relevant geometry is the
radius of the central fiber (2.35 arcsec) and the inner and outer
radii of the surrounding fiber ring (3.28 and 7.97 arcsec,
respectively). We use empirical curves of growth and data-constrained
model profiles to generate aperture corrections for SparsePak.  We
assume the star is centered within the central fiber, but our analysis
can be generalized to handle decentered cases by using the full,
two-dimensional stellar profile information.

\subsection{C.1. Imaging Data}

Imaging data were taken with the Mini-Mosaic CCD camera ($4096 \times
4096$ pixels, 0.14 arcsec pix$^{-1}$) as part of the WIYN Long-Term
Variability Program. Data analyzed here were taken in the $R$-band
selected from 17 runs over four years (1999-2003). The target fields
consist of distant galaxy clusters, but have ample Galactic field
stars for our purpose. We first identified a suite of 9 images that
spanned a wide range of seeing conditions (Table A4). For each image
we used IRAF's {\tt daophot} to detect and identify stars based on
profile size. The initial lists had to be cleaned of compact galaxies
and cosmic rays. Between 18 and 74 unsaturated stars were chosen per
image in a 3-4 mag range, well separated from neighboring sources, and
far from bad columns and detector edges. Sub-rasters around each
sources were cut out, registered, scaled, and then co-added using a
median filter plus sigma-clipping.  The resulting, combined images are
uncontaminated at a measurable level by outlying sources within the
subraster area of roughly 60 arcsec (diameter).

\subsection{C.2. Empirical Curves of Growth}

Accurate curves of growth out to radii of at least 8 arcsec were
determined from multi-aperture photometry on the combined subrasters.
Sky-levels were fine-tuned to provide flat curves of growth between
radii of 15 and 30 arcsec.  The results of our model fitting (below)
indicate the curves of growth asymptote to within 1-2\% of their total
light between radii of 15 and 30 arcsec. Our argument is circular only
insofar as the true profiles have significantly shallower outer wings
to their profiles than the best-fitting models. King (1971) found a
break in the measured PSF profile slope at roughly 10 arcsec radius,
and at the level of 10$^{-4}$ of the peak value. This level is at, or
just below the level of sensitivity of our data. Nonetheless, since
tertiary spectrophotometric standards (e.g., Massey et al. 1988)
typically use slit apertures between 10 and 30 arscec, the absolute
scale of our curves of growth are commensurate with, and therefore
relevant for reference to, such data.

\subsection{C.3. Model Profile Characterization}

The simplest model which adequately fit the observed PSF was
determined from among three, commonly-used functions.  The simplest
and most commonly used is the Gaussian function, but it is known to be a poor
approximation to the true shape of the observed PSF.  The most complex
is the Lorentzian function (Diego, 1985): $$ I(r) = I_0 \ / \ [ 1 +
(r/r_{s1})^{p(1+r/r_{s2})} ],$$ where $r_{s1} = r_{hw}$ is the radius
at the half-width half-maximum of the surface-brightness profile. At
large radii ($r \gg r_{hw}$) $$ I(r) \propto
(r/r_{hw})^{-p(1+r/r_{s2})}.$$ Since in practice $p$ and $r_{s2}$ are
positive quantities, this implies outer profile slopes steepen with
radius and the enclosed light converges within scale-lengths of
several times $r_{s2}$. The Lorentzian model has twice as many degrees
of freedom as a Gaussian.  A PSF model of intermediate complexity was
introduced by Moffat (1969): $$ I(r) = I_0 \ / \ [ 1 +
(r/r_s)^2]^{q},$$ where $$r_s = r_{hw} \ / \ \sqrt{2^{1/q}-1}.$$ At
large radii $$ I(r) \propto (r/r_{hw})^{-2q}.$$ Unlike the Lorentzian
profile, the outer profile of the Moffat function is a power-law of
index $-2q$; the light converges for $q>1$. In general, $p<2q$, but
for very large $r_{s2}$, $p\sim2q$. These conditions are met
independently in our analysis (cf. column 5 of Tables A5 and A6).

The above three functions were fit to our WIYN Mini-Mosaic
PSFs. Figure 25 illustrates the radial surface brightness profiles,
curves of growth, and the various models for all 9 seeing
cases. Tables A4-A6 summarize the fitting parameters. For reference,
each table repeats the empirically measured FWHM, derived directly
from the light profile.

The classic problem with the Gaussian model is seen in Figure
25: the model profile drops too rapidly compared to the observed
profile. Further, a Gaussian profile only yields adequate fits of the
observed profile core (within the FWHM) if the fits are unweighted
(columns 5 and 7 of Table A4, where {\tt psfmeasure} is and IRAF
routine), or if the fits down-weight the outer profile (e.g., IRAF's
{\tt imexamine} routine, column 6 if Table A4). The properly weighted
fits (column 4, Table A4) yield FWHM that are systematically too
large. Because the Gaussian profile is too steep at large radii, in a
$\chi^2$ sense, a properly-weighted fit will tend to enlarge the FWHM
to compensate. This makes the Gaussian profile a non-robust estimator
of the profile shape and scale.

At the other extreme of model complexity, the Lorentz function
provides acceptable fits except at large radii for the poorer seeing
cases.  Our results differ significantly from those of Diego (1985),
from which we graphically estimate: $$p = 0.080 \ r_{hw} + 2.235,$$
$$r_{s1} = 2.5 \ r_{hw},$$ and $$r_{s2} = 21.25 \ r_{hw} - 5.625.$$
First, $r_{s1} = r_{hw}$, and therefore, while Diego finds a very
tight correlation between these quantities, the scaling is
wrong. Using these nominal equations, typically $I(r_{hw})/I_0 \sim
0.9$ instead of 0.5. Second, we do not see a trend in $r_{s2}$ with
FWHM; we find a much larger range of $r_{s2}$ at small FWHM.  Although
our dynamic range in FWHM is smaller, there is no physical motivation
for a correlation between $r_{s2}$ and the FWHM. Third, we find larger
values of $p$.  The data analyzed here extends down to PSF core-widths
a factor of 2 better than Diego's best-seeing case, but spans only the
best-5th of the full range of his seeing cases. We also fit over a
range of radii between 10-30 scale-lengths, as measured in FWHM. The
fitting range by Diego is not specified.

There is some indication that the Lorentz function provides a better
fit in the core than the Moffat function. For our purposes here, this
is unimportant. The large values of the Lorentzian $r_{s2}$
scale-length for the best seeing cases make the Lorentz function
Moffat-like at the outer radii probed by our data. This is relevant.
Therefore we conclude the Moffat function superior because in general
it provides an acceptable fit with the fewest parameters, and in
particular, it yields a more extended outer profile with constant
slope, consistent with observations.

It appears that the outer profile slope is independent of the seeing,
i.e., {\it the outer slope is independent of the core-width of the
profile}. While there is a range in the ``best-fitting'' Moffat outer
slope ($1.9<q<2.65$), and steeper slopes tend to be found for seeing
cases with larger core-widths, the significance of this result is not
large. We find Moffat profiles with $q \sim 2.0$ provide better fits
to the observed FWHM and outer slope.

\subsection{C.4. Aperture Corrections}

We define G$_1$ to be the flux calibration, namely the fraction of the
total flux of a stellar source contained within the centered, central
fiber. The {\it measured} ratio of the flux within the ring of 6
fibers surrounding the central fiber to the flux within this central
fiber is f$_2$/f$_1$. Table A7 contains the relevant values for G$_1$
and f$_2$/f$_1$ based on the empirical profiles, the best-fitting
Moffat-function models, Moffat functions with $q$ set to 2 and 2.6,
respectively (see Table A5). These different values are plotted in
Figure 26, which shows these different calculations yield very similar
values for G$_1$ versus f$_2$/f$_1$. In contrast, the precision of the
relation between G$_1$ versus the FWHM is considerably lower. This is
because the profile core-width, as estimated by the FWHM, is a poorer
measure of growth curve shape than a shape-index, such as
f$_2$/f$_1$. The observed profiles have systematically higher
f$_2$/f$_1$ for a given G$_1$ when the seeing conditions are good
($<1$ arcsec FWHM), but at a very low ($<1$\%) level. This systematic
is a consequence of the observed profiles being slightly shallower in
the core than the Moffat-function fits. The full range of G$_1$ values
for a given measurement of f$_2$/f$_1$ in Table A7 yields a variance
of 1-2\% in this derived aperture correction for any observed value of
f$_2$/f$_1$ between 0.007 and 0.1, or seeing between 0.5 arcsec and
2.3 arcsec FWHM.

An example of how to apply these aperture corrections is illustrated
in Figure 26. Based on the measured values of f$_2$/f$_1$ = $0.0344
\pm 0.0003$ from SparsePak spectra, graphically this corresponds to a
value of G$_1$, which we take conservatively to be the full range of
models and observed data, or $0.885\pm0.015$.

Similar corrections using the PSFs reported here can be developed for
Densepak when used at the Nasmyth imaging port.  Application to fiber
feeds at the ``wide field'' Nasmyth and Cassegrain ports may be
inappropriate since these ports have additional optics and their
delivered PSF has not been optically characterized at the level
described here.

\section{D. SKY SUBTRACTION ALGORITHM}

The specific sky-subtraction algorithm presented in \S 6 is as
follows:

\begin{enumerate}

\item Wavelength calibrated and rectified multi-fiber spectra are put
into a two-dimensional image format, sorted by position along the
slit. These are standard procedures done, for example, with {\tt
dohydra}.

\item Spectral continuum (from both source and sky) is fitted and
subtracted from each fiber: A 2nd-order polynomial is fit to each
spectrum over a limited spectral range (e.g., 511 pixels) centered on
the emission line of interest. In each `sub-spectrum' of 511 pixels
the emission and absorption lines from the sky and object are removed
by iterative sigma clipping.  The initial clipping begins at $\pm 8
\sigma$, where $\sigma$ is determined for each spectrum. In eight
subsequent iterations, recalculating $\sigma$ each time after the
fitted baseline has been subtracted, the clipping levels are slowly
lowered to $\pm 1.5 \sigma$. This leaves enough of the 511 spectral
pixels to get a good baseline fit. The final function is subtracted
from the spectrum. The spectral range over which this fitting process
can be done can be increased; 511 pixels is somewhat arbitrary, and a
practical limit depends on the degree of the continuum curvature and
the order of the adopted fitting function.

\item After continuum subtraction, sky-lines are subtracted at each
wavelength channel again by fitting a low-order function, but this
time in the spatial dimension.  This step is the most critical in the
process, and subsequently we have experimented with several promising
schemes, which we compare in the following section.

\begin{enumerate}

\item ``Direct sky subtraction'' averages a finite sub-set of 
fibers to construct a sky template.  This is similar to
what is done in, e.g., {\tt dohydra}, except that the averaging is
done {\it after} the spectral continuum is subtracted.

\item ``Single clipping using a wavelength-dependent noise estimate''
first identifies those fibers in each spectral channel that deviate
significantly from the mean, prior to fitting and subtracting a
low-order spatial baseline to the 82 fibers in each spectral
channel. The standard deviation due to shot-noise will vary with
spectral channel as a function of background level. The first step is
to determine the relation between the mean background-level in each
spectral channel, $\mu_\lambda$, and the rms within that spectral
channel, $\sigma_\lambda$. This is done empirically by plotting these
two quantities for all (e.g., 511) spectral channels, and eliminating
outliers. The outliers are spectral channels which contain some fibers
contaminated by source flux. The lowest rms value is associated with
the noise in those spectral channels that are entirely free from sky
and object emission lines, and had the lowest combination of sky and
source continuum prior to their subtraction in the previous step. This
lowest rms value is denoted $\sigma_{min}$. We find empirically that
SparsePak spectral channels lying above the
$\mu_\lambda$-$\sigma_\lambda$ relation by more than $2\sigma_{min}$
are contaminated by source emission. Using the measured
$\mu_\lambda$-$\sigma_\lambda$ relation, fibers are clipped within
each spectral channel if they are above or below the mean by more than
$2\sigma(\mu)$. After clipping, a low-order polynomial is then fit and
subtracted from each spectral channel. This scheme works well in the
case where just a few fibers contain modest amounts of source
emission, i.e., if the source flux does not significantly perturb the
estimate of the mean from the true sky value. In practice we find that
a second-order spatial baseline is optimal for SparsePak data.

\item ``Iterative clipping using a wavelength-dependent noise
estimate'' is an attempt to improve upon the previous scheme by
refining the estimate of $\mu_{\lambda}$ (and hence the appropriate
$\sigma_{\lambda}$) at each wavelength. This is done by repeating the
full process in (b) above after eliminating fibers identified as
outliers from the previous step. The iterative clipping proceeds in a
similar, damped fashion as described in step (1) for subtracting the
spectral continuum.

\end{enumerate}

\item Spectral continuum is added back into image (source plus sky).

\item Sky fibers are used to subtract sky continuum from source fibers.
This is done to obtain the proper source continuum levels.

\end{enumerate}

\section{E. THE SPARSEPAK REPACKING TOOL}

Spatial sorting of SparsePak individual fiber spectra provides
astrophysical insight on source extent, kinematics, and geometry.  An
analysis tool\footnote{The IRAF compatible software may be obtained at
http://www.astro.wisc.edu/$\sim$mab/research/sparsepak/.}, illustrated
in Figure 27, reorders the ``ms'' file produced by {\tt dohydra} into
7 alternate arrangements: one sorting by radius, and 6 by PA. The
latter are incremented by 30 deg -- the natural way in which fibers
can be sorted in a hexagonal packing. Radial sorting is useful for a
quick determination of the extent of source emission and continuum
flux. Sorting by PA can be used to gauge the degree of coherence in
the velocity field (e.g., rotation), and to estimate the kinematic
PA. The PA which shows the least gradient in Doppler shift over the
resorted spatial dimension is roughly 90 deg from the kinematic PA.
From Figure 27 we would estimate the kinematic PA for PGC 56010 is
$60^\circ + 90^\circ = 150^\circ$, with an uncertainty of no more than
half the bin width, or $<15^\circ$. In comparison detailed modeling of
the velocity field data presented in Figures 15-17 yields
$146^\circ\pm1.5^\circ$ degrees. The resorting of multi-fiber spectra
also can be performed quickly on raw data to optimize observations.

\end{appendix}

\clearpage

\begin{figure}
\plotfiddle{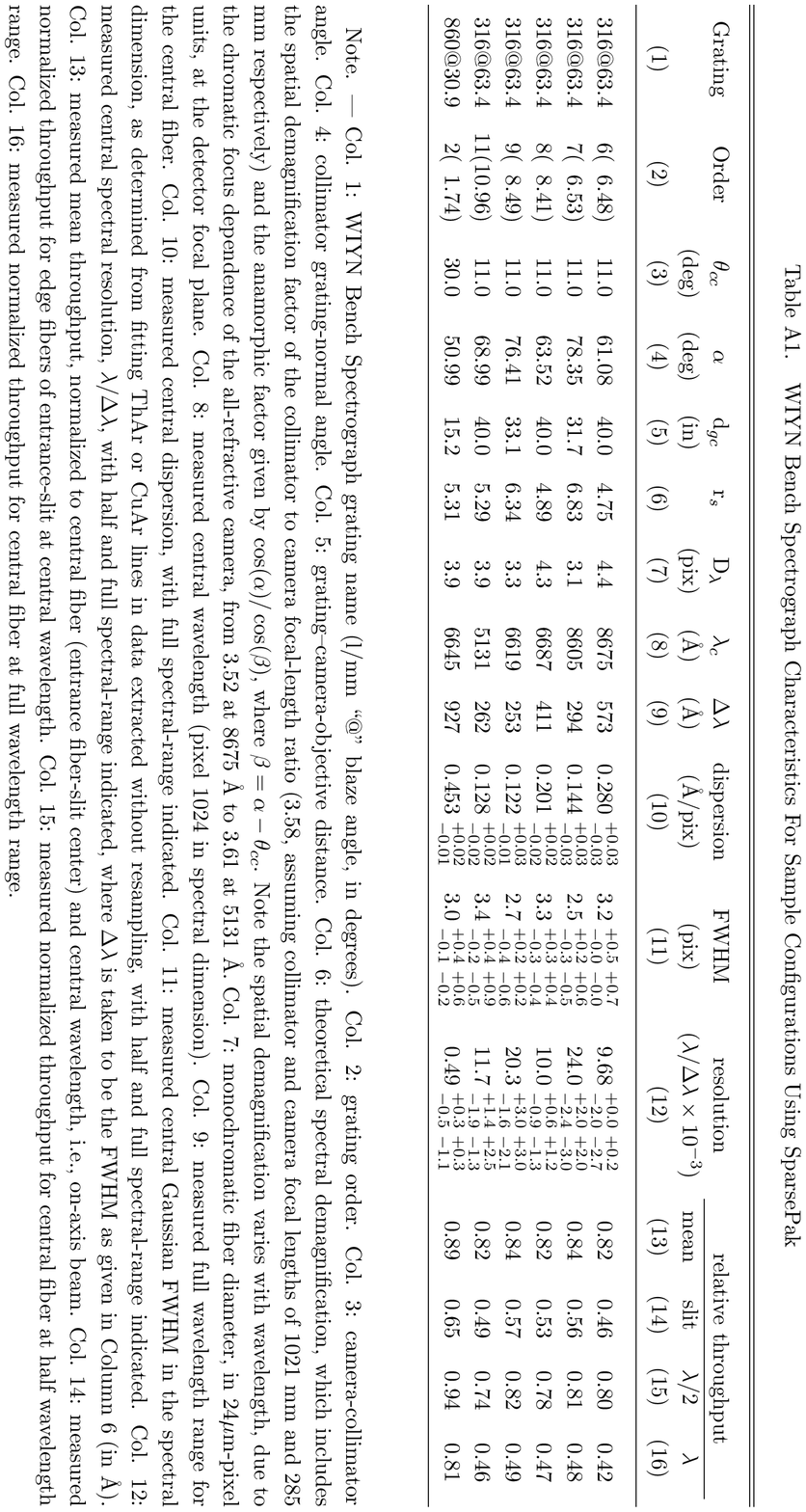}{6in}{-180}{90}{90}{250}{600}
\end{figure}

\clearpage

\begin{figure}
\plotfiddle{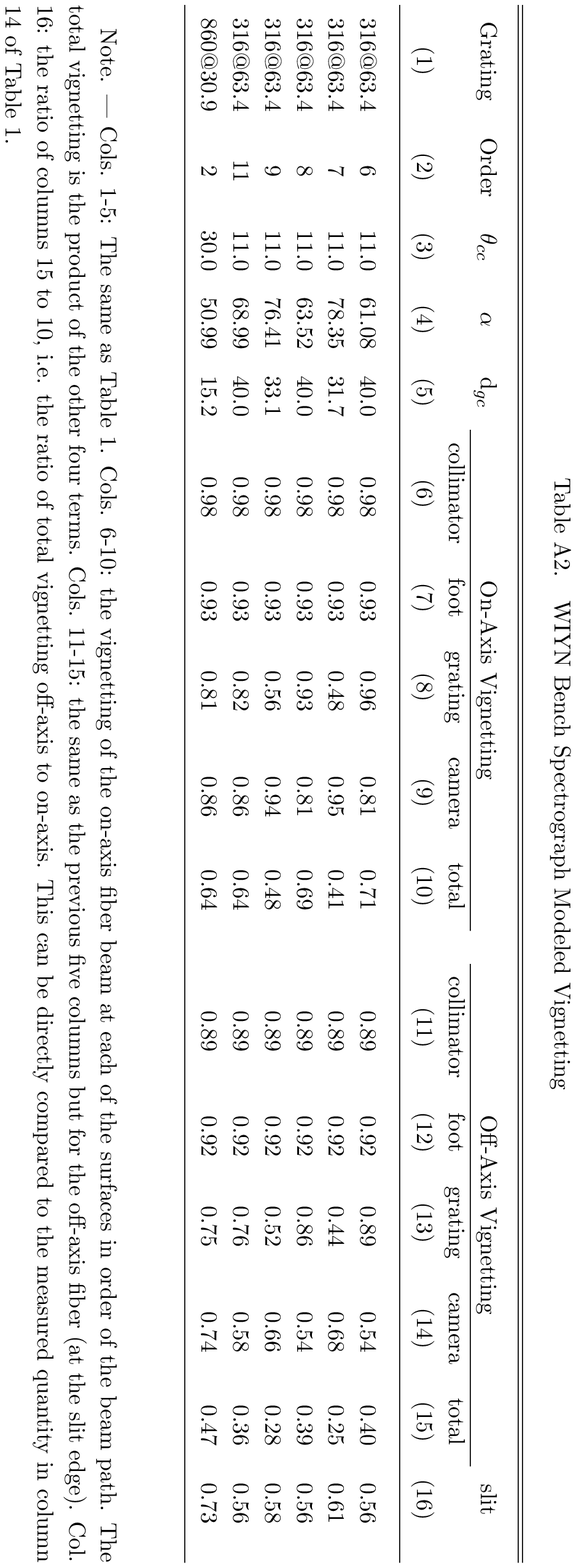}{6in}{-180}{90}{90}{250}{570}
\end{figure}

\clearpage

% TABLE A3.  -------------------------------------------------------------------

\begin{deluxetable}{rlllll}
% \tabletypesize{\scriptsize}
\tablewidth{0pt}
\tablenum{A3}
\tablecaption{WIYN Bench Spectrograph Throughput Budget}
\tablehead{
\colhead{Row \#} &  
\colhead{Component} &  
\colhead{On Axis} & 
\colhead{Off Axis} & 
\colhead{Estimate Quality} & 
\colhead{Note}}
\startdata
\multicolumn{6}{c}{\bf Top-End ``Feed''} \nl \hline
1 & atmospheric transmission & 0.90 & 0.90 & Good      & a \nl
2 & telescope reflectance    & 0.69 & 0.69 & Fair      & b \nl
3 & fiber throughput	       & 0.88 & 0.88 & Excellent & c \nl
4 & fiber ``slit losses''    & 0.91 & 0.91 & Excellent & d \nl
5 & {\it Top-End subtotal}  & 0.50 & 0.50 &           &   \nl\hline
\multicolumn{6}{c}{\bf Spectrograph} \nl \hline
6 & filter transmission      & 0.90 & 0.90 & Good      & e \nl
7 & toes vignetting		 & 1.0  & 1.9  & Excellent & f \nl
8 & collimator reflectance   & 0.89 & 0.89 & Fair      & g \nl 
9 & collimator vignetting    & 0.98 & 0.89 & Excellent & f \nl
10 & pupil obstruction (foot) & 0.93 & 0.92 & Excellent & f \nl
11 & grating efficiency	 & 0.32 & 0.32 & Fair	   & h \nl
12 & grating vignetting	 & 0.93 & 0.86 & Excellent & f \nl
13 & camera transmission	 & T$_{Cam}$ & T$_{Cam}$ & Poor & j \nl
14 & camera vignetting	 & 0.81 & 0.54 & Excellent & f \nl 
15 & ccd window + QE		 & 0.80 & 0.80 & Good      & i \nl
16 & {\it Spectrograph subtotal} & 0.14 T$_{Cam}$ & 0.078 T$_{Cam}$ & &  \nl\hline
\multicolumn{6}{c}{\bf Other} \nl \hline
17 & Spectral Extraction	 & 0.975 & 0.975 & Excellent & k \nl\hline
\multicolumn{6}{c}{\bf Summary} \nl \hline
18 & {\bf Total}              & 0.069 T$_{Cam}$ & 0.038 T$_{Cam}$ & & \nl 
19 & Measurement	         & 0.055 & 0.028 & Excellent & l \nl
20 & T$_{Cam}$                & 0.79	 & 0.74 & & \nl
\enddata
\tablecomments{(a) Extinction for 1.12 airmass at 6687\AA\ at KPNO;
(b) Estimate for 1-2 year old Al in open telescope environment (0.88 per surface);
(c) Laboratory measurement for SparsePak (Paper I);
(d) High-fidelity aperture correction based on the results in Appendix C;
(e) Laboratory measurement from Hydra Manual for X19 filter;
(f) High-fidelity estimate from geometric model of spectrograph (\S4.1.3);
(g) Estimate for 10 year-old Al coating in controlled spectrograph environment;
(h) Peak value from Hydra Manual times theoretical blaze function for 6687\AA\ (0.50$\times$0.63);
(i) Laboratory measurement from Hydra Manual;
(j) Recorded value and witness sample unlocatable.
(k) High-fidelity measurement (\S5).
(l) High-fidelity measurement (\S4.2).
}
\end{deluxetable}

% TABLE A1 -> A4 --------------------------------------------------------------------

\begin{deluxetable}{cccccccc}
% \tabletypesize{\scriptsize}
\tablewidth{0pt}
\tablenum{A4}
% \tablewidth{5.8in}
\tablecaption{WIYN Mini-Mosaic $R$-Band Point-Spread Functions: Gauss
Profile Parameters}
\tablehead{
\colhead{} & 
\colhead{} & 
\colhead{Empirical} & 
\colhead{} &
\colhead{Weight} & 
\colhead{No Weight} & 
\colhead{\tt imexamine} & 
\colhead{\tt psfmeasure} \nl 
\colhead{} & 
\colhead{} & 
\colhead{FWHM} & 
\colhead{} &
\colhead{FWHM} & 
\colhead{FWHM} & 
\colhead{FWHM} & 
\colhead{FWHM} \nl
\colhead{Observation} & 
\colhead{N$_{\rm stars}$} & 
\colhead{(arcsec)} & 
\colhead{} & 
\colhead{(arcsec)} & 
\colhead{(arcsec)} & 
\colhead{(arcsec)} &
\colhead{(arcsec)}
}
\startdata
%     (1)    (2)   (3)     (4)   (5)    (6)    (7)  
10-17-2001 & 49 & 0.45 && 0.92 & 0.55 & 0.49 & 0.59 \nl
12-05-1999 & 47 & 0.69 && 1.20 & 0.75 & 0.69 & 0.79 \nl
11-19-2000 & 63 & 0.83 && 1.19 & 0.87 & 0.82 & 0.95 \nl
11-19-2000 & 74 & 0.97 && 1.38 & 1.05 & 0.96 & 1.16 \nl
03-04-2000 & 18 & 1.35 && 1.56 & 1.45 & 1.34 & 1.63 \nl
11-19-2000 & 53 & 1.43 && 1.95 & 1.50 & 1.38 & 1.71 \nl
11-19-2000 & 64 & 1.61 && 2.01 & 1.73 & 1.59 & 1.98 \nl
11-19-2000 & 19 & 2.12 && 2.56 & 2.20 & 2.06 & 2.44 \nl
11-19-2000 & 24 & 2.50 && 3.00 & 2.61 & 2.45 & 2.88 \nl
\enddata
\end{deluxetable}

% TABLE A2 -> A5 ------------------------------------------------------------------

\def\etal{{\it et~al.}}

\begin{deluxetable}{cccccc}
% \tabletypesize{\scriptsize}
% \tablewidth{3.5in}
\tablewidth{0pt}
\tablenum{A5}
\tablecaption{WIYN Mini-Mosaic $R$-Band Point-Spread
Functions: Moffat Profile Parameters}
\tablehead{
\colhead{Empirical} & 
\colhead{} &
\multicolumn{2}{c}{Best Fit} & 
\colhead{$q=2.0$} & 
\colhead{$q=2.6$} \nl \cline{3-4}
\colhead{FWHM} & 
\colhead{} &
\colhead{FWHM} & 
\colhead{} & 
\colhead{FWHM} & 
\colhead{FWHM} \nl
\colhead{(arcsec)} & 
\colhead{} & 
\colhead{(arcsec)} &
\colhead{$q$} & 
\colhead{(arcsec)} &
\colhead{(arcsec)}
}
\startdata
% (1)    (2)    (3)    (4)    (5)    
0.45 && 0.34 & 1.91 & 0.38 & 0.50  \nl
0.69 && 0.53 & 2.05 & 0.51 & 0.60  \nl
0.83 && 0.68 & 2.06 & 0.66 & 0.73  \nl
0.97 && 0.84 & 1.99 & 0.84 & 0.92  \nl
1.35 && 1.36 & 2.58 & 1.27 & 1.22  \nl
1.43 && 1.30 & 1.97 & 1.32 & 1.37  \nl
1.61 && 1.54 & 2.03 & 1.53 & 1.57  \nl
2.12 && 2.03 & 2.64 & 1.81 & 1.81  \nl
2.50 && 2.39 & 2.50 & 2.13 & 2.19  \nl
\enddata
\end{deluxetable}

% TABLE A3 -> A6 -------------------------------------------------------------------------

\begin{deluxetable}{cccccccc}
% \tabletypesize{\scriptsize}
\tablewidth{0pt}
\tablenum{A6}
% \tablewidth{4.5in}
\tablecaption{WIYN Mini-Mosaic $R$-Band Point-Spread Functions:
Lorentz Profile Parameters}
\tablehead{
\colhead{Empirical} & 
\colhead{} &
\multicolumn{3}{c}{Best Fit} & 
\colhead{} & 
\multicolumn{2}{c}{$r_{s2}=\infty$} \nl \cline{3-5} \cline{7-8}
\colhead{FWHM} & 
\colhead{} &
\colhead{FWHM} & 
\colhead{} &
\colhead{$r_{s2}$} &
\colhead{} &
\colhead{FWHM} & 
\colhead{}  \nl
\colhead{(arcsec)} & 
\colhead{} & 
\colhead{(arcsec)} &
\colhead{$p$} & 
\colhead{(arcsec)} &
\colhead{} & 
\colhead{(arcsec)} &
\colhead{$p$} 
}
\startdata
% (1)    (2)   (3)     (4)          (5)       (6)      
0.45 && 0.58 & 3.58 & $\infty$ && $\cdots$ & $\cdots$ \nl
0.69 && 0.83 & 3.80 & $\infty$ && $\cdots$ & $\cdots$ \nl
0.83 && 0.96 & 3.60 & $\infty$ && $\cdots$ & $\cdots$ \nl
0.97 && 1.16 & 3.52 & $\infty$ && $\cdots$ & $\cdots$ \nl
1.35 && 1.39 & 2.57 & 9.70     && 1.41     & 3.22 \nl
1.43 && 1.46 & 2.90 & 39.37    && 1.56     & 3.29 \nl
1.61 && 1.61 & 2.71 & 29.81    && 1.70     & 3.20 \nl
2.12 && 2.18 & 2.76 & 16.28    && 2.36     & 3.58 \nl
2.50 && 2.57 & 2.86 & 27.91    && 2.71     & 3.55 \nl
\enddata
\end{deluxetable}

% TABLE A4 -> A7 ---------------------------------------------------------------

\begin{deluxetable}{cccccccccccc}
% \tabletypesize{\scriptsize}
\tablewidth{0pt}
\tablenum{A7}
% \tablewidth{6.86in}
\tablecaption{SparsePak Aperture Corrections for Spectrophotometric Calibration}
\tablehead{
\colhead{Empirical} & 
\multicolumn{2}{c}{Empirical} &
\colhead{} & 
\multicolumn{2}{c}{Best Fit} &
\colhead{} & 
\multicolumn{2}{c}{$q=2.0$} &
\colhead{} & 
\multicolumn{2}{c}{$q=2.6$} \nl \cline{2-3} \cline{5-6} \cline{8-9} \cline{11-12}
\colhead{FWHM} &
\colhead{G$_1$} & 
\colhead{f$_2$/f$_1$} & 
\colhead{} & 
\colhead{G$_1$} & 
\colhead{f$_2$/f$_1$} & 
\colhead{} & 
\colhead{G$_1$} & 
\colhead{f$_2$/f$_1$} & 
\colhead{} & 
\colhead{G$_1$} & 
\colhead{f$_2$/f$_1$}
}
\startdata
0.45 & 0.977$\pm$0.032 & 0.007$\pm$0.003 & &  0.982 &  0.005 & &  0.984 &  0.004 & &  0.993 &  0.001 \nl
0.69 & 0.970$\pm$0.018 & 0.013$\pm$0.002 & &  0.974 &  0.007 & &  0.972 &  0.008 & &  0.988 &  0.003 \nl
0.83 & 0.948$\pm$0.011 & 0.017$\pm$0.002 & &  0.959 &  0.011 & &  0.955 &  0.013 & &  0.979 &  0.005 \nl
0.97 & 0.924$\pm$0.008 & 0.025$\pm$0.001 & &  0.927 &  0.021 & &  0.928 &  0.021 & &  0.959 &  0.009 \nl
1.35 & 0.904$\pm$0.010 & 0.032$\pm$0.001 & &  0.913 &  0.023 & &  0.851 &  0.049 & &  0.914 &  0.022 \nl
1.43 & 0.850$\pm$0.003 & 0.052$\pm$0.003 & &  0.837 &  0.055 & &  0.841 &  0.053 & &  0.886 &  0.031 \nl
1.61 & 0.820$\pm$0.003 & 0.066$\pm$0.001 & &  0.800 &  0.071 & &  0.797 &  0.073 & &  0.845 &  0.047 \nl
2.12 & 0.806$\pm$0.003 & 0.066$\pm$0.001 & &  0.794 &  0.069 & &  0.737 &  0.105 & &  0.791 &  0.071 \nl
2.50 & 0.722$\pm$0.003 & 0.108$\pm$0.001 & &  0.700 &  0.126 & &  0.668 &  0.150 & &  0.704 &  0.122 \nl
\enddata
\end{deluxetable}

\clearpage

\begin{figure}
\figurenum{1}
% \epsscale{1.0}
% \plotone{f1.epsi}
\plotfiddle{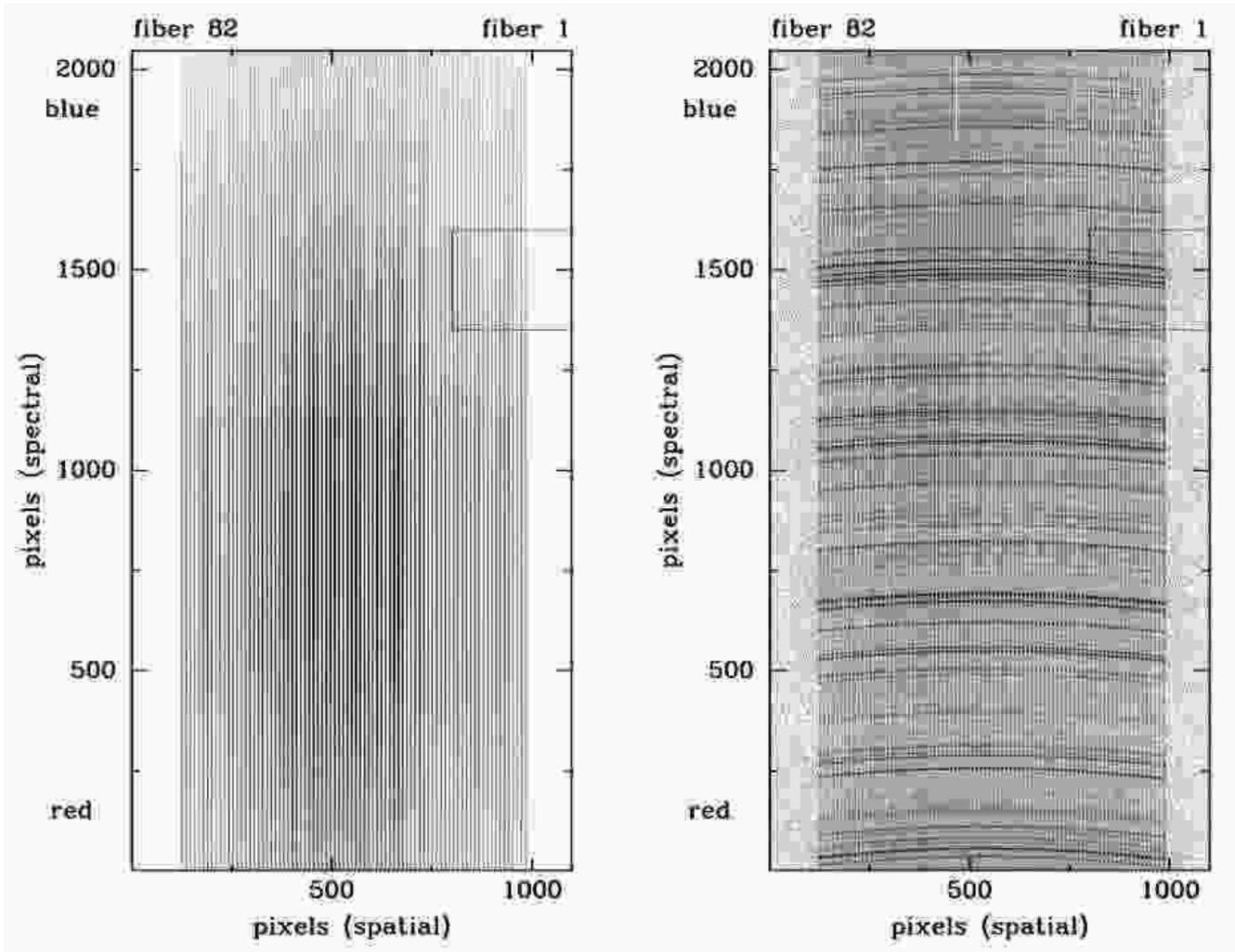}{6in}{0}{90}{90}{-280}{25}
\caption{SparsePak dome-flat (left) and Thorium-Argon arc-lamp (right)
CCD images, bias and over-scan subtracted, and cleaned of cosmic
rays. These data are for the same echelle, 8th order setup centered at
6687\AA\ (Table A1), which is significantly off-blaze.  The small box
in the upper-right of each image shows the extracted region enlarged
in Figures 5 and 6.}
\end{figure}

\clearpage

\begin{figure}
\figurenum{2}
% \epsscale{0.8}
% \plotone{f2.eps}
\plotfiddle{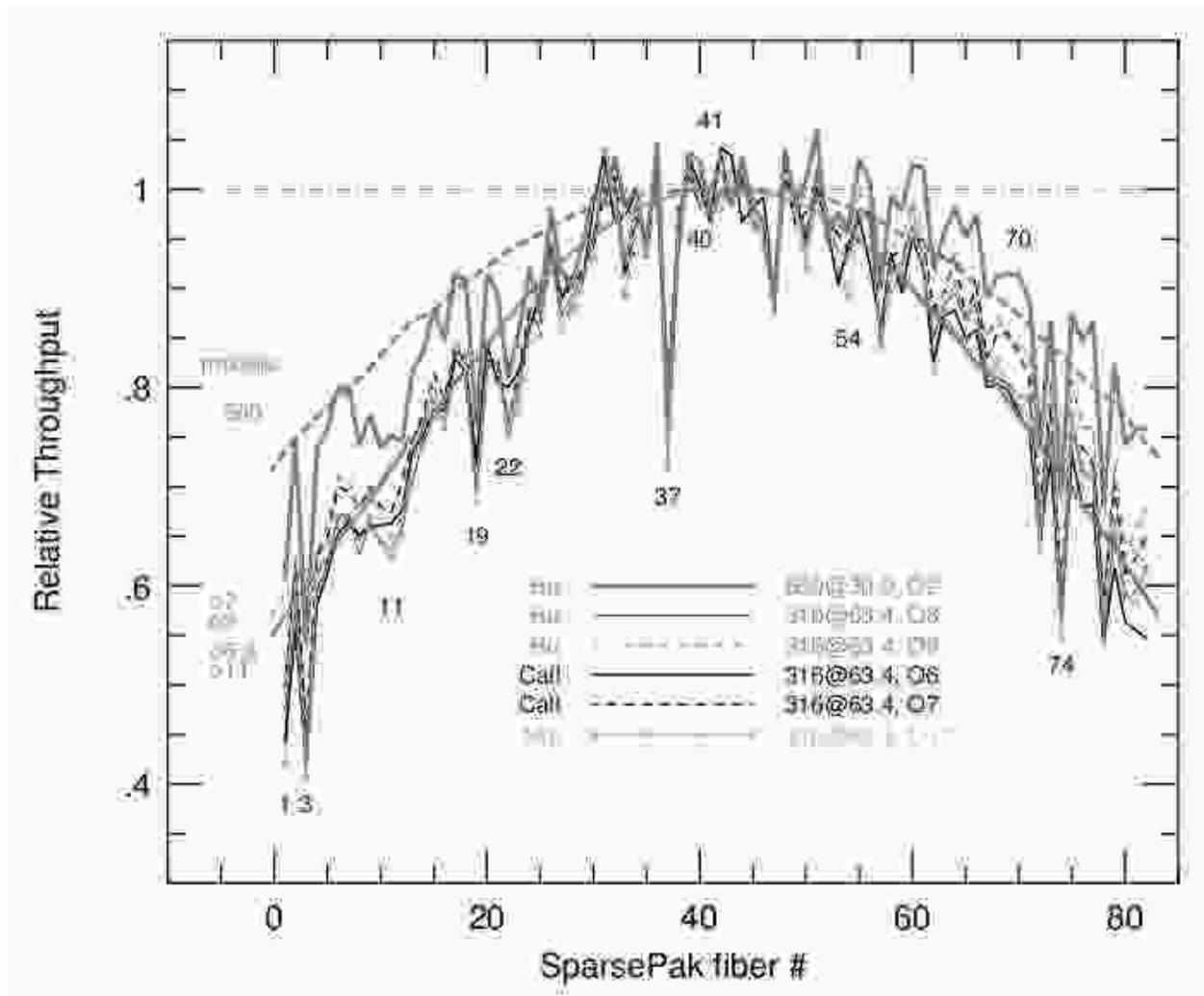}{5in}{0}{85}{85}{-270}{0}
\caption{Relative fiber throughput as a function of fiber number (slit
position) for SparsePak and the Bench Spectrograph, as described in
the text. The horizontal, dotted line at unity is for
reference. Lines for the 6 observed setups are indicated in the figure
key, and can be matched to Table 1. Smooth curves represent the
modeled vignetting profile (see text). The top, heavy, dashed curve is
the model for the 860 l/mm grating setup (order 2); the middle, light,
dotted curve are the models for the echelle orders 7 and 9; the
bottom, light, solid curves are for the echelle orders 6, 8 and 11
(they are indistinguishable). Eleven fibers with uncharacteristically
low or high relative throuhgput are marked, seven of
which have reliable lab throughput measurements from Paper I.
Several important results can be inferred from this plot: Overall, the
model is an excellent match to the data.
The observed vignetting is slightly asymmetric, consistent with the
lower-numbered fibers suffering from more FRD (see Paper 1); this is a
second-order effect which is not in the model.
The fiber-to-fiber variations are consistent between setups at
different wavelengths (orders) for the same (echelle) grating, and at
the same wavelength (H$\alpha$) for different gratings; the variations
are real, and in general the fibers have consistent
wavelength-dependent throughput.
The low-order grating exhibits less spatial vignetting
than the echelle grating due to the 2.6$\times$ smaller grating-camera
distance. Orders 7 and 9 are also slightly less vignetted than order 6
and 8 because the former setups used a 20\% smaller grating-camera
distance.}
\end{figure}

\clearpage

\begin{figure}
\figurenum{3}
% \epsscale{0.6}
% \plotone{f3xx.ps}
\plotfiddle{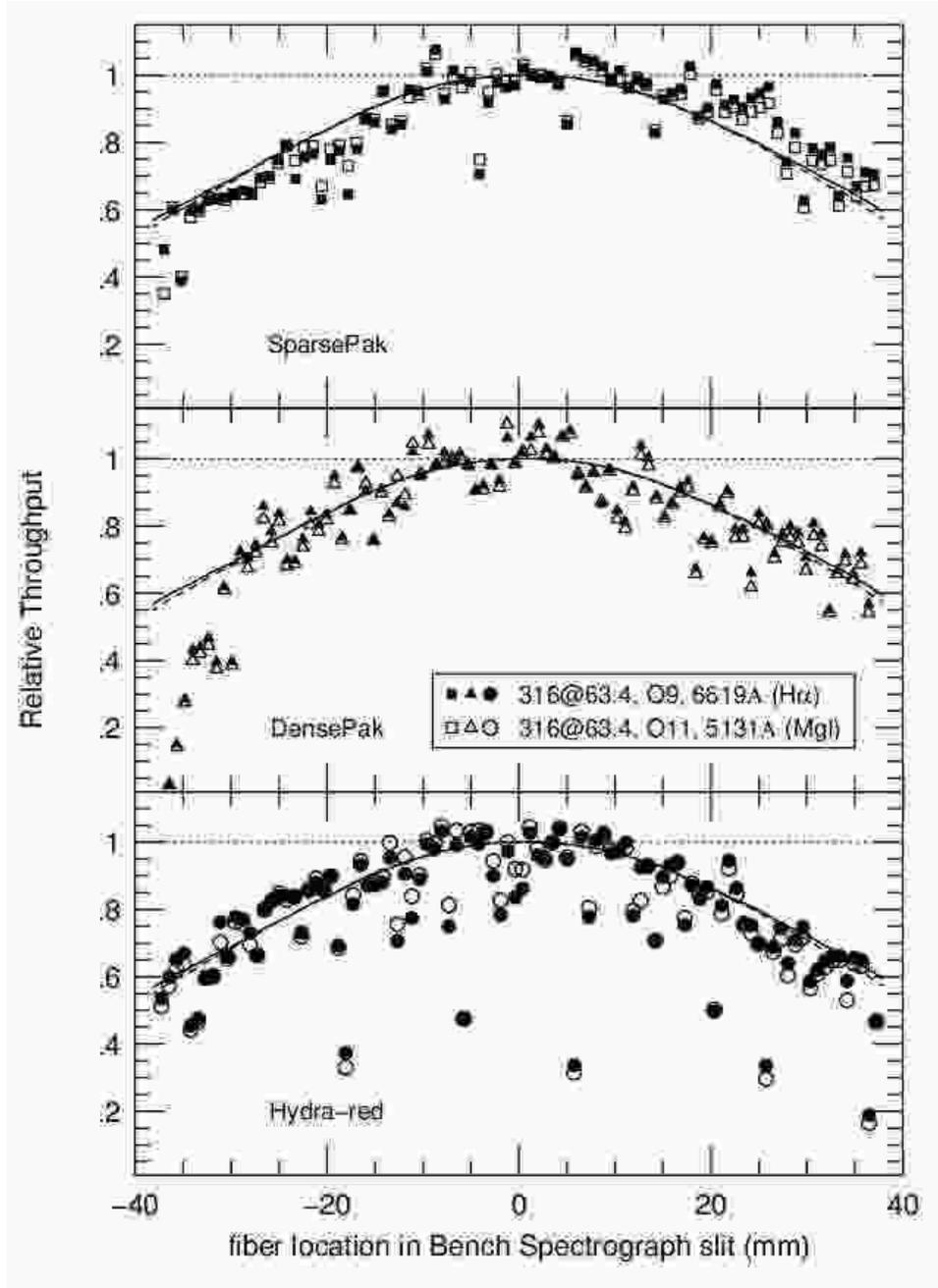}{6in}{0}{65}{65}{-200}{0}
\caption{Relative fiber throughput as a function of slit position
for SparsePak, DensePak, and Hydra-red cables (500$\mu$m, 300$\mu$m
and 200$\mu$ diameter fibers, respectively) and the Bench Spectrograph
for 2 configurations, specified in the key. These were measured
on-telescope and analyzed in the same way as the data presented in
Figure 4. These data represent the slit-function of the spectrograph
convolved with random and systematic variations in the fiber
throughput and/or FRD. The spectrograph configuration for each set up
(orders 9 and 11) were identical for all cables, i.e., the cables were
changed while leaving the spectrograph untouched. Horizontal dotted
lines at unity are for reference. The smooth, solid and dashed curves
represent the modeled vignetting profile for the order 9 and 11
setups, respectively (see text). To first order the vignetting
profiles are identical for all cables, while differences exist in
fiber-to-fiber variations. Second-order variations are seen in the
(low-frequency) pattern of the effective slit function, presumably due
to different dependencies between cables of FRD on slit
position.}
\end{figure}

\clearpage

\begin{figure}
\figurenum{4}
% \epsscale{0.85}
% \plotone{f4xx.ps}
\plotfiddle{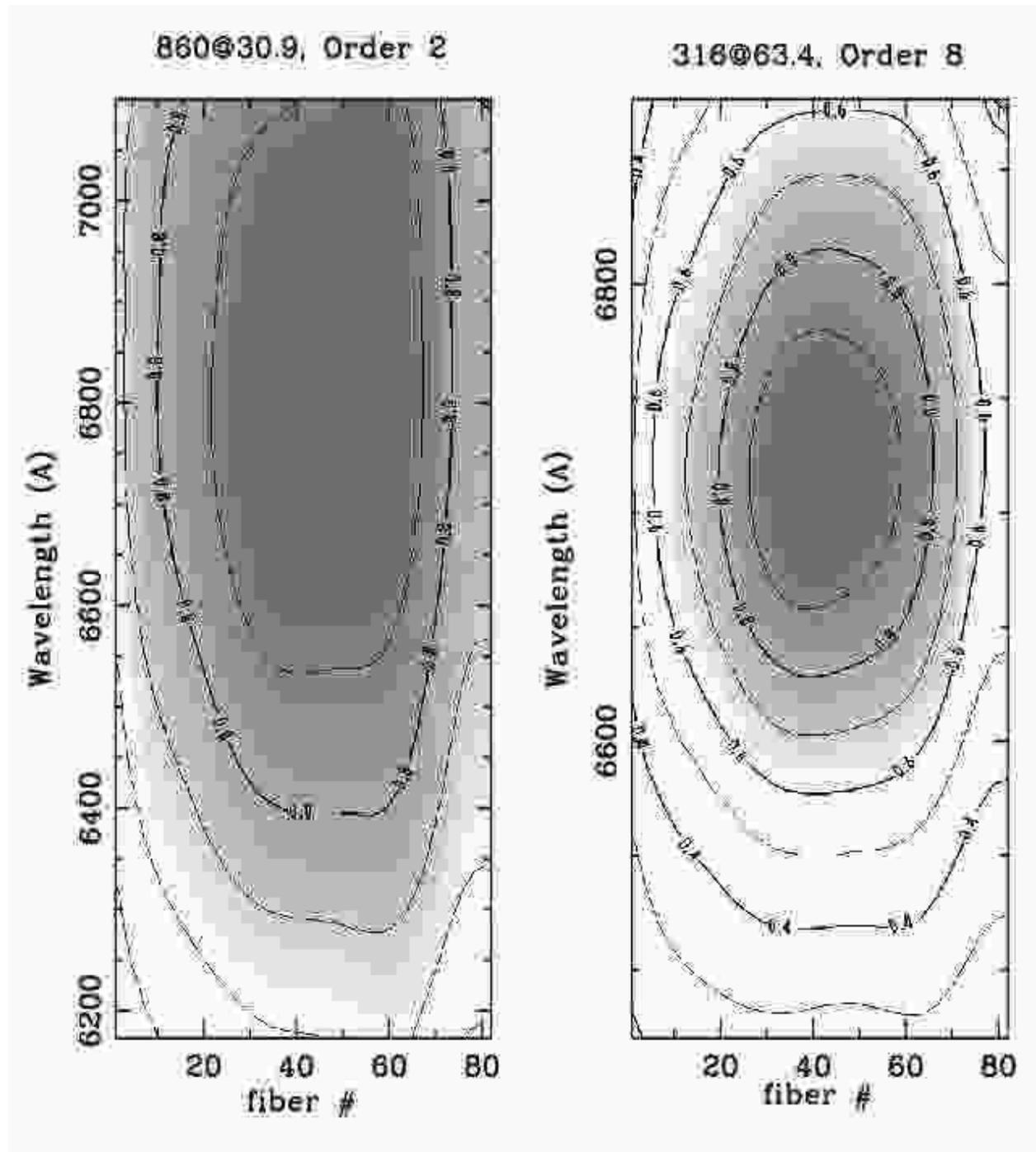}{6in}{0}{85}{85}{-260}{-10}
\caption{Smooth component of spatial and spectral vignetting
function convolved with grating response for the low-order 860 l/mm
grating and the 316 l/mm echelle, both used near rest-frame
H$\alpha$. Fiber-to-fiber variations have been removed by fitting a
low-order spline to the extracted flat-field spectra. No attempt has
been made to correct for color-terms in the lamp spectrum -- a small
effect over the small wavelength range covered. Contours are at 10\%
intervals, normalized to the peak throughput. Both the grating blaze
function (spectral dimension) and the shorter grating-camera distance
(spectral and spatial dimension) serve to provide more uniform
throughput for the low-order grating.}
\end{figure}

\clearpage

\begin{figure}
\figurenum{5}
% \epsscale{0.7}
% \plotone{f5xx.ps}
\plotfiddle{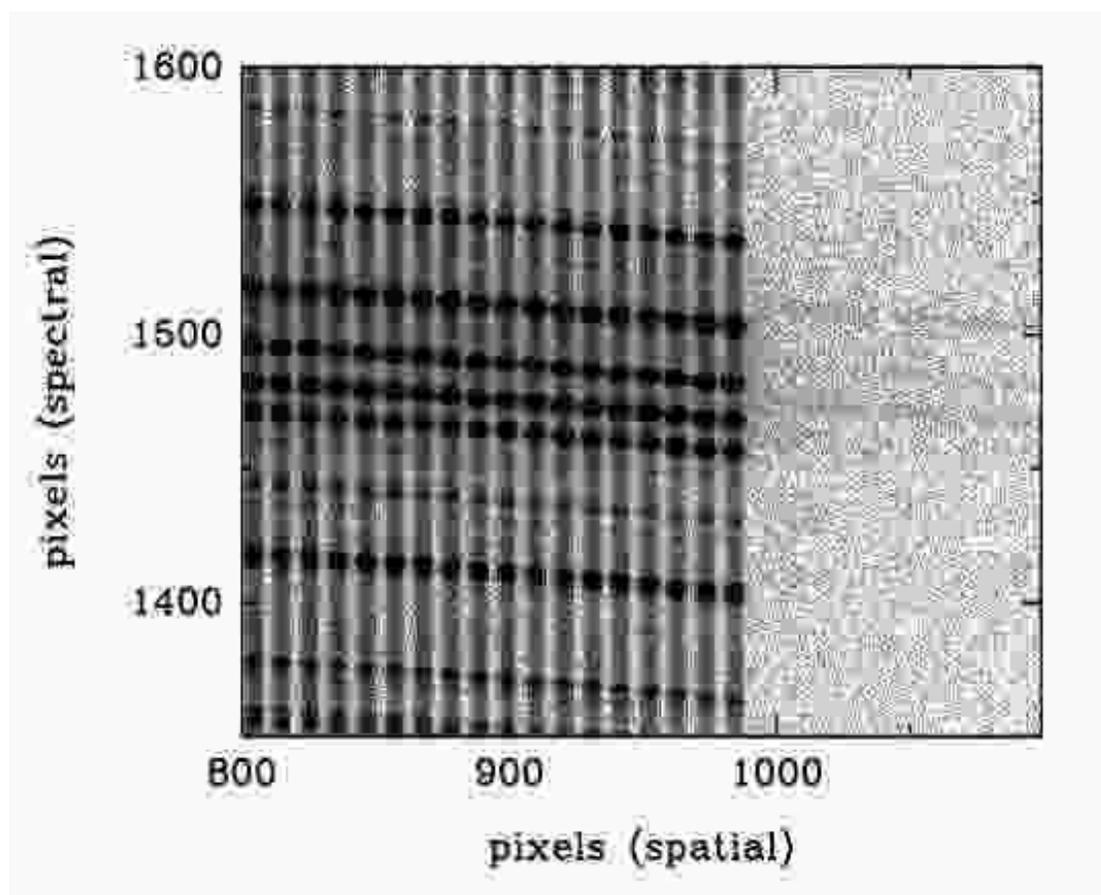}{5in}{0}{75}{75}{-250}{0}
\caption{SparsePak Thorium-Argon arc lamp CCD image for the
region marked in Figure 1. Note the shifted reflection which appears
to the right of the end of the fiber slit. The amplitude of this
reflected signal is $\sim$0.1\% of the primary signal.}
\end{figure}

\clearpage

\begin{figure}
\figurenum{6}
% \epsscale{0.7}
% \plotone{f6xx.ps}
\plotfiddle{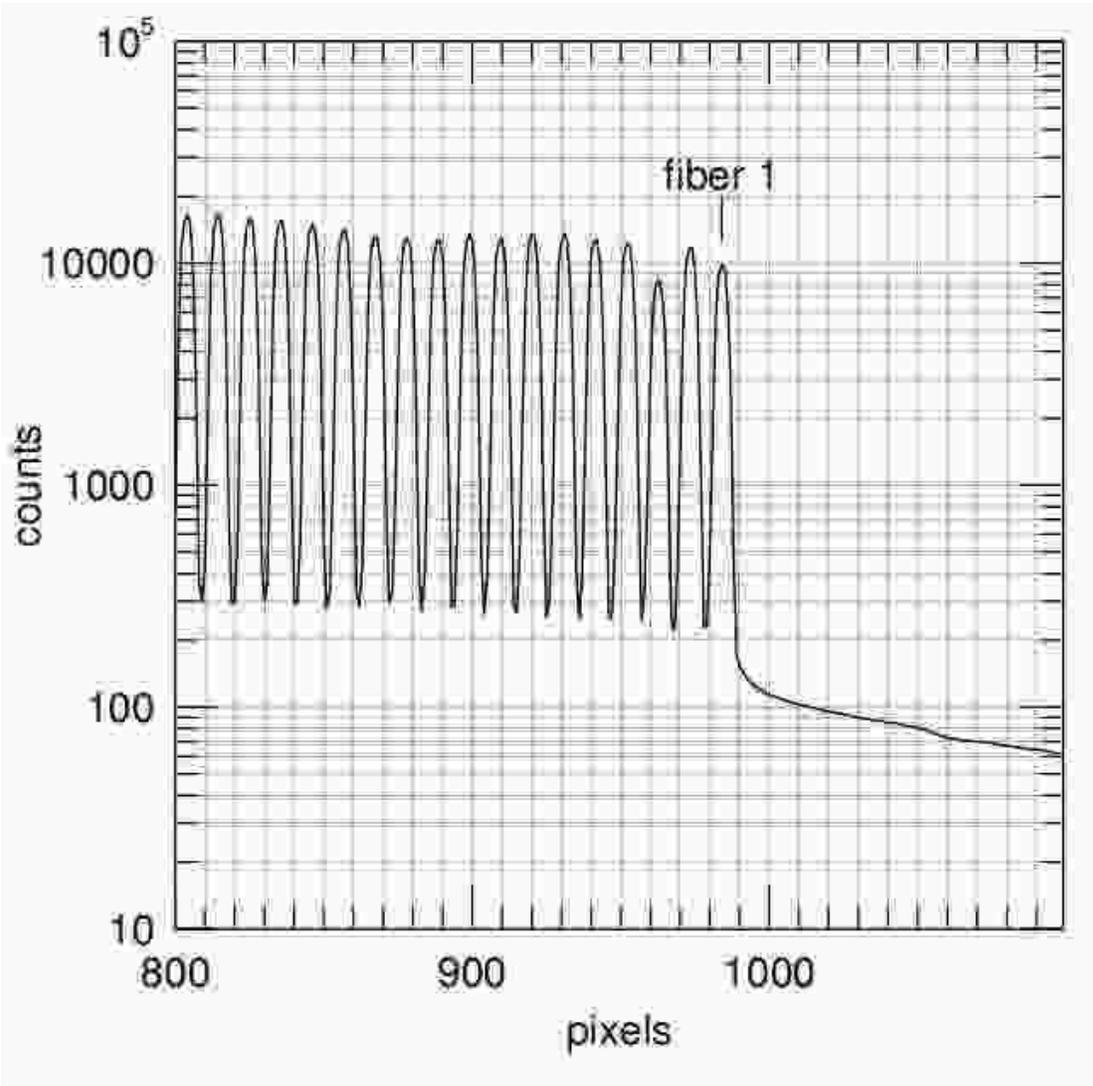}{5in}{0}{75}{75}{-240}{0}
\caption{SparsePak dome-flat cross-section for the region
marked in Figure 1 of the first 18 fibers. Note the even fiber
spacing, well-separated maxima, and low levels of scattered light
($<$2\% at this wavelength).}
\end{figure}

\clearpage

\begin{figure}
\figurenum{7}
% \epsscale{1}
% \plotone{f7xx.ps}
\plotfiddle{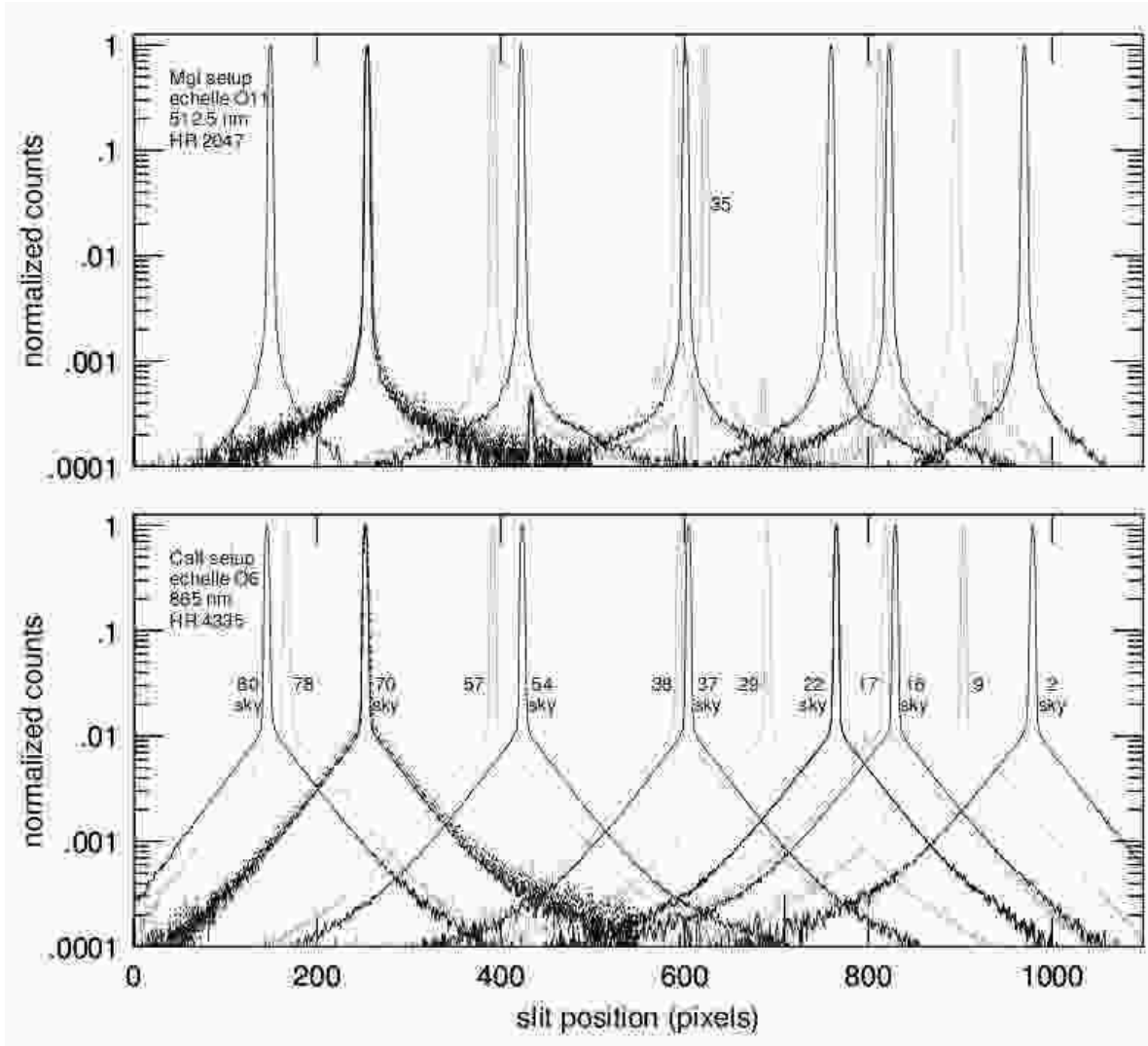}{5in}{0}{80}{80}{-250}{-10}
\caption{Superposition of normalized, spatial profiles of
bright stars observed down individual fibers with SparsePak using the
MgI (top) and CaII (bottom) echelle setups (orders 11 and 6,
respectively, as specified in Table 1). Peak counts are under
half-well. Fiber numbers are labeled, and sky fibers are identified.
Small secondary maxima (at the 0.1-1\% level) are found in non-sky
fibers because of the stellar point-spread-function and the proximity
of adjacent, active fibers. Each cross-section represents an average
over the middle quintile in wavelength. For fiber 70, all 5 quintiles
are shown. In general, the profiles are well characterized by a
Gaussian core and a power-law tail. Note the significantly increased
level of scattering in the CaII region at 8650\AA\ region. The
power-law tail starts at roughly 1\% of the peak value at 8650\AA\ but
only at 0.04\% of the peak at 5125\AA.}
\end{figure}

\clearpage

\begin{figure}
\figurenum{8}
% \epsscale{0.55}
% \plotone{f8xx.ps}
\plotfiddle{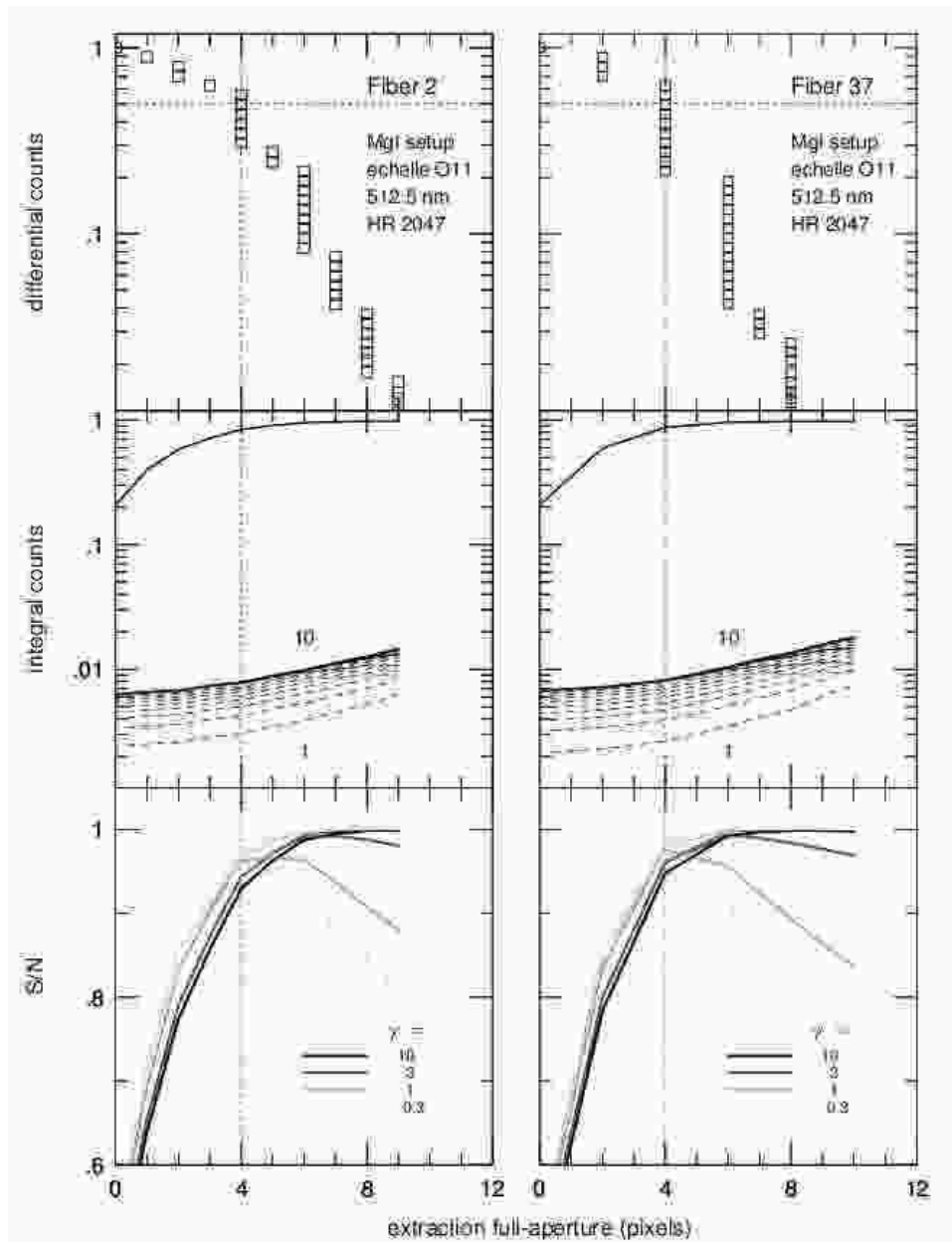}{6in}{0}{65}{65}{-200}{-25}
\caption{Spatial profiles for SparsePak fibers \#2 and \#37
as a function of the extraction aperture using the Bench Spectrograph
with the echelle grating in order 11, centered at a wavelength of 5125
\AA\ (see Table 1). The top panel contains the normalized
surface-brightness profile. The middle panel contains the normalized
curve of growth of the light for the signal (top curve) as well as the
scattered component (bottom set of curves). The dashed lines represent
the cumulative scattering contribution from the nearest up to the
tenth-nearest pair of fibers (bottom to top) for an unweighted
extraction. The scattered light component converges at the distance of
the tenth-nearest fiber pair (assuming all fibers are uniformly
illuminated). The shaded curves are the cumulative scattered light
contribution for a weighted extraction (see text) under different
conditions summarized in the bottom panel. The bottom panel contains
the signal-to-noise profiles of the integrated light assuming a
variety of conditions identified by $\gamma$, the ratio of the source-
to detector-noise contributions for the central pixel. Hence large
values of $\gamma$ are effectively source-limited observations, while
small values of $\gamma$ are detector-limited observations. Unweighted
extractions are shown as heavy lines, while weighted extractions are
shown as thin lines. The weighted and unweighted extractions are
essentially identical for $\gamma\geq3$; the weighted extractions
provide superior performance for $\gamma\leq1$.}
\end{figure}

\clearpage

\begin{figure}
\figurenum{9}
% \epsscale{0.55}
% \plotone{f9xx.ps}
\plotfiddle{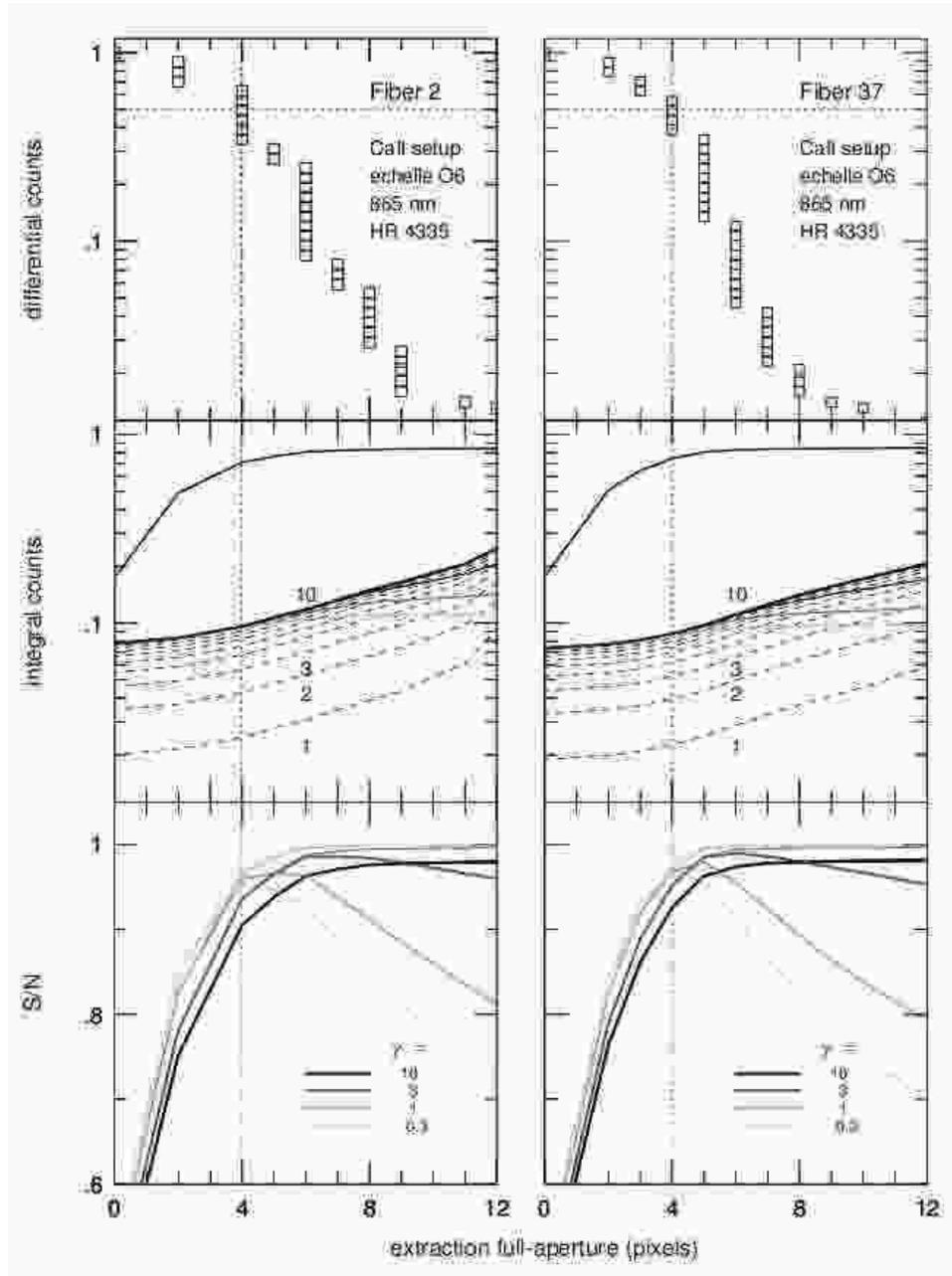}{6in}{0}{65}{65}{-200}{0}
\caption{Same as Figure 8, except for the echelle grating in
order 6, centered at a wavelength of 8650 \AA. Note the significantly
higher scattering component than in Figure 8.}
\end{figure}

\clearpage

\begin{figure}
\figurenum{10}
% \epsscale{0.65}
% \plotone{f10xx.ps}
\plotfiddle{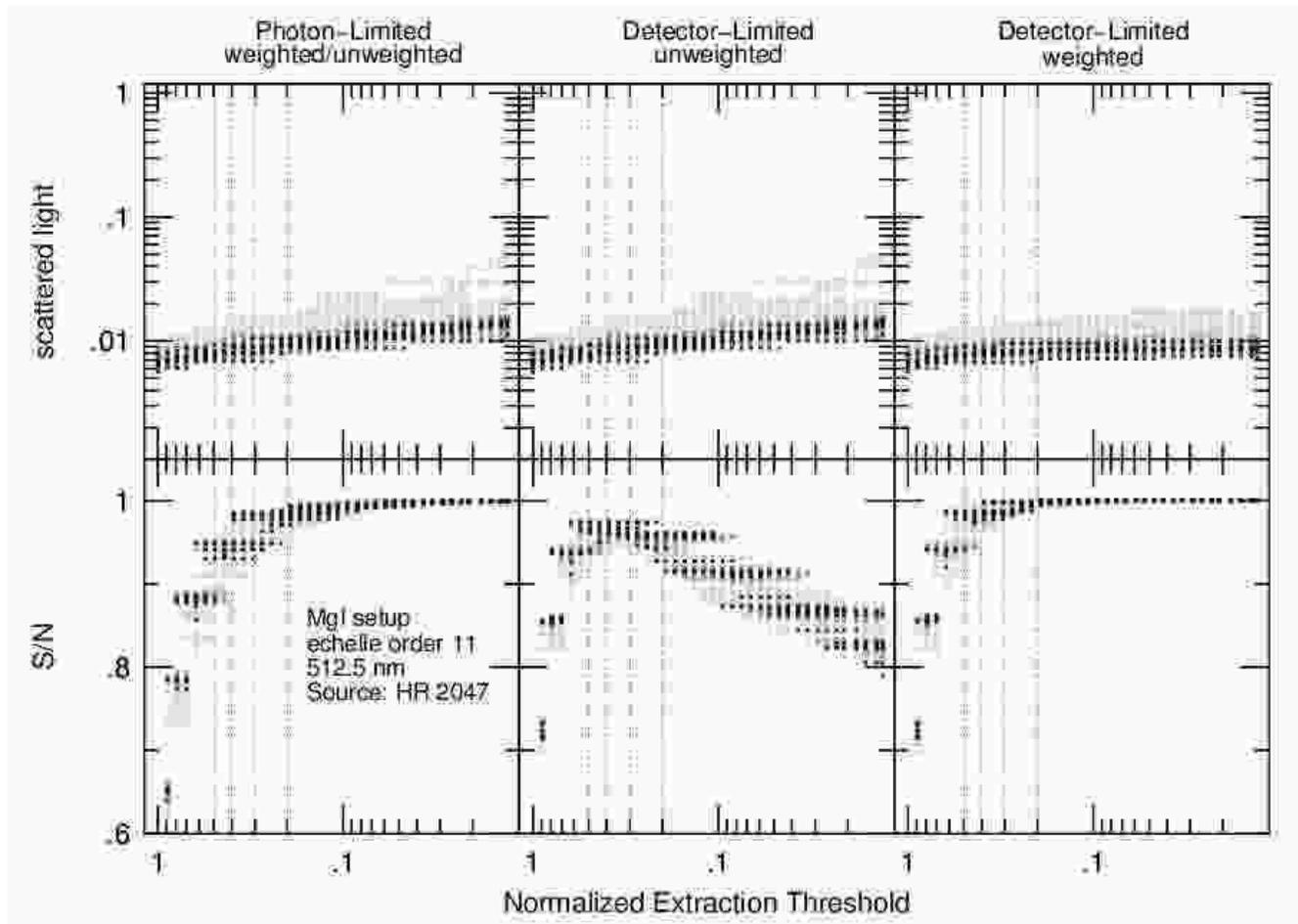}{5in}{0}{90}{90}{-275}{05}
\caption{Scattered light and S/N profiles as a function of
normalized extraction-threshold defining the spatial apertures for
SparsePak (echelle grating, order 11, 5125 \AA; see Table 1). The
normalized extraction-threshold is the same as the ``differential
counts'' in the top panels of Figures 8 and 9. Profiles are given for
the photon-limited case ($\gamma=10$), and the detector-limited case
for unweighted and weighted extractions, as labeled. Data is shown for
all seven sky fibers (\#2, 16, 22, 37, 54, 70, and 80) in five
wavelength regions spanning the full detected range on the 2048 pixels
of the CCD. The different symbols represent the central quintile in
wavelength (dark, small squares), the 2nd and 4th quintiles in
wavelength (medium shaded and sized squares), and the 1st and 5th
quintiles in wavelength (light, large squares). Dotted lines indicate
a range of normalized extraction thresholds between 0.5 and 0.2 which
appear to be suitable for optimizing the trade between maximizing S/N
and minimizing scattered light.}
\end{figure}

\clearpage

\begin{figure}
\figurenum{11}
% \epsscale{0.65}
% \plotone{f11xx.ps}
\plotfiddle{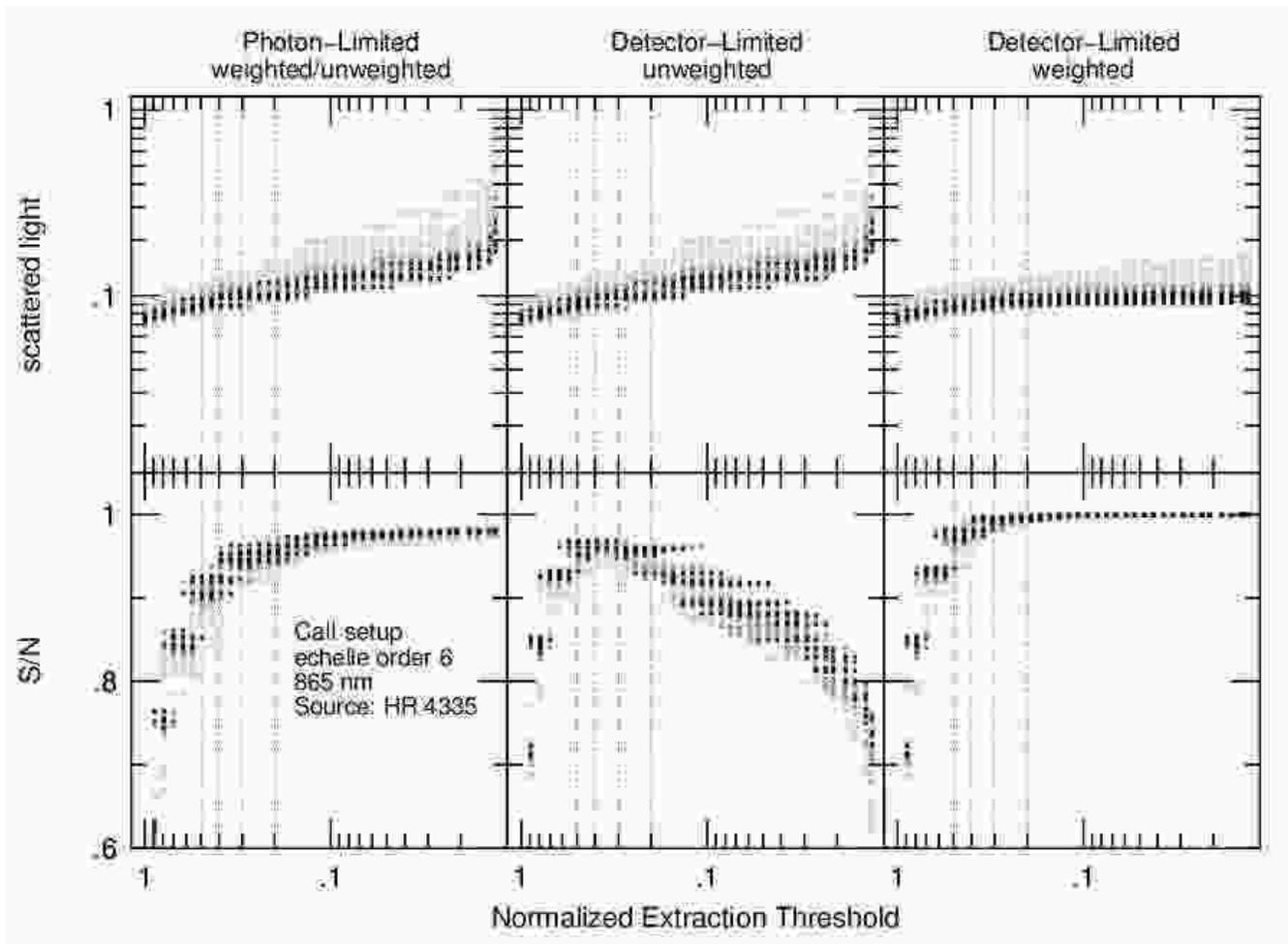}{5in}{0}{90}{90}{-275}{05}
\caption{Same as Figure 10 except for the echelle grating in order 6,
centered at a wavelength of 8650 \AA. Note that the scattered light is
roughly $10\times$ worse at these red wavelengths than for the visible
region, for reasons which are presently unknown. Greater than 90\% of
peak S/N can be obtained while keeping the total scattered light to
$\sim 10$\% of the signal. Even in the presence of the greater
scattered light, a weighted extraction offers marginal gain at
thresholds of relevance.}
\end{figure}

\clearpage

\begin{figure}
\figurenum{12}
\plotfiddle{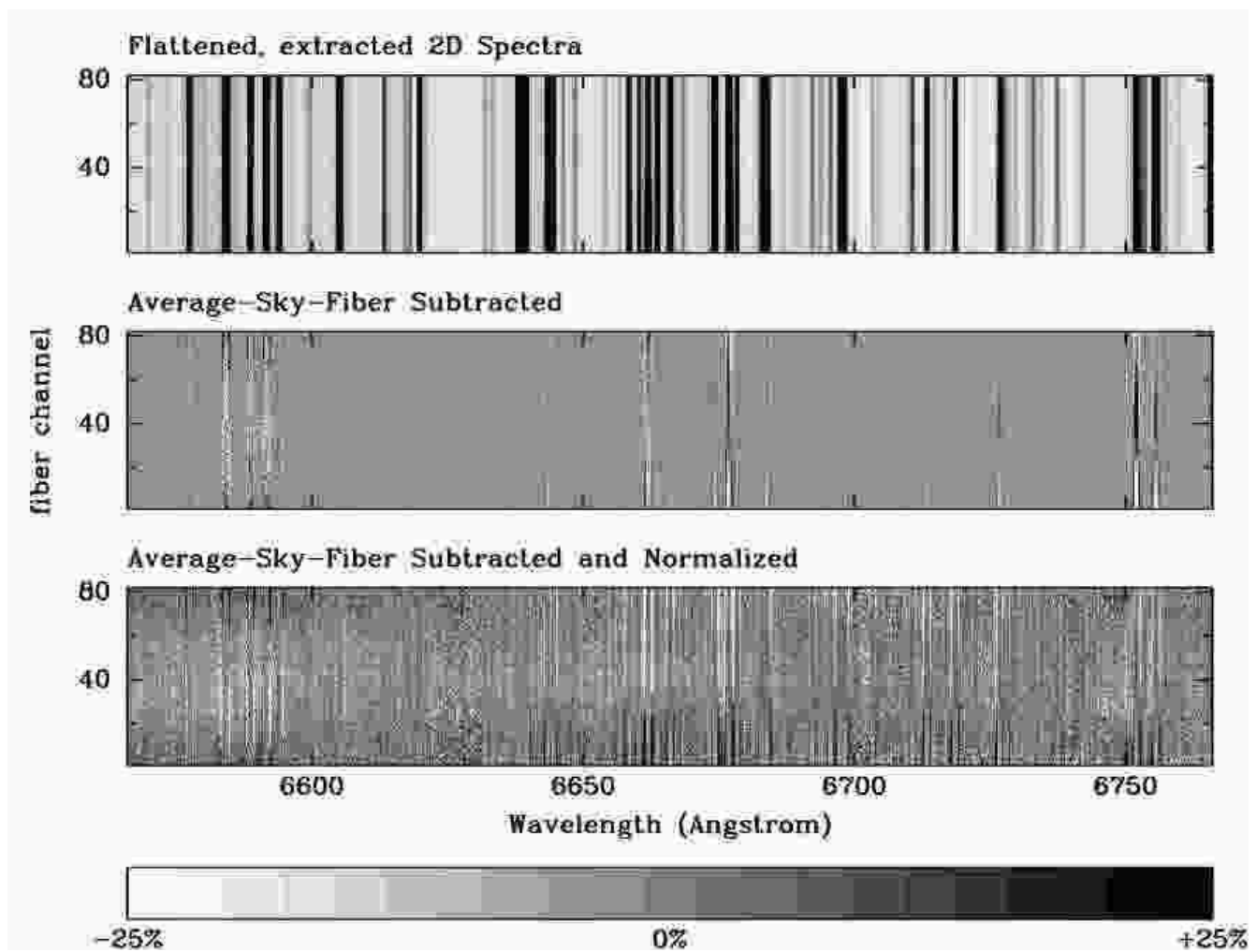}{6in}{0}{90}{90}{-270}{0}
% \epsscale{0.65}
% \plotone{f12xx.ps}
\caption{Illustration
of why sky subtraction is poor with fiber spectrographs when using
mean sky spectra. This example uses a Thorium-Argon line-lamp spectrum
observed with SparsePak and the Bench Spectrograph using the echelle
grating in order 8 centered at 6687\AA. All three panels contain the
two-dimensional spectra for all 82 fibers, which have been flattened,
wavelength calibrated, and extracted (the output of {\tt
dohydra}). The spectra are ordered by fiber \#, which run contiguously
along the slit; this is, in effect, like a long-slit spectrum, with
the inter-fiber gaps taken out. The top panel is the original,
extracted spectrum. The middle panel has had the average sky-fiber
spectrum subtracted, revealing significant residuals at the location
of the lines. The bottom panel is the same sky-subtracted spectrum,
normalized by the original spectrum in the top panel. The grey-scale
bar applies to the bottom panel. Note the significant residuals around
(unresolved) lines change shape as a function of slit position and
wavelength. The residuals are like what is seen in
poorly-sky-subtracted data, and are due to the changing, or
field-dependent optical aberrations within spectrographs.}
\end{figure}

\clearpage

\begin{figure}
\figurenum{13}
\plotfiddle{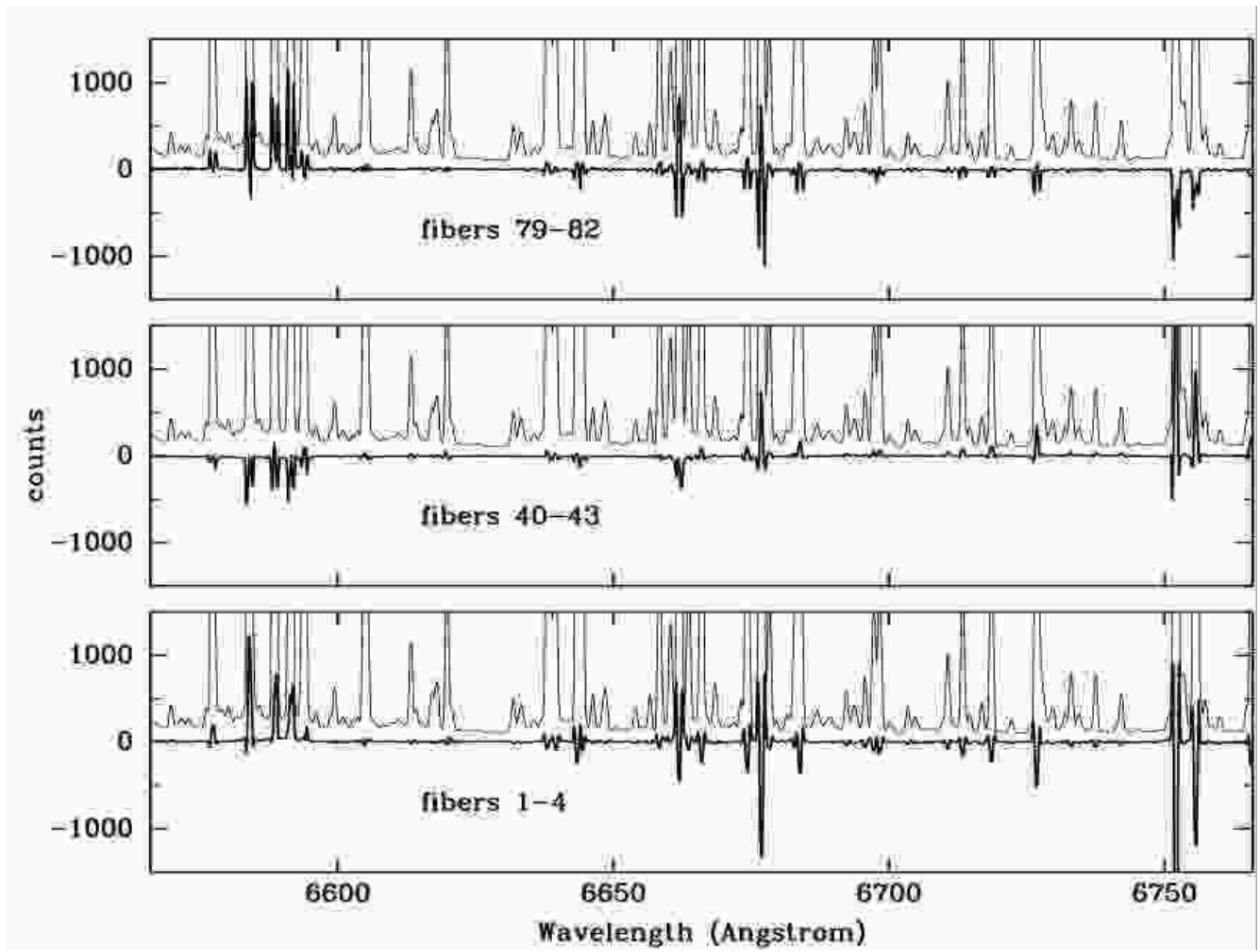}{6in}{0}{90}{90}{-270}{0}
% \epsscale{0.7}
% \plotone{f13xx.ps}
\caption{Mean Thorium-Argon spectra shown in (a) for three groups of
fibers at the edges and the center of slit. The thin lines are the
original spectra, while the thick lines represent the sky-fiber
subtracted spectra. The counts are arbitrary units, but have the same
scale in all panels. Again note the significant residuals around
(unresolved) lines change shape as a function of slit position and
wavelength.}
\end{figure}

\clearpage

\begin{figure}
\figurenum{14}
\plotfiddle{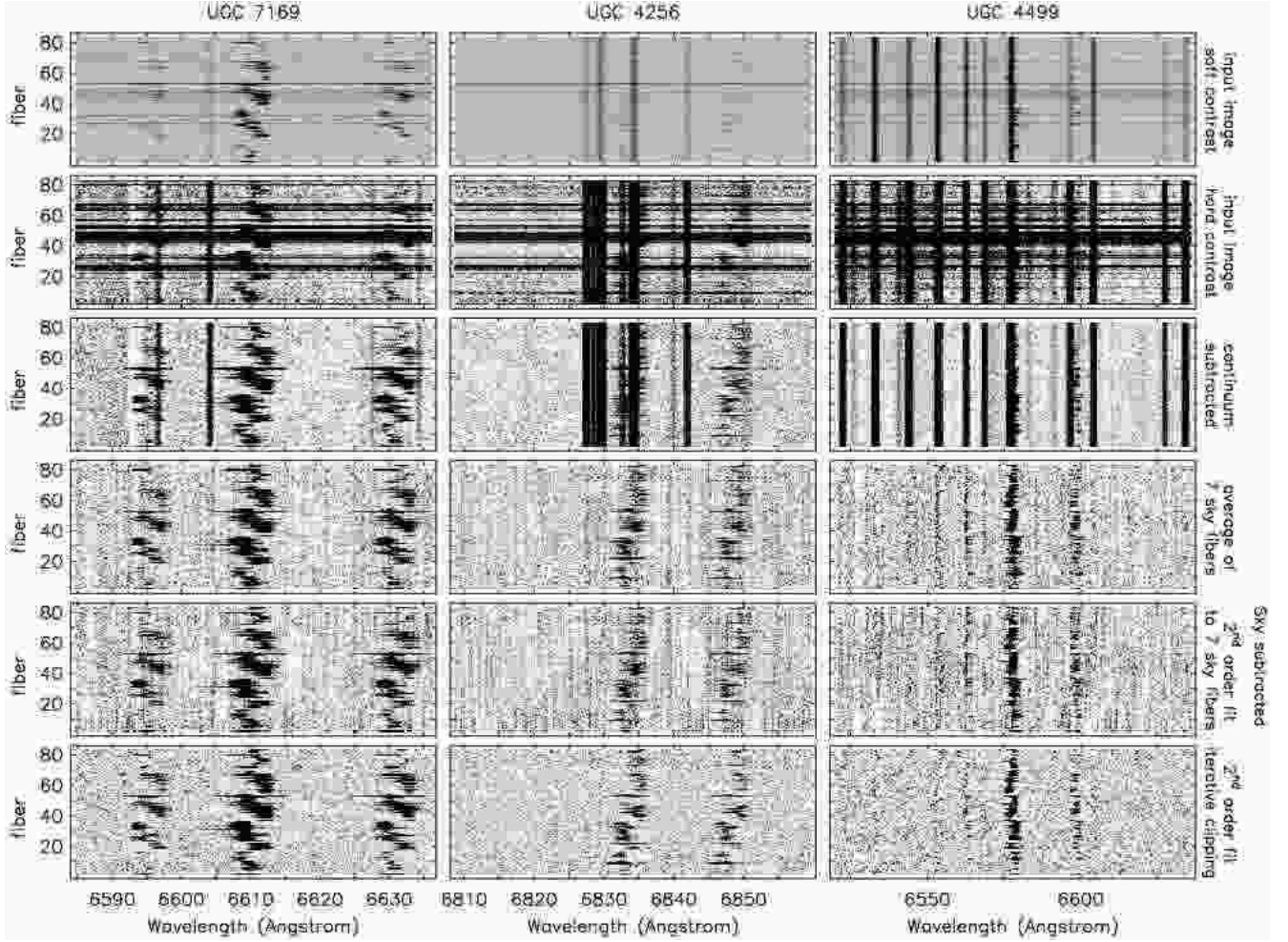}{6in}{0}{90}{90}{-270}{0}
% \epsscale{0.6}
% \plotone{f14xx.ps}
\caption{Illustration
of superior sky subtraction methods for fiber spectrographs using our
fitting and iterative clipping algorithm.
Shown are the flattened, extracted, and wavelength
calibrated spectra from {\tt dohydra} for a portion of the spectra
around H$\alpha$ (visible as well are the two [NII] lines at 6548 and
6584 \AA) for UGC 7169 and UGC 4499, and around [SII] at 6717 and 6731
\AA\ for UGC 4256. 
For the two galaxies observed with the echelle
grating (UGC 7169 ad UGC 4256), their internal velocities are well
resolved despite the fact that these galaxies are nearly face-on. For
the low-surface-brightness galaxy UGC 4499, observed with SparsePak
and the 860 l/mm grating (order 2), the internal velocity structure is
not well-resolved. 
Panels 1 and 2 (starting from the top) show the
original, processed spectrum at two different contrast levels. The
bottom 5 panels are displayed at the same levels with respect to the
inter-line background. Panel 3 shows the effect of subtracting the
source (and background) continuum, fiber-by-fiber. Panels 4-6 show
three different schemes for subtracting the sky lines, starting with
the same continuum-subtracted spectrum in panel 3 (see text): the
``mean-sky'' method; low-order polynomial fitting to sky fibers;
low-order polynomial fitting to all fibers. The iterative fitting
scheme (bottom panel) yields the most visibly clean spectrum, and best
signal-to-noise. This method does subtract some of the source signal
for UGC 4499, but for the other two galaxies, the bottom panel yields
the best results.}
\end{figure}

\clearpage

\begin{figure}
\figurenum{15}
% \epsscale{0.7}
% \plotone{f15xx.ps}
\plotfiddle{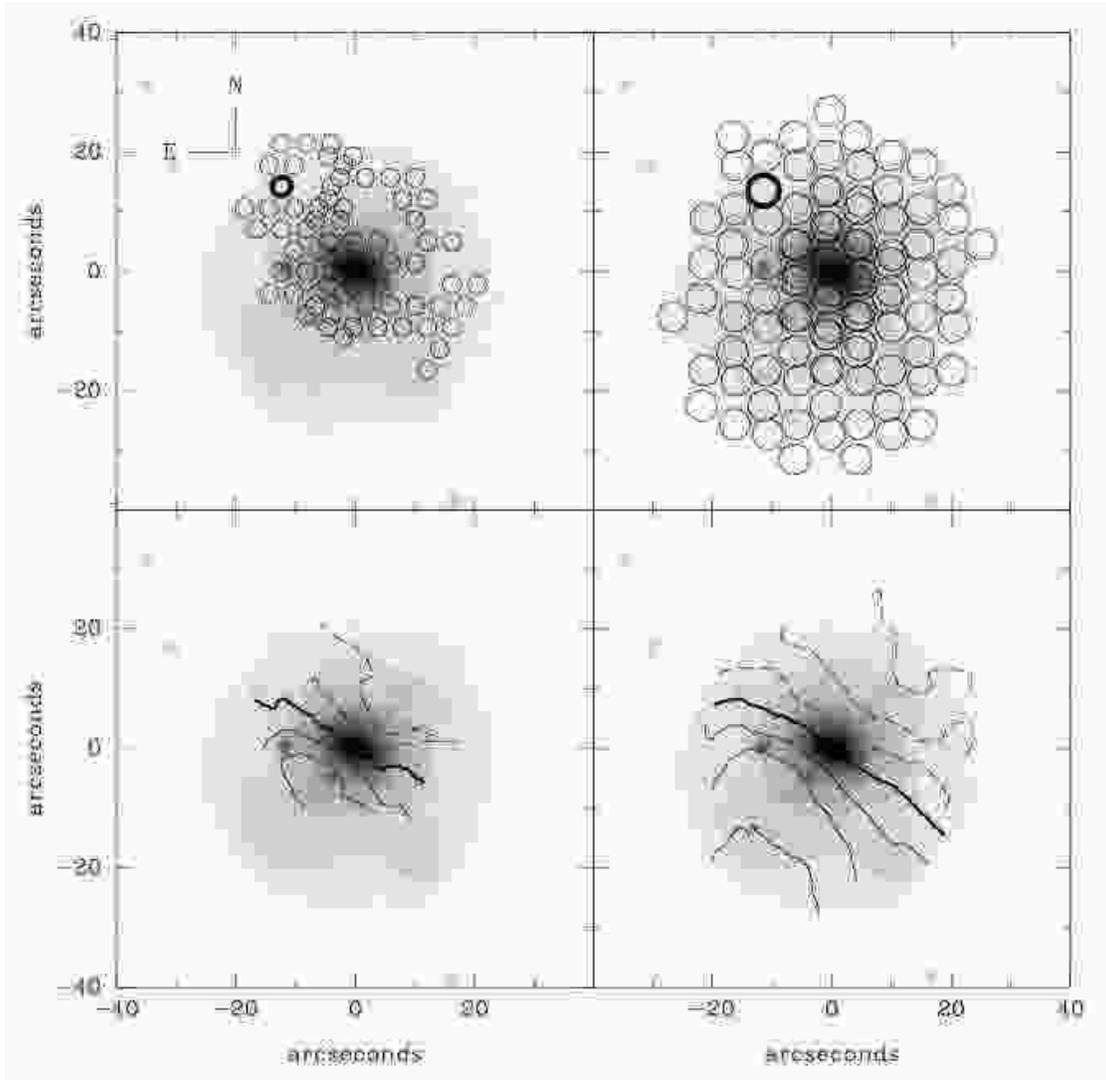}{5in}{0}{75}{75}{-230}{0}
\caption{SparsePak and DensePak H$\alpha$ echelle
observations of PGC 56010, a moderately low-surface-brightness, nearly
face-on disk galaxy (SBc) with the following properties: M$_B$ = -18.8
(H$_0$=75  km s$^{-1}$/Mpc), recession velocity of 4468  km s$^{-1}$, $I$ = 13.7 mag,
disk central surface-brightness of $\mu_0(I)$ = 20.6 mag
arcsec$^{-2}$, disk scale length $h_R$ = 9.9 arcsec, half-light radius
r$_{1/2}$ = 14.5 arcsec, and $\eta$-radius at $\eta=0.2$ of 29.0
arcsec (Andersen, 2001). For reference, a Freeman disk would have
roughly a value of $\mu_0(I)$ = 20 mag arcsec$^{-2}$ for a typical
$B-I$ color of 1.7 mag. In the top panels, the fiber footprints of
Densepak (left) and SparsePak (right) are overlayed on 1.8 arcsec FWHM,
$I$-band image observed in May, 1999 (WIYN, S2KB CCD imager). Only
fibers with measured signal are shown; the spectra of the
heavy-weighted fibers are used for comparison in Figure 16. Densepak
and SparsePak observations both consisted of two pointings of
2$\times$1800s and 2$\times$1200s, respectively. DensePak observations
were taken as part of D. Andersen's thesis observations (Andersen,
2001) on March 29, 1999. SparsePak observations were taken as part of
the early commissioning on June 9, 2001. The lower panels show the
extracted H$\alpha$ velocity fields from these observations. Contours
are at 10  km s$^{-1}$ intervals, with solid lines representing the
approaching projection. Note the greater extent of the SparsePak
velocity field even in areas (e.g., WSW) where both arrays cover the
source. The SparsePak velocity field is smoother due to higher S/N and
larger beam.}
\end{figure}

\clearpage

\begin{figure}
\figurenum{16}
% \epsscale{0.6}
% \plotone{f16xx.ps}
\plotfiddle{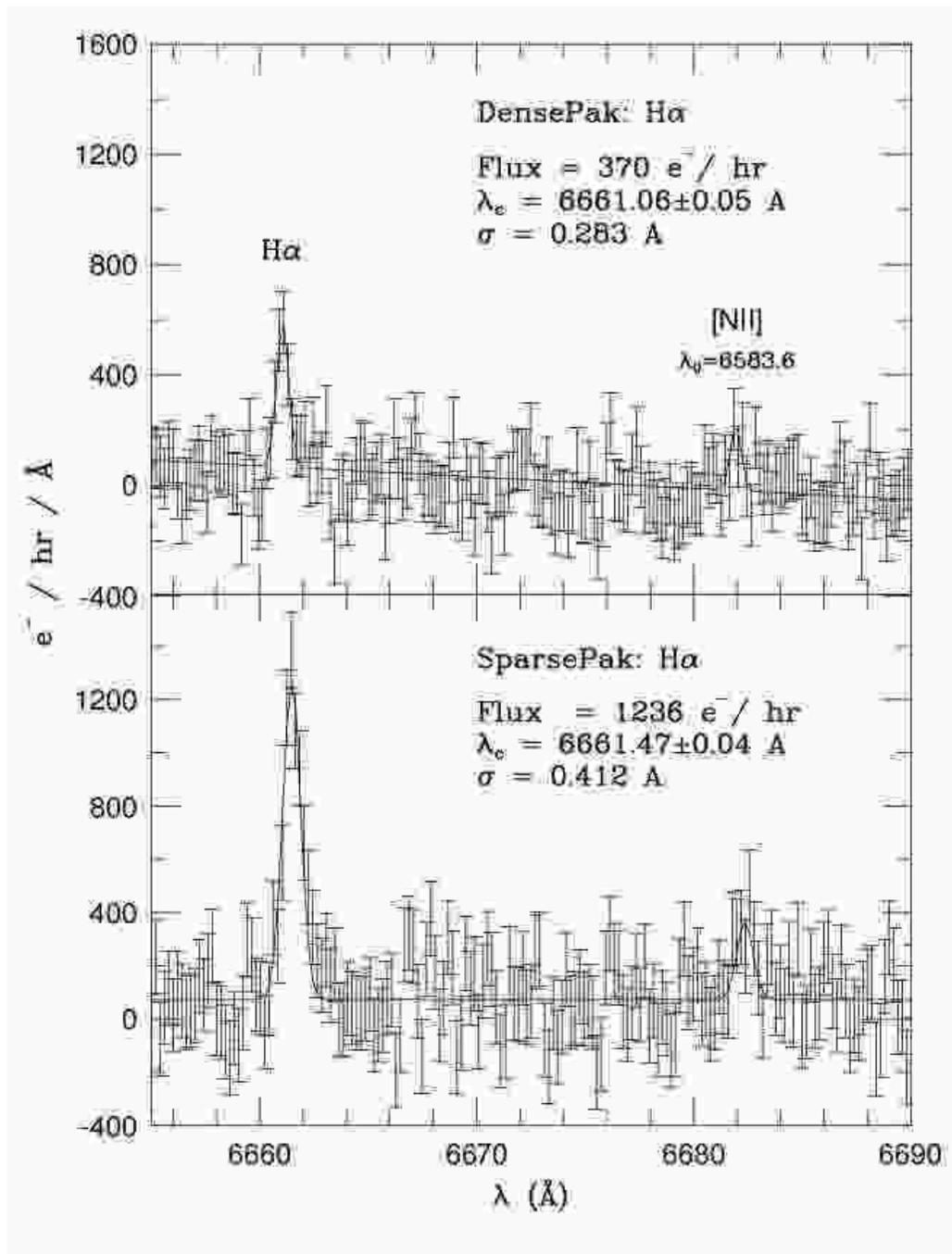}{6in}{0}{70}{70}{-210}{-10}
\caption{Comparison of Densepak (top) and SparsePak
(bottom) single-fiber spectra for PGC 56010 at almost exactly the same
position $\sim$20 arcsec to the NW of the nucleus (see Figure
15). Both spectra were taken in clear conditions with the same
spectrograph setup (yielding mean dispersions of 0.2\AA/pixel), have
been sky-subtracted, and have had heliocentric corrections
applied. Errors are estimated for each extracted ``pixel'' based on
the known detector read-noise and photon shot-noise from source plus
sky. (The SparsePak spectra were on-chip binned by 2 in the spatial
dimension.) The instrumental resolution with SparsePak at this
wavelength with this setup is 0.28\AA\ ($\sigma$), i.e. a spectral
resolution of $\sim$10,000. The flux units are such that differences
between SparsePak and DensePak flux levels are attributable only to
the fiber area and throughput. Note the ratio of the integrated line
fluxes (SparsePak:DensePak) is greater than the ratio of their
areas. Also note the slope in the DensePak continuum, which is
attributable to improper sky-subtraction.}
\end{figure}

\clearpage

\begin{figure}
\figurenum{17}
% \epsscale{0.7}
% \plotone{f17xx.ps}
\plotfiddle{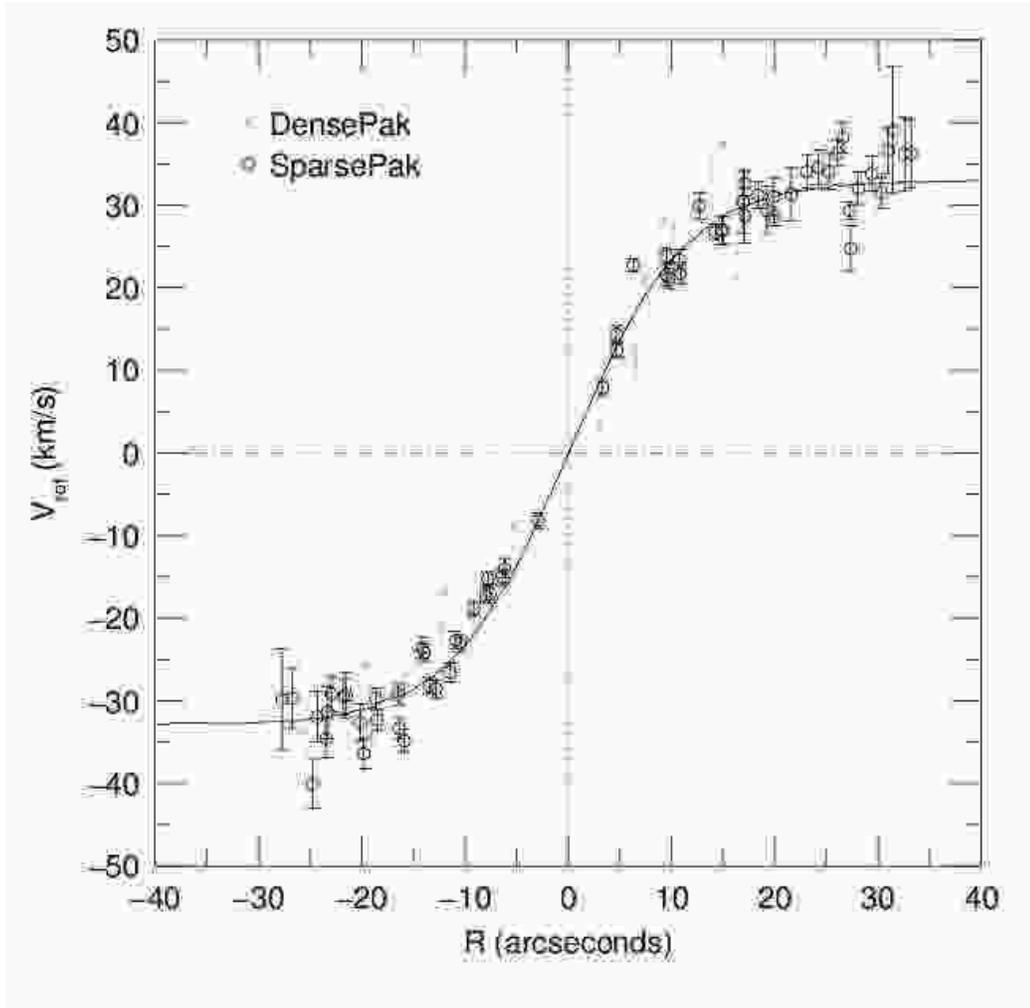}{5in}{0}{70}{70}{-200}{0}
\caption{Comparison of the derived rotation curves for
Densepak and SparsePak observations of PGC 56010 using fibers within
60$^\circ$ of the derived kinematic major axis. The SparsePak data is
at higher precision out to larger radii. The formal single-disk model
fits to these data yield inclinations and position angles (in degrees)
of $14.4^{+12.6}_{-14.4}$, $143.6\pm1.3$ for DensePak and
$24.8^{+6.6}_{-10.0}$, $146.1\pm1.5$ for SparsePak. While the position
angles are in good agreement, the disk inclination measurement is
significantly improved with the SparsePak observations which go deeper
and cover more area.}
\end{figure}

\begin{figure}
\figurenum{18}
% \epsscale{0.7}
% \plotone{f18xx.ps}
\plotfiddle{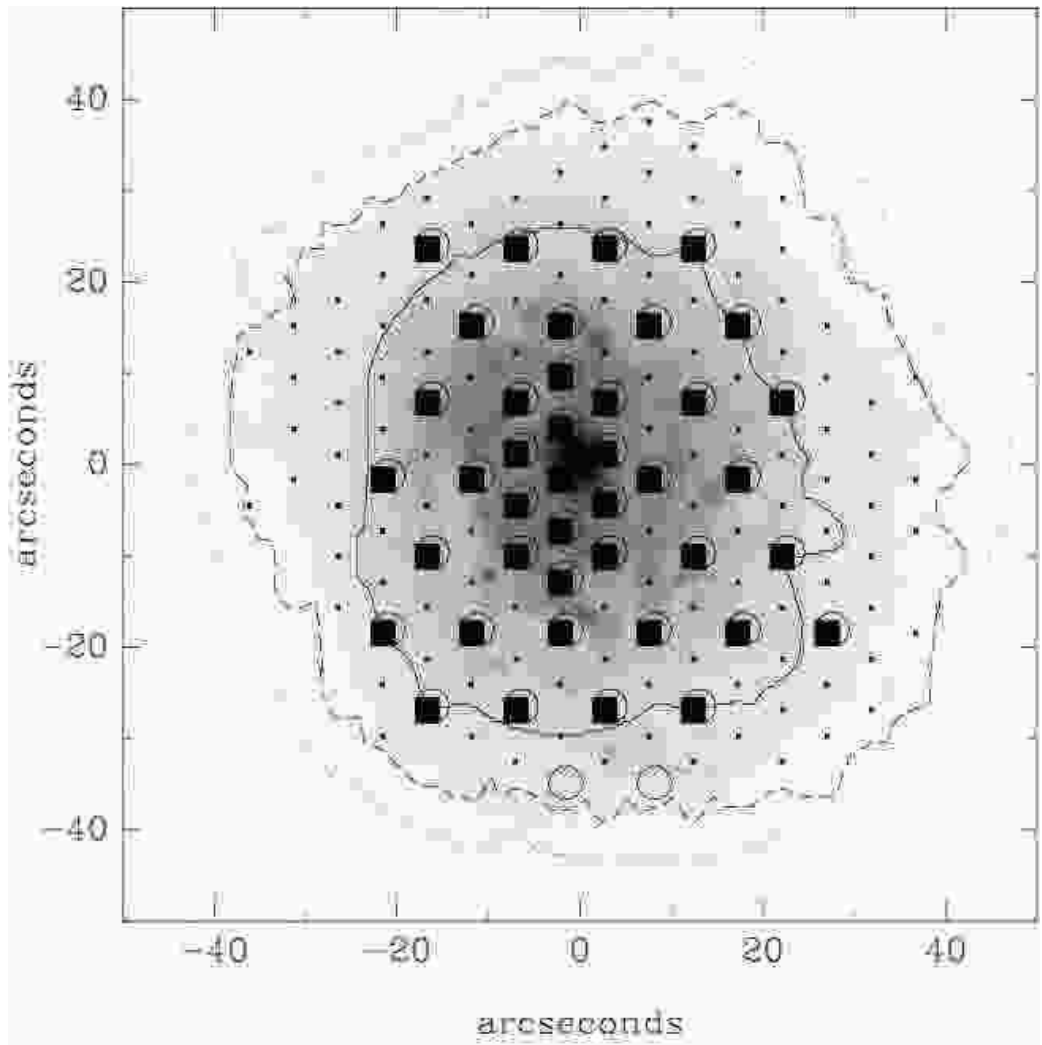}{5in}{0}{70}{70}{-210}{0}
\caption{Overlay of SparsePak footprints on an R-band image of
NGC~3982. This galaxy has an inclination of $26\pm2^\circ$, a blue
$B-I$ color of 1.5 mag, high central surface-brightness of $\mu_0(B) =
19.3$ mag arcsec$^{-2}$, but a small radial scale-length of h$_{\rm
R}$=0.9 kpc or 10.4 arcsec in the $I$-band (Verheijen, 1996).
Footprints represent measurement positions for H$\alpha$ line-emission
(small squares) and Ca II triplet stellar absorption (open
circles). H$\alpha$ measurements contain three pointings which fill
the array, while the Ca II measurements were for one pointing
only. Large squares represent H$\alpha$ measurements most closely
aligned with Ca II measurements. Contours correspond to the limits of
interpolated and smoothed velocity information for Ca II measurements
(solid), and H$\alpha$ with fine (dashed) and coarse (dotted)
interpolation and smoothing (refer to Figure 19).}
\end{figure}
\clearpage

\begin{figure}
\figurenum{19} 
% \epsscale{0.7} 
% \plotone{f19xx.ps}
\plotfiddle{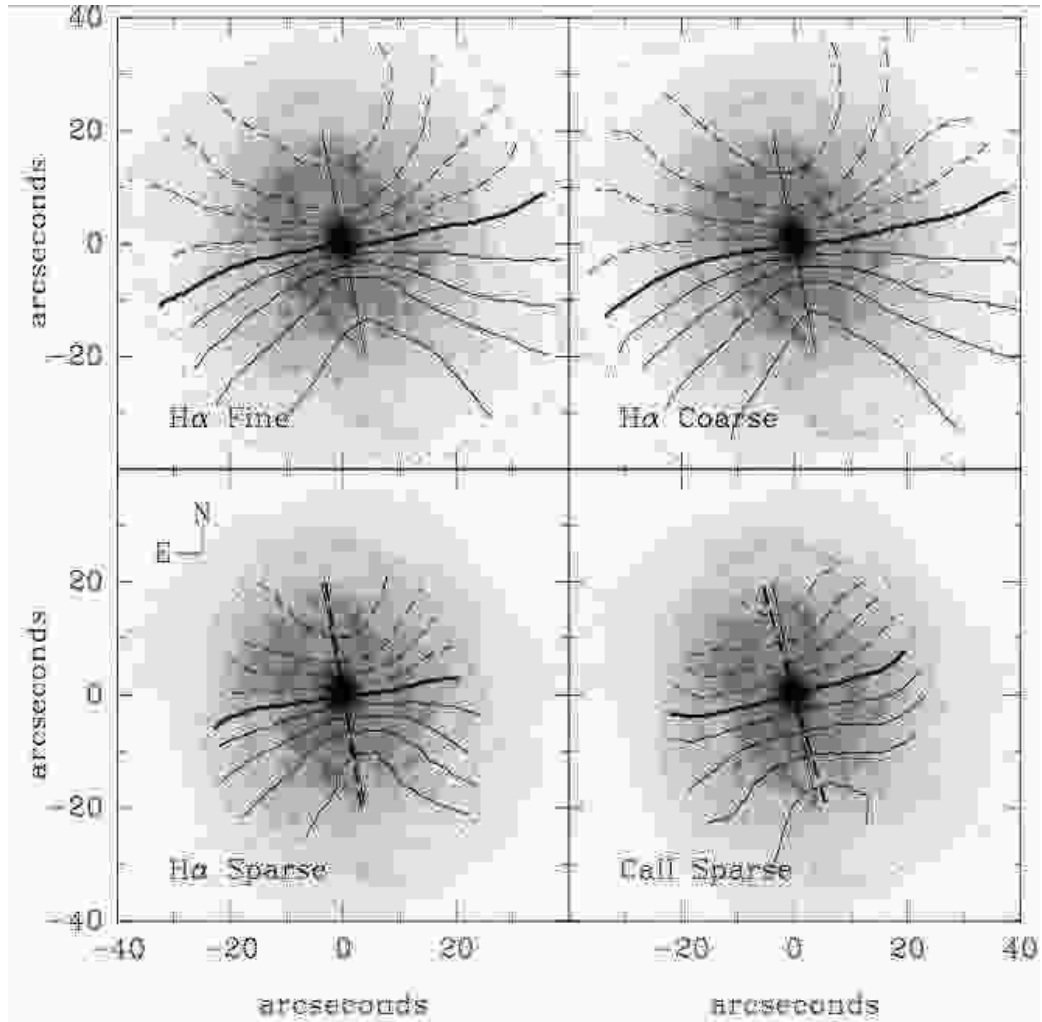}{5in}{0}{70}{70}{-200}{0}
\caption{SparsePak velocity fields for NGC 3982 measured in ionized
gas (H$\alpha$) and stellar absorption (CaII triplet).  The top two
panels show the emission-line velocity field using all fibers with
measured signal (Figure 18). ``Fine'' and ``Coarse'' refer to the
interpolation scale used for creating the velocity field. The bottom
two panels use the same, sparse fiber sampling (Figure 18), and the
coarse interpolation of the top-right panel.  Iso-velocity contours
are spaced at 20 km s$^{-1}$, with solid representing approaching
projection, and bold the line of nodes. Kinematic major axes are
indicated as solid lines. In the bottom left the dashed line
represents the kinematic major axis for the H$\alpha$ ``Fine''
sampling, while in the bottom right this line represents the kinematic
major axis for the H$\alpha$ ``Sparse'' sampling.}
\end{figure}
\clearpage

\begin{figure}
\figurenum{20}
% \epsscale{0.8}
% \plotone{f20xx.ps}
\plotfiddle{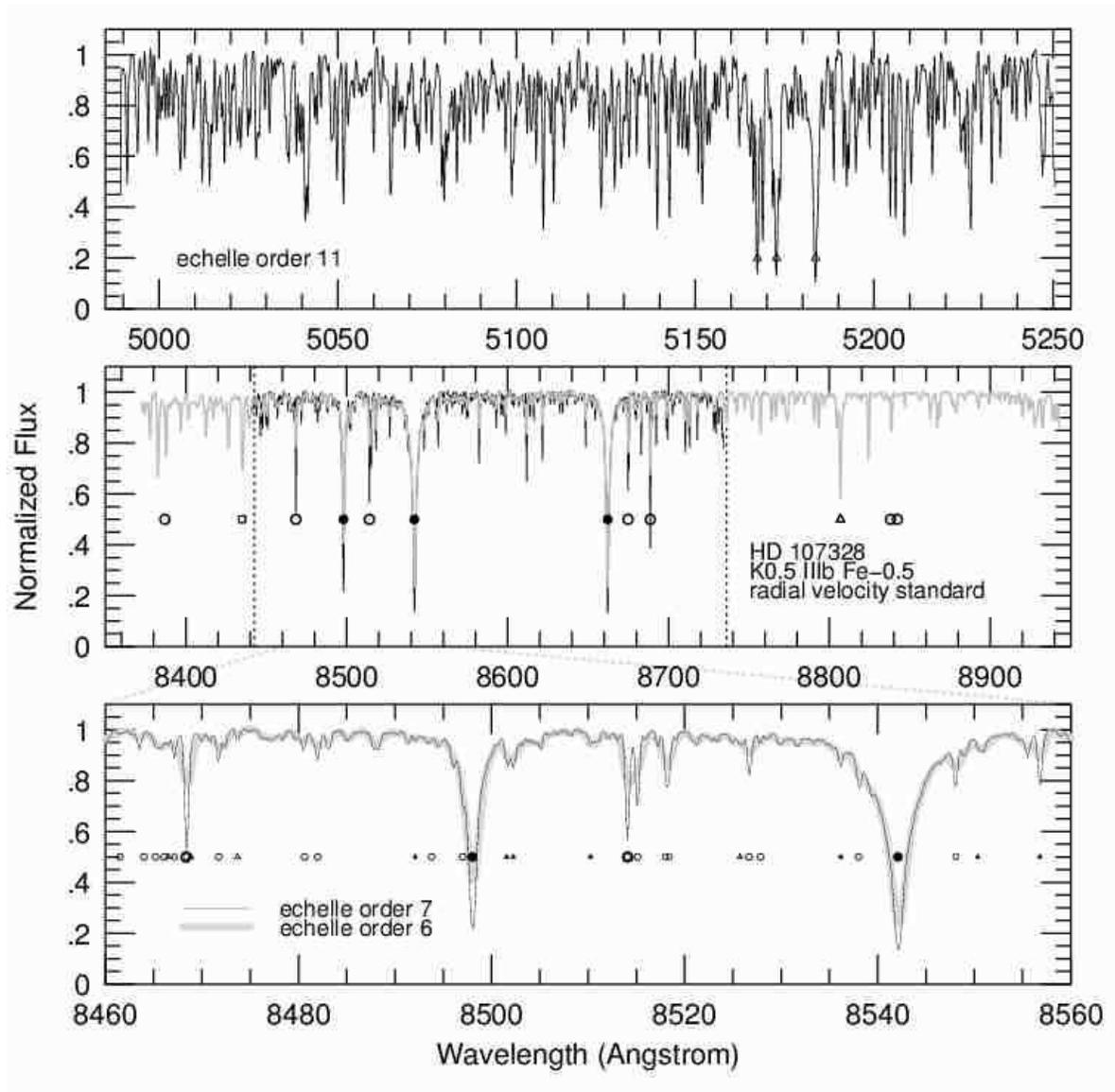}{6in}{0}{80}{80}{-230}{0}
\caption{SparsePak echelle spectra of HD 107328, a $V \sim 5$ mag K
giant, in orders 11 (Mg I region), 6 and 7 (CaII triplet region). The
effective integration times (the transit across one fiber) was 9.4
seconds in the order-7 and 11 configurations, and 4.6 seconds in the
order-6 configuration. The signal-to-noise in these spectra is
$\sim$300 per resolution element of 2.5-3.5 pixels.  The top two
panels contrast the differences between the line frequency, strength,
and shape in the two regions. The middle panel contrasts the higher
and lower resolution CaII-triplet setups; the vertical, dotted lines
indicates the ends of the higher (and higher resolution) order. The
bottom panel is a blow-up of the middle panel showing the detailed
changes in line-depth and shape between these two Ca~II-triplet
orders. Note the significantly deeper lines of the order-7 setup. In
all panels lines from Ca~II, Fe~I, Mg~I, and Ti~I are marked,
respectively as filled circles, open circles, triangles, and squares,
based on a solar line list (Pierce \& Breckinridge, 1973). For
simplicity, only the Mg~I lines ((5175, 5169, and 5183 \AA)) are
marked in the top panel; most of the power in other lines is
contributed from Fe~I.}
\end{figure}

\begin{figure}
\figurenum{21}
% \epsscale{0.8}
% \plotone{f21xx.ps}
\plotfiddle{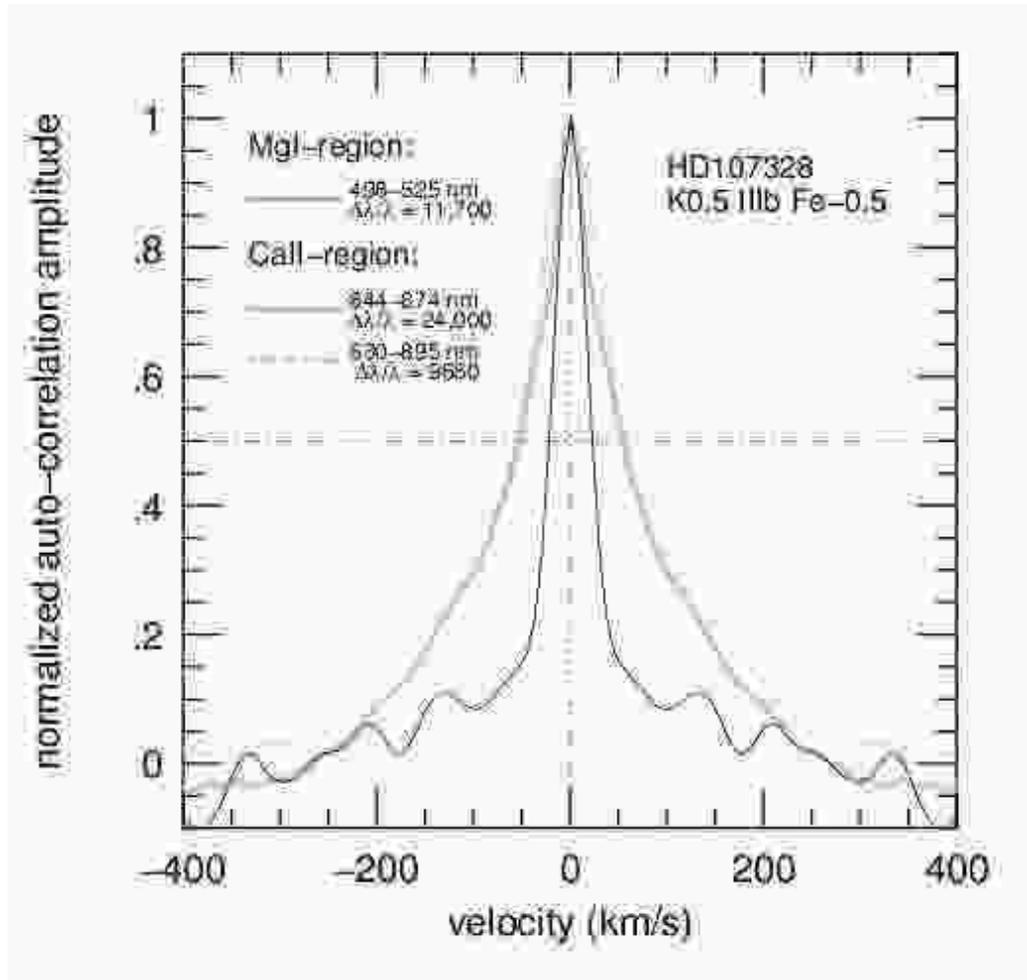}{5in}{0}{70}{70}{-210}{10}
\caption{Auto-correlation function of the spectra shown in
Figure 18. Note the relative narrowness of the MgI-region spectrum,
which is dominated by the many weak lines in this regions, primarily
from Fe I lines. In contrast, the CaII-region is dominated by the
broad CaII lines. Also note that the higher-resolution (order 7)
spectrum only alters the core of the auto-correlation function (above
65\% of the peak).}
\end{figure}

\clearpage

\begin{figure}
\figurenum{22}
% \epsscale{1.0}
% \plotone{f22xx.ps}
\plotfiddle{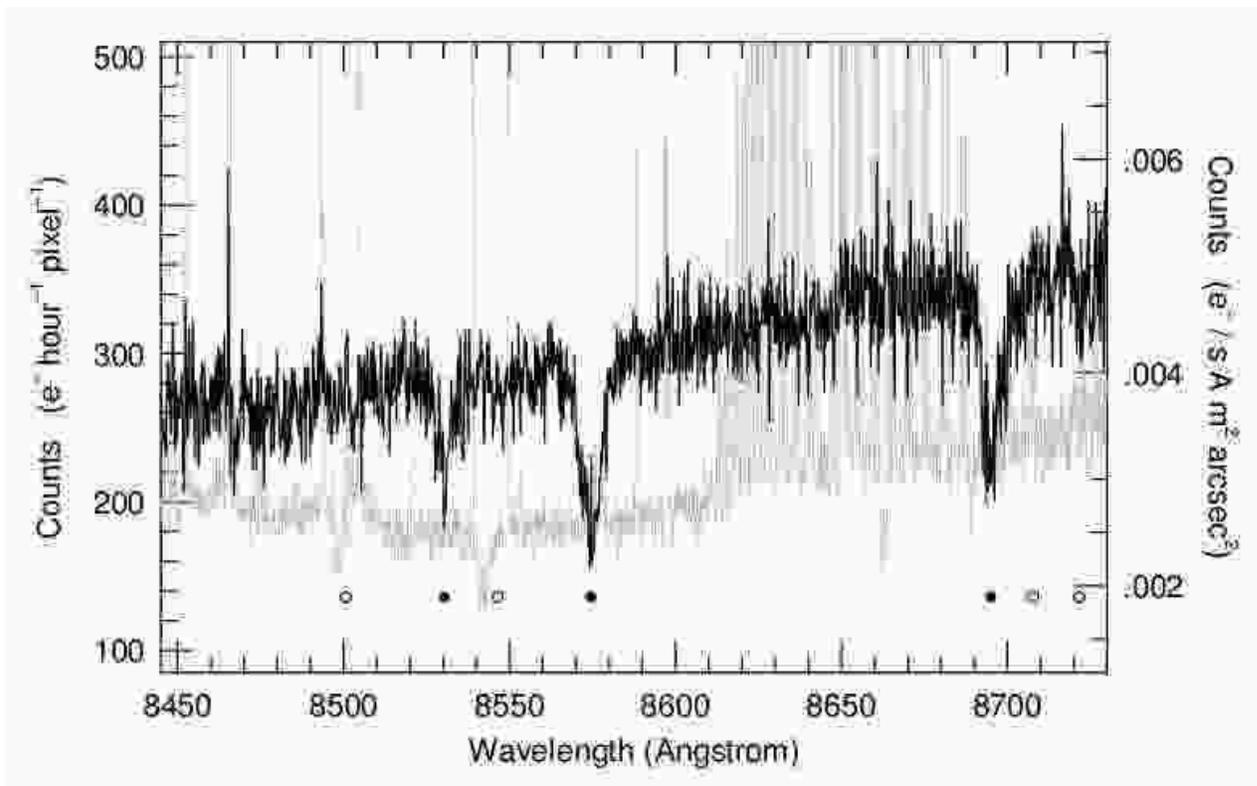}{5in}{0}{85}{85}{-240}{10}
\caption{SparsePak echelle spectra of the central regions of NGC
3982 and the night sky in the CaII-triplet region (order 7). NGC 3982
is a high-surface-brightness, blue star-forming galaxy in Ursa Major
(see text). CaII and Fe I lines are marked as in Figure 18. Two flux
scales are provided, relevant for calibration and exposure-time
calculation, as calibrated using the registration of the spectral
continuum map with a broad-band image (see text and next two
figures). The source and sky continua are equivalent to 17.75 and 18.2
mag arcsec$^-2$, respectively. Note the narrow, but strong and
frequent sky-lines.}
\end{figure}

\clearpage

\begin{figure}
\figurenum{23}
% \epsscale{0.9}
% \plotone{f23xx.ps}
\plotfiddle{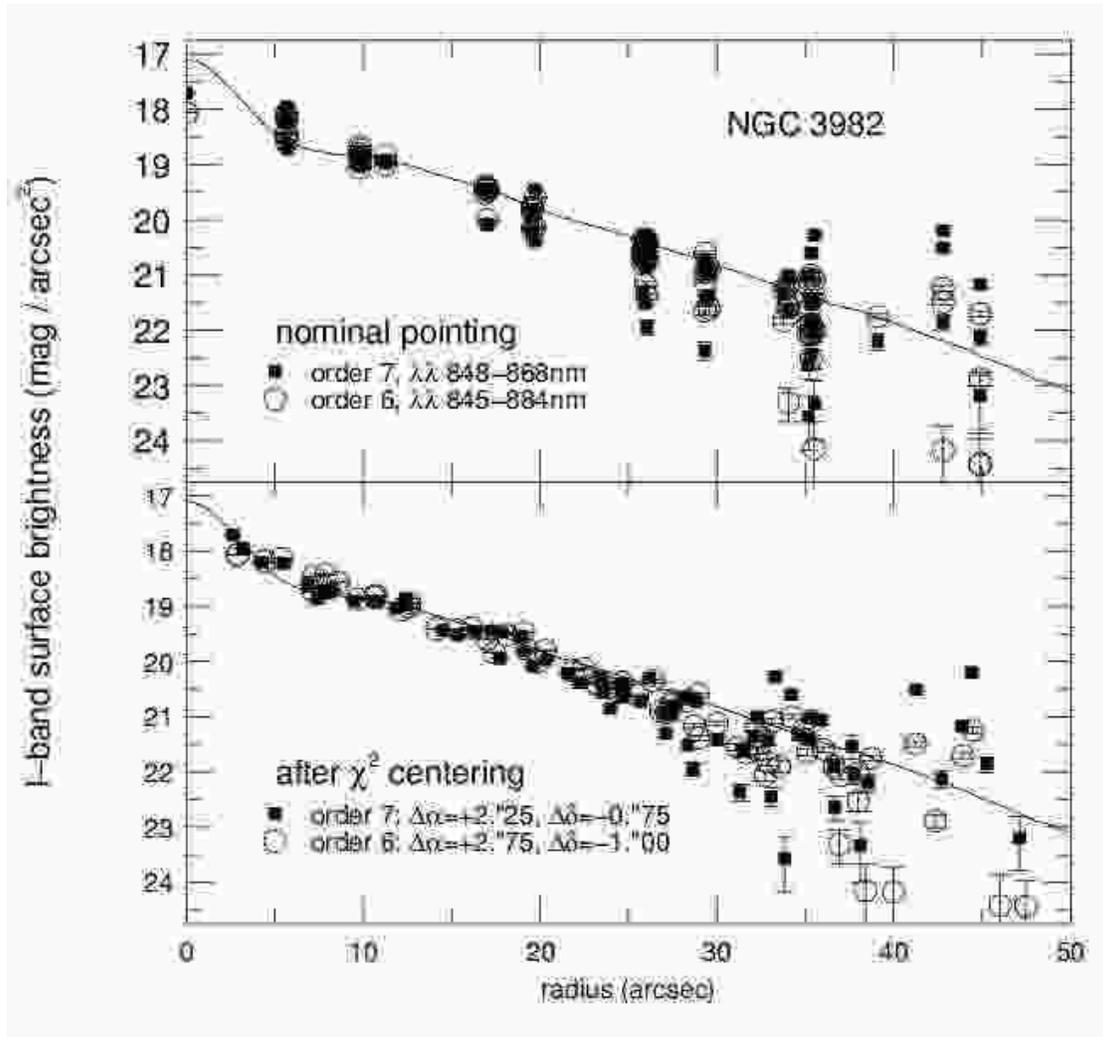}{5in}{0}{75}{75}{-230}{0}
\caption{I-band surface-brightness profile of NGC 3982 as derived
from I-band imaging (solid line, Verheijen, 1996) and SparsePak
spectra (points) in the CaII triplet region as observed
in orders 6 and 7. The spectral estimates are from source continuum flux
measurements of individual fibers. The top panel compares the data
adopting the nominal SparsePak centering from the recorded telescope
coordinates, normalized in magnitude in a least-squares sense to the
broad-band data. The bottom panel has been shifted, based on a
registration of the continuum map from SparsePak data to the
broad-band profile.}
\end{figure}

\clearpage

\begin{figure}
\figurenum{24}
% \epsscale{0.7}
% \plotone{f24xx.ps}
\plotfiddle{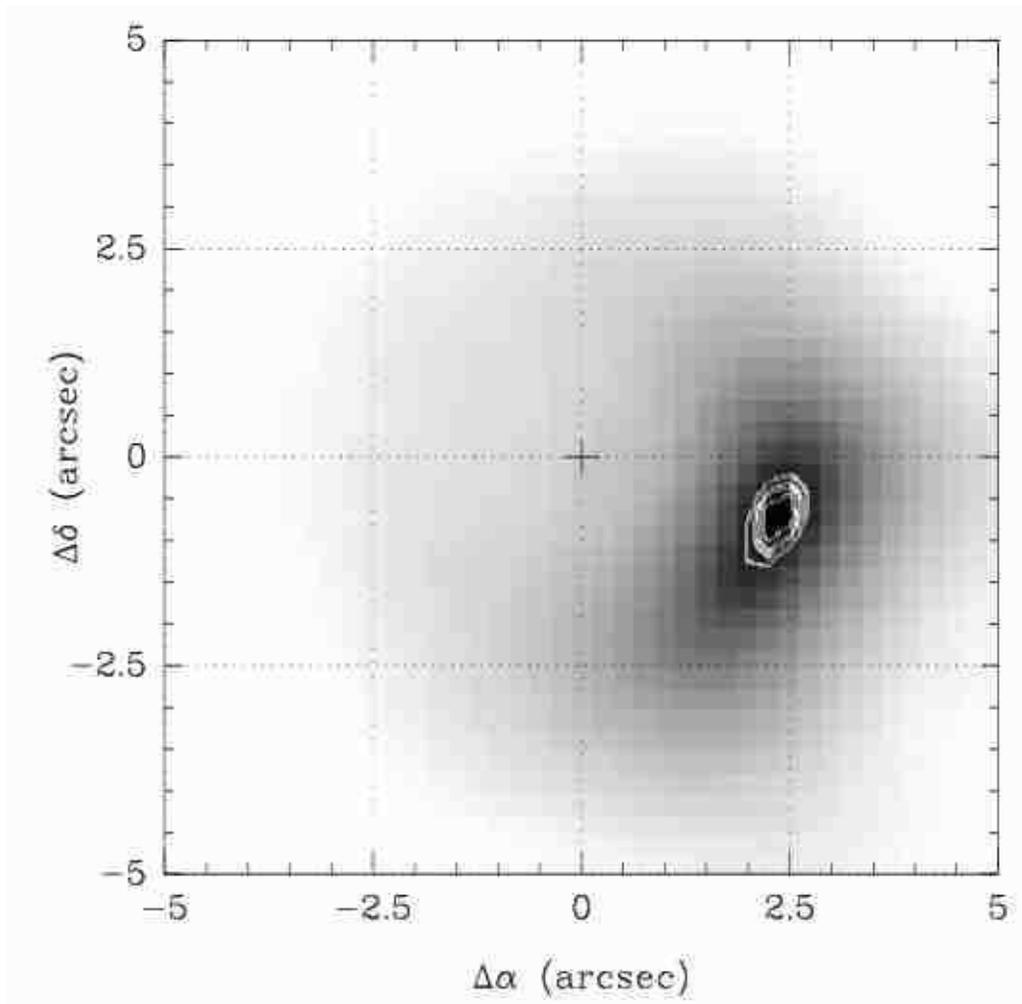}{5in}{0}{70}{70}{-210}{0}
\caption{$\chi^2$ surface, shown using a logarithmic stretch,
defined by the fitting the observed I-band profile to the SparsePak
continuum map as a function of telescope offset in R.A. and
Declination. Contours are plotted at the formal 68, 90, 95, and 99\%
confidence levels. Since the minimum reduced-$\chi^2$ is significantly
greater than 1, these are not rigorously correct, but do illustrate
the steepness of the surface near the minimum.}
\end{figure}

\clearpage

\begin{figure}
\figurenum{25}
% \epsscale{0.8}
% \plotone{f25xx.ps}
\plotfiddle{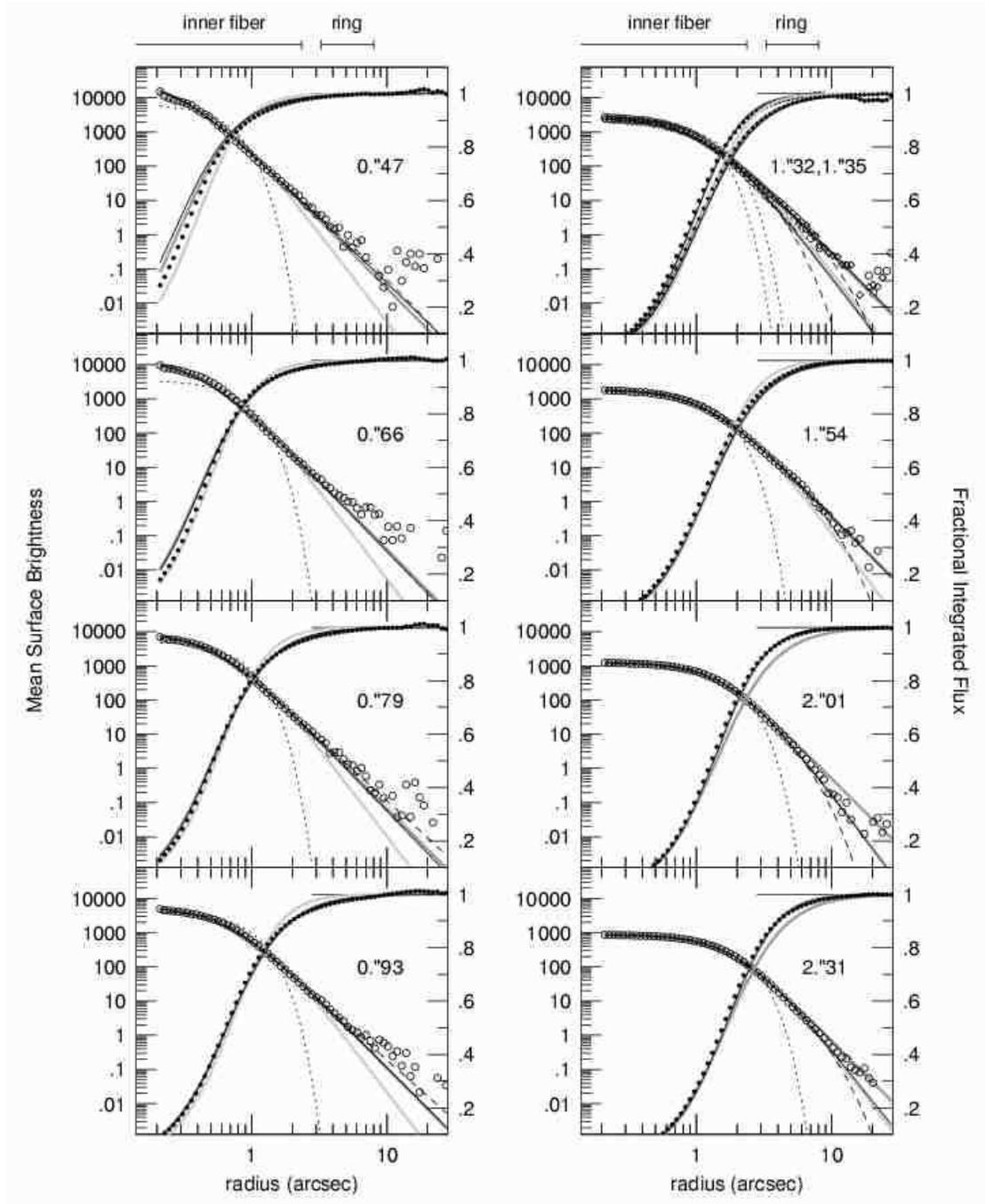}{6in}{0}{75}{75}{-230}{0}
\caption{Observed radial profiles of surface-brightness (open
circles, left scale) and normalized integrated light (filled circles,
right scale) for the 9 seeing cases studied, labeled by FWHM in each
panel. The surface-brightness profiles are normalized to have the same
number of total counts for all profiles.  Moffat models are
represented by solid lines: best fit (dark), $q=2$ (dark grey),
$q=2.6$ (light grey). Gaussian and Lorentz fits to our data are
represented by dotted lines and dashed lines, respectively.  A partial
horizontal line (at unity, right axis) is shown for reference for the
integrated light profiles. At the top of each column of figures, the
radial range of the inner fiber and the annular fiber ring is
indicated (see text for explanation).}
\end{figure}

\clearpage

\begin{figure}
\figurenum{26}
% \epsscale{1.0}
% \plotone{f26xx.ps}
\plotfiddle{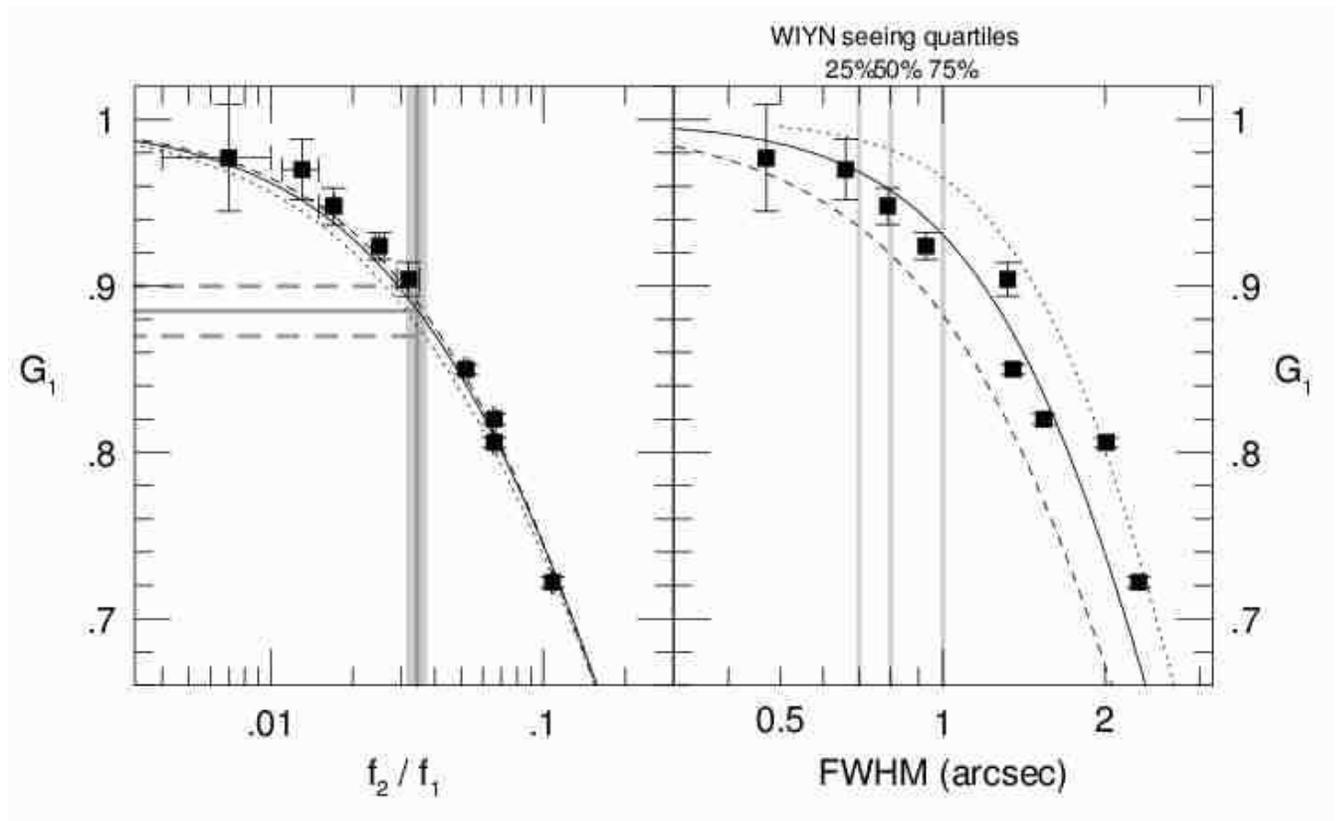}{5in}{0}{90}{90}{-260}{0}
\caption{SparsePak aperture correction, G$_1$, as a function of
the flux ratio f$_2$/f$_1$ (left) and as a function of the seeing FWHM
(right). Points are measured values. Curves are for Moffat functions
with three outer slopes, $q$ which bracket the observations: $q=2.2$
(mean value, solid lines), $q=1.9$ (minimum value, dashed lines), and
$q=2.6$ (maximum value dashed lines). Note the tightness of the
relation for f$_2$/f$_1$ compared to the FWHM. In the left panel, the
grey filled and dashed lines refer to the flux calibration measurement
described in the text. In the right panel, the grey vertical lines
represent the quartiles in WIYN seeing during August 2002 through
January 2003 (C. Corson, private communication).}
\end{figure}

\clearpage

\begin{figure}
\figurenum{27}
% \epsscale{0.65}
% \plotone{f27xx.ps}
\plotfiddle{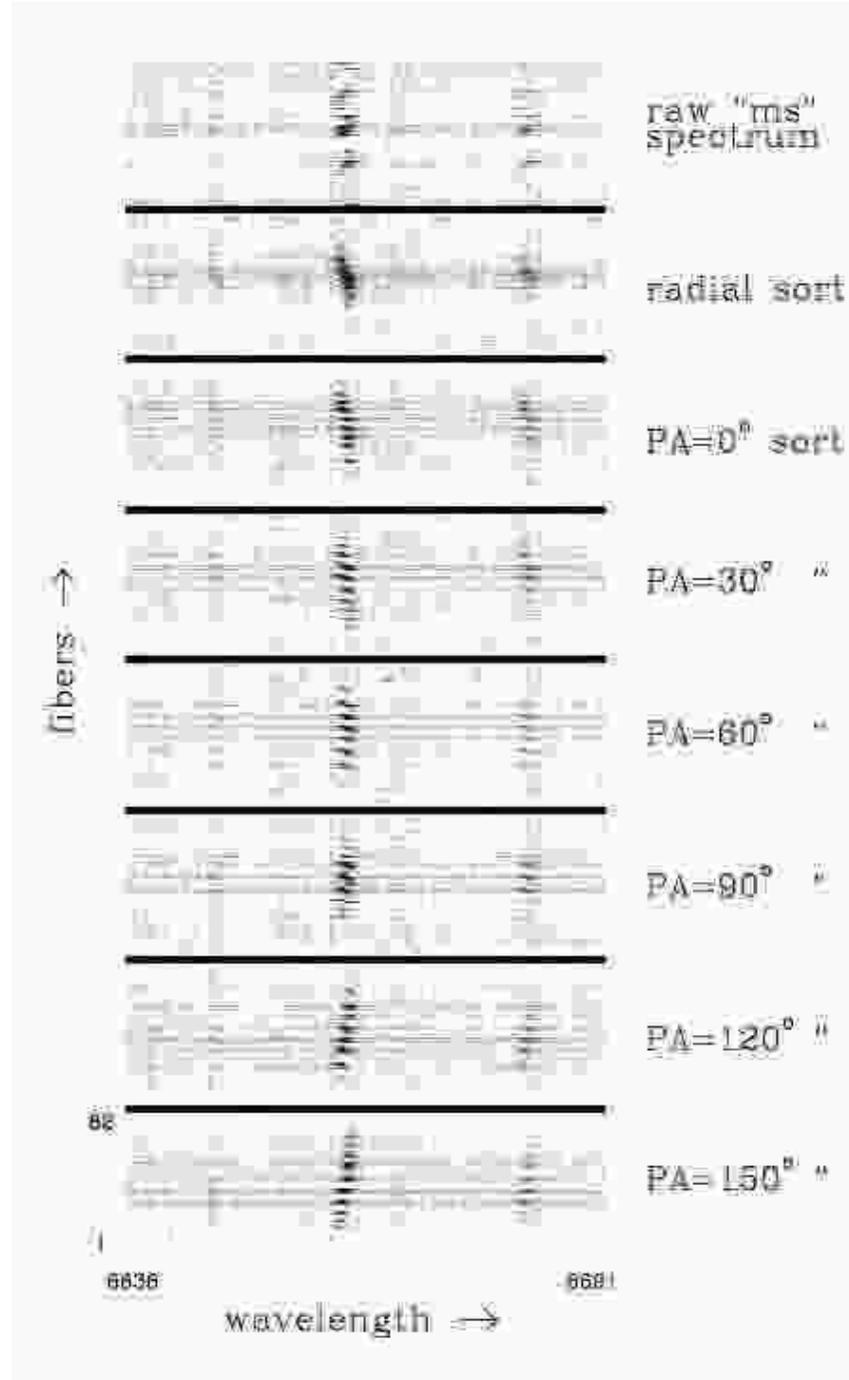}{6in}{0}{65}{65}{-140}{-10}
\caption{Montage of the output of the re-packing tool for
extracted SparsePak spectra. The 8 panels show 275 pixels of spectra
for PGC 56010 for all fibers, centered near H$\alpha$ (the bright
emission feature; [NII]$\lambda$6583.6 is also visible). Each panel
has the fiber spectra sorted in a different fashion, into different
``pseudo-slits'' as noted in the key and explained in the text. The
``radial sort'' clearly shows the central concentration and limits of
the extent of the stellar continuum and nebular emission in this
galaxy. For ``PA = 60$^\circ$'' there is clearly the least coherent
velocity gradient across the fiber pseudo-slit; this is close the
major axis, offset by 90$^\circ$, in agreement to the model fitting
which finds a major axis of $146.1\pm1.5$.}
\end{figure}

\end{document}